\documentclass[12pt]{article}
\usepackage{latexsym}
\usepackage{graphicx}

\usepackage{amsmath}
\numberwithin{equation}{section}

\usepackage{hyperref}
\usepackage{cleveref}

\def\floatcaption#1#2{ \caption{#2 \label{#1}} }

\def\a{\alpha}
\def\b{\beta}
\def\c{\chi}
\def\d{\delta}
\def\e{\epsilon}                
\def\f{\phi}                    
\def\g{\gamma}

\def\j{\psi}

\def\l{\lambda}
\def\m{\mu}
\def\n{\nu}
\def\o{\omega}
\def\p{\pi}                     
\def\r{\rho}                    
\def\s{\sigma}                  
\def\t{\tau}

\def\D{\Delta}

\def\G{\Gamma}
\def\J{\Psi}
\def\L{\Lambda}
\def\O{\Omega}

\def\S{\Sigma}
\def\U{\Upsilon}

\def\ca{{\cal A}}

\def\cc{{\cal C}}
\def\cd{{\cal D}}
\def\ce{{\cal E}}

\def\ch{{\cal H}}   

\def\cm{{\cal M}}
\def\cn{{\cal N}}

\def\cp{{\cal P}}

\def\car{{\cal R}}

\def\ct{{\cal T}}
\def\cu{{\cal U}}
\def\cv{{\cal V}}

\def\cbo{{\,\raise-.15ex\Sc [\,}}                       

\def\sl#1{\rlap{\hbox{$\mskip 1 mu /$}}#1}      
\def\vev#1{\Big\langle #1 \Big\rangle}           

\def\sbra#1{\left\langle #1\right|}             
\def\sket#1{\left| #1\right\rangle}             
\def\svev#1{\left\langle #1\right\rangle}       

\def\ddt#1{{\buildrel {\hbox{\LARGE .\kern-2pt.}} \over {#1}}}

\def\ie{\mbox{\it i.e.}}
\def\eg{\mbox{\it e.g.}}
\def\etc{\mbox{\it etc.}}

\def\leqx{\,\raisebox{-1.0ex}{$\stackrel{\textstyle <}{\sim}$}\,}
\def\geqx{\,\raisebox{-1.0ex}{$\stackrel{\textstyle >}{\sim}$}\,}

\def\half{\frac{1}{2}}
\def\Re{{\rm Re\,}}

\def\det{{\rm det}}

\usepackage[dvipsnames]{xcolor}

\thicklines                         
\def\mynnnext{\bigskip\noindent }

\def\Tab#1{Table~\ref{#1}}
\def\Fig#1{Fig.~\ref{#1}}

\def\bj{\overline\psi}
\def\bJ{\overline\Psi}

\def\bq{\overline{q}}

\def\bp{\bar{p}}
\def\bl{\bar\l}

\def\tT{\tilde{T}}

\def\tW{\widetilde{W}}
\def\tD{\widetilde{D}}
\def\tH{\widetilde{H}}
\def\tG{\tilde\G}
\def\tm{\tilde{m}}
\def\tb{\tilde{b}}
\def\te{\tilde\e}
\def\hH{\hat{H}}

\def\hcH{\hat\ch}
\def\hcT{\hat\ct}

\def\U{{\rm U}}
\def\SU{{\rm SU}}
\def\DWF{DW}
\def\DDWF{D_{\DWF}}
\def\HDWF{H_{\DWF}}
\def\cHDWF{\ch_{\DWF}}
\def\GDWF{G_{\DWF}}
\def\ZDWF{Z_{\DWF}}
\def\ODWF{\O_{\DWF}}
\def\DGW{D_{GW}}
\def\DOV{D_{ov}}

\def\DBNO{D_{BNO}}
\def\mres{m_{\rm res}}

\def\Deff{D_{\rm eff}}
\def\ceff{\c_{\rm eff}}
\def\etamax{\eta_{\rm max}}
\def\lmin{\l_{\rm min}}
\def\lmax{\l_{\rm max}}

\def\Msub{\overline{M}}

\begin{document}

\begin{center}
\vspace{10mm}
{\large\bf Domain wall fermions}\\[8mm]
Thomas Blum$^a$ and Yigal Shamir$^b$\\[8 mm]
$^a$Department of Physics, University of Connecticut,
Storrs, CT 06269, USA\\
$^b$Raymond and Beverly Sackler School of Physics and Astronomy,\\
Tel~Aviv University, 69978, Tel~Aviv, Israel\\[10mm]
\end{center}

\renewcommand{\thefootnote}{*}

\begin{quotation}
We introduce the formulation of domain wall fermions in the context of
lattice QCD.  We prove the recovery of exact chiral symmetry in the limit of
an infinite fifth direction, and derive the effective four-dimensional
operator satisfying the Ginsparg-Wilson relation obtained in this limit.
We discuss the residual breaking of chiral symmetry
for finite extent of the fifth direction, and how it is affected by
spectral features of the Wilson kernel.  We also discuss various improvements
of domain wall fermions including notably M\"obius fermions.
These notes are a chapter contributed to the on-line book ``Lattice QCD
at 50 years'' ({\em LQCD@50}).\footnote{
  {\em LQCD@50} is edited by Tanmoy Bhattacharya, Maarten Golterman,
  Rajan Gupta, Laurent Lellouch, and Steve Sharpe.
}
\end{quotation}

\setcounter{footnote}{0}
\renewcommand{\thefootnote}{\arabic{footnote}}

\newpage
\section{\label{intro} Introduction}
From the beginning of lattice field theory in the mid 1970s it was realized
that putting fermions on the lattice is a difficult problem.
The simplest way to see this is to note that continuum fermions
obey a first-order differential equation, which is linear in the momentum.
For example, in one spatial dimension
the dispersion relation of a single-component massless fermion is $E=\pm p$.
If we discretize space (keeping the time coordinate continuous),
we have to replace the continuum derivative
by a difference operator.  Using the nearest-neighbor lattice
difference operator yields the dispersion relation
$E=\pm(1/a)\sin(ap)$ where $a$ is the lattice spacing.
This dispersion relation conforms to the periodicity of
the lattice's Brillouin zone, $0\le p \le 2\p/a$.  For $ap\ll 1$,
the lattice dispersion relation is $E=\pm p(1+O(ap)^2)$,
which reproduces the linear continuum dispersion relation
up to small discretization effects.  The problem is that
now $E(p)$ vanishes not only for $p=0$ but also for $p=\p/a$.
In the vicinity of the latter point, $E=\mp q(1+O(aq)^2)$ where the momentum
is effectively $q=p-\p/a$.
Hence the dispersion relation is again linear up to discretization effects,
but with an opposite slope, hence, opposite chirality.\footnote{%
  In one spatial dimension, the $\pm$ chirality corresponds to the fermion
  being a right-mover or a left-mover.
}
Instead of the single massless fermion of definite chirality
we initially had in the continuum theory, the lattice theory describes
two massless fermions with opposite chiralities.  This is the well-known
{\em doubling problem}.

Alternatively, one can understand fermion doubling by considering
the implications of anomalies for a lattice theory.
The basic rule is very simple.  The lattice is a non-perturbative regulator.
Hence, any symmetry that is exact on the lattice will remain so
in the continuum limit.  It follows that the lattice theory {\em must not}
have exact chiral symmetry (if the chiral transformation has the same form
as in the continuum) otherwise it will be impossible to reproduce
the axial anomaly in the continuum limit!

The doubling problem was formalized in a series of
``no go'' theorems \cite{Karsten:1980wd,Nielsen:1980rz,Nielsen:1981xu,Pelissetto:1987ad,Karsten:1981gd}, which clarify the conditions under which
doubling is unavoidable.  We comment that fermion doubling can be avoided
by using non-local lattice difference operators.  However, this invariably leads
to various inconsistencies when the fermions are coupled to a gauge field
(see, \eg, Ref.~\cite{Golterman:2000hr} and references therein).

Early on,
two basic strategies emerged for dealing with the doubling problem.
The first strategy is to live with it but to try to minimize it.
This leads to staggered fermions, a single-component lattice field which,
in four dimensions, yields four Dirac fermions in the continuum limit.
Apart from describing several, and not just one,
Dirac fermions, also the flavor symmetry structure of the resulting
continuum theory is not manifest in the lattice theory.
For an introduction to staggered fermions we refer to a companion chapter
of the {\em LQCD@50} book by Golterman \cite{Golterman:2024xos},
as well as to Ref.~\cite{MILC:2009mpl}.

The alternative strategy avoids doubling by breaking chiral symmetry explicitly
on the lattice, and restoring it only in the continuum limit.
This is how Wilson fermions work.
The advantage over staggered fermions is that vectorial flavor symmetries
remain manifest, the same as in the continuum.  The price is that
in the absence of a chiral symmetry to protect it at the lattice level,
the bare mass suffers a large $O(1/a)$ additive renormalization.
Other chirally sensitive
observables suffer similarly large discretization effects.
For an introduction to Wilson fermions we refer to a companion chapter
of the {\em LQCD@50} book by Dalla Brida.

Domain wall fermions were introduced by Kaplan much later, in the early 1990s
\cite{Kaplan:1992bt}.  A posteriori, one might loosely describe the
{\em raison d'etre} of domain wall fermions
as ``maximum chiral symmetry at minimal doubling.''
To what extent this goal has been achieved depends on whether
the target continuum gauge theory is a vector-like theory like QCD;
or, alternatively, a chiral gauge theory.

For QCD the goal has been fully achieved.
Using domain wall fermions for QCD, or for QCD-like theories,
is the subject of this introduction.
In this case a single domain-wall fermion field gives rise to a single
quark field in the continuum limit.  The vector flavor symmetry
is fully preserved as for Wilson fermions.
Chiral symmetry is also fully preserved when a suitable limit is taken,
leaving ``just enough'' room to reproduce the axial anomaly,
as in the continuum.
The main price to pay is that numerical simulations of domain wall fermions
are more costly.

For chiral gauge theories (Kaplan's original motivation)
the situation is much more complicated. Chiral gauge theories are outside
the scope of the {\em LQCD@50} book, and so we will only comment briefly
on this topic in the concluding section.

Earlier texts on fermion methods, including in particular
domain wall fermions, can be found for example
in the book by DeGrand \& DeTar \cite{DeGrand:2006zz}.\footnote{%
  See also Refs.~\cite{Smit:2002ug,Gattringer:2010zz}.
}
Domain wall fermions are also discussed in a set of
lecture notes by Kaplan \cite{Kaplan:2009yg}.
For the closely related overlap fermions, once again we refer
to the DeGrand \& DeTar book, as well as to a review talk by Niedermayer
\cite{Niedermayer:1998bi}.  Overlap fermions are also covered in
a companion chapter of the {\em LQCD@50} book,
written by DeGrand \cite{DeGrand:2025ttq}.
We refer to his chapter in particular for a more detailed discussion
of the Ginsparg-Wilson relation \cite{Ginsparg:1981bj} and its consequences.

\section{\label{cont} Why domain wall fermions}
Domain wall fermions are a system in which a light fermion field
is attached to a $d$-dimensional ``defect'' of a massive fermion theory
in $d+1$ dimensions.
For $d<4$, domain wall fermions play a role in condensed matter physics.
In this introduction we will consider only the case $d=4$ relevant for
elementary particle physics.  We begin in this section by introducing
domain wall fermions in continuum $(4+1)$-dimensional euclidean space.
In the next section we will turn
to the lattice formulation of domain wall fermions.

Consider the following Dirac equation
\begin{equation}
\label{cont5}
(i{\sl{p}} + \g_5 \partial_5 + m(s)) \c = 0 \ ,
\end{equation}
where we work in momentum space for the four physical coordinates,
and in coordinate space for the fifth coordinate $s$.
At this stage the fifth coordinate can take any value, $-\infty<s<\infty$.
We assume that the mass profile is
\begin{equation}
\label{ms}
m(s) = \left\{ \begin{array}{ll}
    -M'\qquad & s\le 0 \ , \\
     M\qquad & s\ge 0 \ ,
     \rule{0ex}{3ex}
     \end{array}\right.
\end{equation}
where $M,M'>0$.  The discontinuity of the mass profile at $s=0$ is
called a domain wall.

The Dirac equation~(\ref{cont5}) has a right-handed (RH) zero mode.
This homogeneous solution of the equation is a normalizable function of
the fifth coordinate $s$, which is bound to the ``domain wall'' at $s=0$,
\begin{equation}
\label{zmcont}
\c(s) = e^{-m(s)} = \left\{ \begin{array}{ll}
    e^{M' s} \qquad & s\le 0 \ , \\
    e^{-M s} \qquad & s\ge 0 \ .
     \rule{0ex}{3ex}
     \end{array}\right.
\end{equation}
Notice the absence of a similar left-handed (LH) zero mode:
the homogeneous solution $e^{+m(s)}$ is not normalizable.
Had we flipped the signs of $M$ and $M'$ simultaneously,
the LH homogeneous solution would be normalizable, hence, a zero mode,
whereas the RH homogeneous solution would be non-normalizable.

The five-dimensional fermion system has a conserved fermion number symmetry.
This symmetry is neither vectorial nor axial: these notions exist only
for an even number $d$ of dimension, where one can construct
a chirality matrix $\propto \g_1 \g_2 \cdots \g_d$, which anti-commutes
with each of the individual $d$-dimensional Dirac matrices.
However, let us consider the effective low-energy theory associated
with a five-dimensional domain wall fermion.
For the (free) domain-wall fermion system introduced above,
this will be a (free) four-dimensional theory
consisting of a single RH massless fermion.  The effective four-dimensional
theory inherits the fermion number symmetry; but since this symmetry has now
only a RH fermion to act on, it effectively becomes a chiral symmetry!

Kaplan's original idea was the realization that, since chiral symmetry
does not exist in a continuum five-dimensional system,
it should be possible to put this theory on the lattice using Wilson fermions
without giving up on any (internal) symmetry of the continuum theory
and without encountering fermion doubling.  As long as the transition
to the lattice preserves the zero mode that is bound to the domain wall,
one expects that chiral symmetry will emerge in the low-energy effective theory
on the lattice, just as it does in the continuum.

Infinities can be elusive, however.  Before we turn to the lattice,
let us try to recover the fermion system with an unbounded $s$ coordinate
introduced above as a limit of a finite-extent fifth dimension.
To this end we will assume that the fifth coordinate takes values
$-L_5\le s\le L_5$ with periodic boundary conditions in the fifth direction.
The question is whether the limit $L_5\to\infty$ reproduces the
system discussed above, or not.  The mass profile introduced in
Eq.~(\ref{ms}) describes a single domain wall at $s=0$.  But,
once we impose periodic boundary conditions in the $s$ direction
on the fermions, this effectively introduces {\em another} domain wall
at $s=L_5=-L_5$.  It is easy to see that, if we disregard
exponentially small effect on the order of $e^{-ML_5}$ or $e^{-M'L_5}$,
we now have not only a RH zero mode bound to the domain wall at $s=0$,
but also a LH zero mode bound to the new domain wall at $s=L_5=-L_5$.
For $L_5\to\infty$ both zero modes become exact.  The $L_5\to\infty$ system
thus has one massless Weyl field of each chirality.  This is different from
the result of the earlier, formal treatment of the infinite-$s$ case,
in which only one of the two zero modes, the one at $s=0$, was ``visible.''

As we will briefly discuss in the concluding section, the existence of an
opposite-chirality zero mode within a careful treatment of the fifth direction
is a major obstacle if the goal is to construct a chiral gauge theory.
By contrast, for QCD this is not a problem at all, since each
quark field consists of a RH and a LH component---just the light
field content of the finite-$L_5$ system.  Moreover, as we will see later on,
the geometric separation of the RH and LH fields in the fifth direction
still allows for global chiral symmetry to emerge in the limit $L_5\to\infty$.

There is one more observation to make before we turn to the lattice theory.
The parameters $M$ and $M'$ in the mass profile~(\ref{ms}) are independent.
Since we are ultimately interested in the physics of the light mode(s)
bound to the domain wall(s), we can obtain a more economic formulation
as follows.  Focusing first on the RH zero mode near the $s=0$ domain wall,
let us examine the effect of increasing $M$ or $M'$.  For $M'\to\infty$,
the zero mode will be supported on $s\ge 0$ only,
whereas for $M\to\infty$ it will be supported on $s\le 0$ only.
As for the LH zero mode, in the first case it will be supported
on $s\leqx L_5$, while in the latter case it will be supported on $s\geqx -L_5$.

Either way, we can discard half of the range of the fifth coordinate.
We end up with fifth coordinate $0\le s\le L_5$, or alternatively,
$-L_5\le s\le 0$, with a RH and a LH zero modes that are now attached
to the two four-dimensional boundaries.  In the continuum,
sending $M\to\infty$ or $M'\to\infty$ are equally valid choices.
But as we will soon see, when we latticize the system
only one of these choices works.

In the next section we turn to the lattice formulation of
domain wall fermions.  For a more extensive discussion of continuum
domain wall fermions, see the lecture notes of Ref.~\cite{Kaplan:2009yg}.

\section{\label{free} Lattice domain wall fermions: free theory}
In this section we discuss free domain wall fermions in $4+1$ dimensions,
following Ref.~\cite{Shamir:1993zy}.  As a first step, in Sec.~\ref{semi5} we consider
a five-dimensional lattice with a semi-infinite fifth coordinate.
We show that a single massless Weyl fermion exists
on the four-dimensional boundary, for a suitable range of parameters.

When using a non-perturbative regularization such as the lattice,
one must consider carefully
not just the continuum limit but also the infinite volume limit.
In particular, a semi-infinite fifth direction is an idealization,
which must be reached as a suitable limit of a finite but large fifth direction.
In Sec.~\ref{finite5} we discuss a finite fifth direction,
and show that a second Weyl fermion with the opposite chirality
exists on the ``far'' boundary.  This is similar to what we already found
in Sec.~\ref{cont} for the continuum theory.  We then confirm that
the limit of a semi-infinite fifth direction
can be reached without a ``penalty'' in the free theory.
This situation will change in Sec.~\ref{interacting},
where we consider the interacting theory.
We also explain how to combine the chiral fermions that reside
on the two boundaries into a light Dirac fermion with an adjustable mass.

The domain-wall fermion operator is a five-dimensional Wilson operator
designed to produce light chiral fields on four-dimensional ``defects,''
which, in our constructions, are just the four-dimensional boundaries
of the five-dimensional bulk.  When we speak about the Wilson (or
Wilson-Dirac) operator, we will always refer to the four-dimensional
operator unless explicitly stated otherwise.

\subsection{\label{semi5} Semi-infinite fifth direction}
The domain-wall fermion matrix on a semi-infinite fifth direction is
\begin{equation}
\label{semiinf}
\DDWF = \left(\begin{array}{cccccc}
D-1 & P_R & 0   & 0   & 0   & \ldots \\
P_L & D-1 & P_R & 0   & 0   & \ldots \\
0   & P_L & D-1 & P_R & 0   & \ldots \\
0   & 0   & P_L & D-1 & P_R &  \\
\vdots & \vdots & & \hspace{-4ex}\ddots\hspace{4ex}
& \hspace{-2.5ex}\ddots\hspace{2.5ex} & \ddots
\end{array}\right) ,
\end{equation}
This matrix structure corresponds to the fifth coordinate $s$, which
we assume to take values $s=1,2,\cdots$.
We will mostly work in units of the four-dimensional lattice spacing,
equivalently, set $a=1$.  For now, we keep the lattice spacing for the
fifth direction $a_5$ equal to the four-dimensional one.\footnote{%
  For generalizations that allow for $a_5/a\ne 1$, see Sec.~\ref{improve}.
}
Each entry is a $4\times 4$ matrix in Dirac space.
In addition, if the lattice dimensions are
$L_1\times L_2\times L_3\times L_4$ then each entry is also a $V\times V$ matrix
in four-dimensional space, with $V=L_1 L_2 L_3 L_4$.
$P_{R,L}=\half(1\pm\g_5)$ denote the chiral projectors.
$D$ is the four-dimensional Wilson-Dirac operator
\begin{equation}
\label{DW}
   D = D_K + M-W \ ,
\end{equation}
where in the free theory
\begin{eqnarray}
\label{DK}
D_K(x,y) &=& \half \sum_\m \left[\d_{x+\hat\m,y} - \d_{x-\hat\m,y} \right] \g_\m\ ,
\\
\label{W}
W(x,y) &=&
  \half \sum_\m \left[2\d_{xy} - \d_{x+\hat\m,y} - \d_{x-\hat\m,y} \right] \ .
\end{eqnarray}
We use Greek indices for the four physical dimensions,
and the sums are over $\m=1,\ldots,4$.  The $\g_\m$ matrices
satisfy the euclidean Dirac algebra $\{\g_\m,\g_\n\}=2\d_{\m\n}$.
The four-dimensional coordinates are labeled $x,y$, while $\pm\hat\m$
moves to the next site in the positive or negative $\m$ direction.
Here $D_K$ is the usual nearest-neighbor discretization of the massless
continuum Dirac operator $\sl\partial$, while the Wilson term $W$
is a lattice discretization of the continuum laplacian.
In momentum space
\begin{eqnarray}
\label{DKp}
D_K(p) &=& i\sum_\m \g_\m \sin(p_\m) \ ,
\\
\label{Wp}
W(p) &=& \sum_\m(1-\cos(p_\m)) \ .
\end{eqnarray}
We will also use the notation $D_K(p)=i{\sl\bp}$,
with $\bp_\m = \sin(p_\m)$.  Taken alone, the operator $D_K(p)$ exhibits
the familiar doubling problem, as it vanishes in all 16 ``corners'' $p_0$
of the Brillouin zone, defined by $\sin(p_{0,\m})=0$ for all $\m$.
The Wilson term is a momentum dependent mass term that gives the 15 doublers
a mass proportional to the lattice cutoff at the price of
breaking chiral symmetry explicitly, while the fermion at the origin
of the Brillouin zone remains massless in the free theory.

Apart from the presence of boundaries, the domain-wall fermion operator
is just the Wilson-Dirac operator in five dimensions, whose structure
follows immediately by taking the coordinates and the index $\m$ in
Eqs.~(\ref{DK}) and~(\ref{W}) to be five-dimensional.  In any dimension,
the most general form of the Wilson-Dirac operator is $D_K + M - rW$
where $r$ is the Wilson parameter.
For $r=1$, hopping in the fifth direction\footnote{
  Hopping in the physical $\m$ direction is similarly constrained
  by the projectors $(1\pm\g_\m)/2$ for $r=1$.
}
is constrained by the chiral projectors $P_{R,L}$,
as seen in Eq.~(\ref{semiinf}).
When $r\ne 1$ the hopping in the fifth direction is not constrained
by the chiral projectors and the structure of the zero modes
gets more complicated \cite{Shamir:1993zy}.
In this introduction we set $r=1$ unless otherwise stated.

In the context of domain wall fermions, the Wilson-Dirac operator $D$
of Eq.~(\ref{DW}) is often called the {\em Wilson kernel}.
Notice that the Wilson term $W$ is a positive operator.
The parameter $M$, called the {\em domain wall height},
is given by $M=-m_0$, where $m_0$ is the bare Wilson fermion mass
in the convention where $W+m_0 \ge 0$ when $m_0 \ge 0$.
We recall that because of the explicit breaking of chiral symmetry by the
Wilson term, the bare mass $m_0$ suffers a large additive renormalization
when lattice QCD is formulated using Wilson fermions.
Features of domain wall fermions which are related
to the way the Wilson-fermion bare mass renormalizes will be discussed
later on, in Sec.~\ref{wavef} and Sec.~\ref{quench}.

As we will see next, for domain wall fermions we need positive $M$,
or equivalently, negative $m_0$.  This implies that the operator $W-M=W+m_0$
does not have a definite sign, a property called {\em super-criticality}
of the Wilson kernel.
Apart from playing a crucial role in the very existence
of the light chiral modes, super-criticality has additional
important consequences that we will encounter later in this introduction.\footnote{%
  Super-criticality of the Wilson kernel is equally crucial
  for overlap fermions \cite{Neuberger:1997fp,Niedermayer:1998bi}
}

In order to find the spectrum of domain wall fermions we go to momentum space
for the four physical dimensions.
The domain-wall fermion matrix~(\ref{semiinf}) becomes
\begin{equation}
\label{semip}
\DDWF(p) = \left(\begin{array}{cccccc}
-b(p)+i{\sl\bp} & P_R & 0 & 0 & 0 & \ldots   \\
P_L & -b(p)+i{\sl\bp} & P_R & 0 & 0 & \ldots  \\
0 & P_L & -b(p)+i{\sl\bp} & P_R & 0 & \ldots  \\
0 & 0 & P_L & -b(p)+i{\sl\bp} & P_R &         \\
\vdots & \vdots & & \hspace{-9ex}\ddots\hspace{9ex}
& \hspace{-6ex}\ddots\hspace{6ex} & \ddots
\end{array}\right) ,
\end{equation}
where
\begin{equation}
\label{bp}
  b(p) = 1-M+W(p)\ .
\end{equation}
The 16 corners of the Brillouin zone are given explicitly by
$p_0 = \p (n_1,n_2,n_3,n_4)$ where $n_\m=0,1$.
At each corner of the Brillouin zone $\DDWF$ commutes with $\g_5$,
and the equation $P_R \DDWF\, \c_R = 0$ has two linearly independent
RH homogeneous solutions\footnote{
  It is easy to verify that there are no LH homogeneous solutions
  at any corner of the Brillouin zone.
  Consider the equation $P_L \sum_{s'} \DDWF(s,s') \c_L(s') = 0$.
  For $s=1$ the only solution is $\c_L(1)=0$.
  Using this, we next find $\c_L(2)=0$, and so on.
}
$\c_{(\a)}$ given explicitly by
\begin{equation}
\label{RHsol}
\c_{(\a)\b}(s) = (P_R)_{\a\b}\, b^s(p_0) \ , \qquad b(p_0) = 1+2n-M \ ,
\end{equation}
where $n$ is the number of components of the momentum $p_0$ equal to $\p$.
The spinor indices take values $\a,\b=1,2$, consistent with the
two degrees of freedom of a four-dimensional Weyl field.
Notice the double role of the chiral projector $P_R$.
The notation $(P_R)_{\a\b}$ refers to the nonzero
diagonal block of $P_R$ proportional to the $2\times 2$ identity matrix.
In order for the homogeneous solution to be a zero mode that belongs to
the physical spectrum, it must be normalizable.
Remembering that the range of the fifth coordinate is semi-infinite,
this will be true if and only if $|1+2n-M|<1$,
which in turn constrains the range of the domain wall height $M$
as a function of $n$ \cite{Jansen:1992tw}.

The zero modes spectrum is summarized in \Tab{zms}.
The first column is $n$, which ranges from 0 to 4.
The second column gives the range of domain wall height where,
as a function of $n$, the homogeneous solution~(\ref{RHsol}) is normalizable.
In particular, for $n=0$ this range is $0<M<2$.
The next column gives the number of corners,
hence the number of (pairs of) linearly independent zero modes
with the given value of $n$.  The last column gives
the {\em physical} chirality of the zero modes.
Notice that while Eq.~(\ref{RHsol}) has a fixed chiral projector $P_R$, the
physical chirality alternates between RH and LH,
depending on whether $n$ is even or odd.
In brief, the reason is that the way to determine the continuum-limit
properties of the massless fermion near a given corner $p_0$ is to first
apply a discrete transformation that brings $p_0$ to the origin of the
Brillouin zone. These discrete transformation are constructed from
the Dirac matrices, and act on the wave function of the fermions as well.
Hence, the chirality of the transformed wave function can be flipped.
For a detailed explanation see Refs.~\cite{Karsten:1980wd,DeGrand:2006zz}.

\begin{table}[t]
\vspace*{2ex}
\begin{center}
\begin{tabular}{ c|ccc } \hline
  $n$ & $M$ range & \#(z.m.) & chirality \\ \hline
  0 & (0,2)  & 1 & RH \\
  1 & (2,4)  & 4 & LH \\
  2 & (4,6)  & 6 & RH \\
  3 & (6,8)  & 4 & LH \\
  4 & (8,10) & 1 & RH \\
\hline
\end{tabular}
\end{center}
\begin{quotation}
\caption{\label{zms}
The zero modes spectrum on a semi-infinite lattice
as a function of $n$, the number of momentum components equal to $\p$.
See text for further explanations.
}
\end{quotation}
\vspace*{-4ex}
\end{table}

Before we continue, we digress to compare the situation on the lattice
with the continuum zero mode of Eq.~(\ref{zmcont}).  Instead of a semi-infinite
fifth direction, we momentarily consider an infinite fifth direction,
with domain-wall height $M$ for $s\ge 1$ as before, and a different
domain-wall height $-M'$ for $s\le 0$, with $M'>0$.
Omitting the spinor indices, the RH
homogeneous solutions at the corners of the Brillouin zone are now\footnote{%
  For $n=0$, the continuum wave function~(\ref{zmcont}) can be reproduced
  by reintroducing the lattice spacing in the fifth direction $a_5$
  and taking the limit $a_5\to 0$.  We leave this exercise to the reader.
}
\begin{equation}
\label{dwlatt}
\c_R(s) = \left\{ \begin{array}{ll}
   (1+2n+M')^{s-1} \qquad & s\le 0 \ , \\
   (1+2n-M)^{s-1} \qquad & s\ge 1 \ .
     \rule{0ex}{3ex}
     \end{array}\right.
\end{equation}
Since $M'>0$, the wave function for negative $s$ is always normalizable.
As for the wave function for positive $s$, it will be normalizable under
the same condition as before, namely, if and only if $|1+2n-M|<1$.
Adding the negative-$s$ region has thus not changed the fact
that the zero modes spectrum of \Tab{zms} is controlled by the
positive-$s$ region only.  The negative-$s$ half-space carries
no benefit with it, and we will thus avoid it.

Returning to the semi-infinite fifth direction with $s\ge 1$,
in lattice QCD with domain wall fermions one always dials
the domain wall height to have a single zero mode at the origin of the
Brillouin zone, the case $n=0$ in \Tab{zms}.  Disregarding the overall
normalization, the zero mode wave function is then $\c_R(s) = P_R (1-M)^s$.
While as already noted, this solution is normalizable for $0<M<2$,
the region $1<M<2$ gives rise to an oscillatory behavior as a function of $s$,
which we would like to avoid.  Hence we will further restrict $0<M\le 1$.
The special case $M=1$ is peculiar: the wave function becomes completely
localized on the boundary layer, $\c_R(s) = \d_{s,1}$.
This ideally localized wave function is only possible in the free theory.
In Sec.~\ref{resid} we will see how the wave function gets modified
in the interacting theory, as well as how this affects the optimal choice
of the domain wall height $M$.

We next show that the pair of zero modes attached to the boundary that
we have at $p=0$ when $0<M\le 1$ correspond
to a four-dimensional Weyl fermion.  One way to demonstrate this is
to consider momentarily the domain wall hamiltonian
\begin{equation}
\label{HDW}
\HDWF(\vec{p}) = \g_4 \DDWF(\vec{p};p_4=0) \ .
\end{equation}
We seek the bound states of $\HDWF$ localized near the boundary.
We parametrize an eigenstate as
\begin{equation}
\label{bswf}
\J(\vec{p},s) = P_R\, \c_R(s;\vec{p}) \f(\vec{p}) \ ,
\end{equation}
where $\f(\vec{p})$ is a two-component spinor.\footnote{
  Strictly speaking, $\f(\vec{p})$ is the RH part of
  a four-component spinor whose LH part vanishes identically.
}
We now take $\c_R(s;\vec{p})=b^s(\vec{p})$, requiring that
this wave function is normalizable.  In this introduction we use the
chiral representation of the Dirac matrices (in $2\times 2$ block notation)
\begin{equation}
\label{dirac}
\g_k = \left(\begin{array}{cc} 0 & i\s_k \\ -i\s_k & 0 \end{array} \right)\ ,
\qquad
\g_4 = \left(\begin{array}{cc} 0 & 1 \\ 1 & 0 \end{array} \right)\ ,
\qquad
\g_5 = \left(\begin{array}{cc} 1 & 0 \\ 0 & -1 \end{array} \right)\ ,
\end{equation}
with $\s_k$ the Pauli matrices.  In particular,
\begin{equation}
\label{HK}
\g_4 D_K(\vec{p};p_4=0) = i \sum_{k=1}^3 \g_4\g_k \bp_k \ ,
\end{equation}
where
\begin{equation}
\label{sigmap}
i\g_4\g_k \bp_k =
\left(\begin{array}{cc} \s_k \bp_k & 0 \\ 0 & -\s_k \bp_k \end{array} \right)
= \g_5
\left(\begin{array}{cc} \s_k \bp_k & 0 \\ 0 & \s_k \bp_k \end{array} \right) \ ,
\end{equation}
exhibiting the familiar connection between chirality and helicity eigenstates.
The part of $\HDWF(\vec{p})$ that anti-commutes with $\g_5$
annihilates $\c_R(s;\vec{p})$, hence also $\J(\vec{p},s)$,
and it follows that
the three-dimensional wave function is a RH helicity eigenstate,
\begin{equation}
\label{helicity}
\sum_{k=1}^3 \s_k \bp_k \, \f(\vec{p}) = E(\vec{p}) \f(\vec{p}) \ .
\end{equation}
This interpretation is valid for $p_k\ll 1$ in lattice units,
which allows us to approximate $\bp_k = p_k$.

The chiral bound states localized near the boundary exist
for a range of lattice momenta which is constrained by $|b(\vec{p})|<1$.
As long as we take $M=O(1)$, this condition defines an open subset of
the Brillouin zone which includes the origin, and which is also $O(1)$
in lattice units; thus it becomes an infinite range in the continuum limit.

For any $\vec{p}$, in addition to the single bound state,
the hamiltonian $\HDWF(\vec{p})$ has a continuous spectrum as well.
Because of complete reflection at the boundary of the semi-infinite
fifth dimension, the continuum eigenstates are standing waves.
When $|b(\vec{p})|=1$, the wave function $\c(s;\vec{p})$ of the bound state
becomes delocalized, and its energy reaches the threshold
of the continuous spectrum.

Returning to euclidean space, we can alternatively establish the existence
of the massless RH Weyl fermion near the four-dimensional boundary
by examining the singularity structure of the domain wall propagator.
We start with
the five-dimensional Wilson operator on an infinite fifth direction,
\begin{equation}
\label{D5}
D_5 = P_R\,\d_{s+1,s'} + P_L\,\d_{s-1,s'} + (-b(p)+i{\sl\bp})\d_{s,s'} \ .
\end{equation}
As before, we work in momentum space for the four physical dimensions.
As an operator acting on the fifth coordinate,
the inverse of $D_5$ can be expressed as $D_5^\dagger G_5$,
where $G_5$ is the inverse of the (discretized) second-order operator
\begin{equation}
\label{D5D5dag}
\O_5 \equiv
D_5 D_5^\dagger = -b(p)(\d_{s+1,s'} + \d_{s-1,s'}) + (b^2(p)+\bp^2+1)\d_{s,s'} \ ,
\end{equation}
Explicitly
\begin{eqnarray}
\label{G5}
G_5 &=& \cn e^{-\a(p)|s-s'|} \ ,
\\
\cn^{-1} &=& 2 b(p) \sinh\a(p) \ ,
\nonumber
\end{eqnarray}
where $\a(p)$ is the positive solution of
\begin{equation}
\label{alphap}
2\cosh\a(p) = \frac{1+b^2(p)+\bar{p}^2}{b(p)} \ .
\end{equation}
Notice that $e^{\pm\a(p)s}$ are the homogeneous solutions of
the equation $D_5 D_5^\dagger\,\c = 0$,
while $G_5$ is the Green function that satisfies physical
boundary conditions: it vanishes when the separation $|s-s'|$
tends to infinity.  We have required $0<M\le 1$, hence $b(p)\ge 0$.
For $p\to 0$, both $b(p)$ and $e^{-\a(p)}$ tend to $1-M$.
Below, we will need the $O(p^2)$ corrections,
\begin{subequations}
\label{abpsq}
\begin{eqnarray}
\label{bpsq}
b(p) &=& 1-M + \frac{p^2}{2} +\cdots \ ,
\\
\label{apsq}
e^{-\a(p)} &=& 1-M - \frac{M^2-4M+2}{2M(2-M)}\,p^2 +\cdots \ .
\end{eqnarray}
\end{subequations}
Expanding $b(p)$ is trivial.
For $e^{-\a(p)}$, we write $x \equiv e^{-\a} = 1-M+cp^2$,
and solve Eq.~(\ref{alphap}) to $O(p^2)$ for $c$,
noting that the left-hand side is $x+x^{-1}$.

We now turn to the domain wall propagator $\GDWF$.
As before, it can be expressed as
$\GDWF = \DDWF^\dagger G$, where now $G$ is the inverse of
\begin{equation}
\label{DDdag}
\ODWF \equiv \DDWF \DDWF^\dagger = P_R \O_+ + P_L \O_- \ .
\end{equation}
The explicit form of the operator $\ODWF$ on a finite lattice
is given in Eq.~(\ref{DDdagN5}) below.
The operators $\O_\pm$ carry no Dirac indices,
and so $G$ admits a similar decomposition
\begin{equation}
\label{G}
G = P_R G_+ + P_L G_- \ .
\end{equation}
Several considerations help us in constructing $G_\pm$.
First, in the absence of a boundary we already know the inverse
of the second-order operator $\O_5$, it is $G_5$.
Hence we expect $G_\pm = G_5 + \cdots$, where the ellipsis stands for
an additional term that corrects for the existence of the boundary.
Now, for $s$ away from the boundary we have
\begin{equation}
\label{propG}
\sum_{s''} \ODWF(s,s'') G_5(s'',s') =
\sum_{s''} \O_5(s,s'') G_5(s'',s') = \d_{s,s'} \ .
\end{equation}
This relation is true except when $s$ is right on the boundary,
namely except for $s=1$.  The reason is that both $\DDWF$
and $\ODWF=\DDWF\DDWF^\dagger$ contain only nearest-neighbor couplings.
In order not to spoil the already correct behavior, the new term must somehow
be formed from the homogeneous solutions of the translationally covariant
second-order operator $\O_5 = D_5 D_5^\dagger$.
Next, as follows from Eqs.~(\ref{DDdag}) and~(\ref{DDdagN5}),
the operators $\O_\pm=(\O_\pm)_{ss'}$ are symmetric in the
indices $s$ and $s'$, hence the same must be true
for the new term.  Last, this term must exhibit sensible physical behavior
when the fifth coordinate tends to infinity.
Suppressing the $p$ dependence, this implies\footnote{%
  The slight difference between Eq.~(\ref{Gpm}) and the corresponding equation
  in Ref.~\cite{Shamir:1993zy} is because here we assume $s=1,2,\cdots$,
  while Ref.~\cite{Shamir:1993zy} assumes $s=0,1,\cdots$.
}
\begin{equation}
\label{Gpm}
G_\pm(s,s') = G_5(s,s') + A_\pm \, e^{-\a(s+s'-2)} \ ,
\qquad s,s'\ge 1 \ .
\end{equation}
where the above considerations determine the form of the new term
up to the proportionality constants $A_\pm$.

It remains to determine the amplitude $A_\pm$ by requiring
that $G_\pm$ is the inverse of $\O_\pm$, including in particular
at the boundary.  The result is
\begin{eqnarray}
\label{Ap}
A_+ &=& -\cn\,e^{-2\a} \ ,
\\
\label{Am}
A_- &=& \cn\,e^{-2\a}\,\frac{e^{\a}-b}{b-e^{-\a}} \ .
\end{eqnarray}
The amplitudes $\cn$ (Eq.~(\ref{G5})) and $A_+$ are regular for all values of $p$,
but $A_-$ has a pole.  Using Eq.~(\ref{abpsq}) we find
\begin{equation}
\label{Amsngl}
A_-(p) = \frac{M(2-M)}{p^2}+ \cdots\ , \qquad p \to 0 \ .
\end{equation}
The Ellipsis stands for regular terms.
This leads to a pole in the domain wall propagator as well,
\begin{eqnarray}
\label{GF}
\GDWF &=& \DDWF^\dagger G \ = \ \DDWF^\dagger P_L G_- + \cdots
\\
&=& \rule{0ex}{4ex}
-i P_R \,\frac{\sl{p}}{p^2}\, M(2-M)(1-M)^{s+s'-2} + \cdots \ .
\nonumber
\end{eqnarray}
Note that $(M(2-M))^{1/2}(1-M)^{s-1}$ is the normalized zero mode for $p\to 0$.
The singularity of the domain wall propagator clearly exhibits the presence
of a massless RH chiral fermion near the boundary, a result which is valid
for the semi-infinite fifth direction.  In the next subsection,
we will see how this result changes when the fifth direction is finite.

\subsection{\label{finite5} Finite fifth direction}
We next consider a finite fifth direction, $s=1,2,\cdots,N_5$,
always assuming that $N_5$ is even.  The domain-wall fermion matrix is
\begin{equation}
\label{finites}
\hspace{-3ex}
\DDWF = \left(\begin{array}{ccccccccc}
D-1 & P_R & 0   & 0   &\hspace{1ex}\cdots\hspace{1ex}& 0 & 0 & 0 & -m P_L \\
P_L & D-1 & P_R & 0   & \cdots & 0 & 0 & 0 & 0 \\
0   & P_L & D-1 & P_R & \cdots & 0 & 0 & 0 & 0 \\
\rule{0ex}{4ex}
\vdots&\vdots& & \hspace{-3ex}\ddots\hspace{3ex} & \ddots &
\hspace{3ex}\ddots\hspace{-3ex} & &\vdots&\vdots \\
\rule{0ex}{4ex}
0   & 0   & 0   & 0   & \cdots & P_L & D-1 & P_R & 0   \\
0   & 0   & 0   & 0   & \cdots & 0   & P_L & D-1 & P_R \\
-m P_R & 0 & 0  & 0   & \cdots & 0   & 0   & P_L & D-1
\end{array}\right)
\end{equation}
The Wilson operator satisfies $\g_5$-hermiticity, namely,
$\g_5 D \g_5 = D^\dagger$, and thus $\g_5 D$ is hermitian.
Related, a hermitian version of the domain-wall fermion operator is
$\cHDWF=\car\g_5\DDWF$,
where $\car$ is the reflection on the 5th coordinate: $\car(s) = N_5+1-s$.
Equivalently, $\DDWF^\dagger = \car\g_5\DDWF\car\g_5$.
It follows that $\det(\DDWF)$ is real.

In addition to the RH zero mode $\c_R(s) \propto P_R (1-M)^s$ localized
near the $s=1$ boundary, now there is also a LH zero mode
$\c_L(s) \propto P_L (1-M)^{N_5-s}$ localized near the $s=N_5$ boundary.
Neither of these modes are exact: the boundary
at $s=N_5$ somewhat disrupts the RH zero mode, and vice versa.
But the effect is exponentially small, proportional to $(1-M)^{N_5}$,
and vanishes for $N_5\to\infty$.
This can be established via a variational argument.
One uses $\c_R(s)$ and $\c_L(s)$ as trial wave functions
for the second-order operator $\cHDWF^2$.
Introducing $\sket{\xi}=\cHDWF\sket{\c_R}$, one has $\xi(s)=0$
for $1\le s<N_5$.  The reason is that the domain wall operator contains only
nearest-neighbor couplings, and so for this range of the fifth coordinate
the action of $\cHDWF$ on $\c_R$ coincides with that
of the domain wall operator on a semi-infinite lattice. In other words,
$\cHDWF\sket{\c_R}$ does not vanish only because of the new boundary
at $s=N_5$, hence $\sbra{\c_R}\cHDWF^2\sket{\c_R} \sim (1-M)^{2N_5}$,
with a similar result for $\c_L$.  Since $\cHDWF^2$ is a positive operator,
it must therefore have two eigenvectors with a common eigenvalue
$E^2$ where $E \sim (1-M)^{N_5}$.  This, in turn, implies that $\cHDWF$
has two eigenvectors with eigenvalues $E_\pm = \pm E$, where
both eigenvectors are, approximately, linear superpositions
of $\c_R(s)$ and $\c_L(s)$.
For now we will neglect such exponentially small effects,
but we will return to the question of their role at the end of this section.

In the case of a semi-infinite fifth direction (Sec.~\ref{semi5}), the RH zero mode
signals the presence of a RH chiral field.  Similarly, we now have
both RH and LH chiral fields, localized on opposite boundaries.
Together, they form a very light Dirac fermion (if the exponentially small
mixing between the two boundaries is not neglected), which becomes
massless for $N_5\to\infty$.  We may introduce an effective four-dimensional
``domain wall quark'' field for the light Dirac fermion,
simply by picking the appropriate components of the five-dimensional
domain-wall field $\j(x,s)$.  The domain wall quark field is defined as
\begin{eqnarray}
\label{dwq}
q(x) &=& P_R\,\j(x,1) + P_L\,\j(x,N_5) \ ,
\\
\bq(x) &=& \bj(x,1)P_L + \bj(x,N_5)P_R \ .
\nonumber
\end{eqnarray}
This four-dimensional effective field will turn out to be very useful.

Notice the off-diagonal terms proportional to $m$
in Eq.~(\ref{finites}).  These terms couple the opposite-chirality zero modes,
and thus play the role of a mass term for the domain wall quark,
as we will soon confirm explicitly.  This allows the mass of the
domain wall quark to be controlled independently of the extent of
the fifth direction.
In later sections, we will see that in the interacting theory
there are additional, quantum effects that contribute to the mass
of the domain wall quark.

The mass term was first introduced in Ref.~\cite{Shamir:1993zy},
but with a sign convention opposite to what we are using here.
The present convention follows Ref.~\cite{Furman:1994ky}.
As was shown in that paper,
when $m>0$ in Eq.~(\ref{finites}), one has $\det(\DDWF)>0$ for arbitrary
four-dimensional gauge fields (see Sec.~\ref{interacting} below).

Our next task is to obtain the domain wall propagator
for a finite fifth direction,
and from it, the propagator of the effective domain wall quark field.
The (free) second order operator $\O=\DDWF\DDWF^\dagger$ in momentum space
is now (compare Eq.~(\ref{D5D5dag}))
\begin{equation}
\label{DDdagN5}
\O = \left(\begin{array}{ccccccccc}
X_{++} & -b & 0 & 0 & \ldots & 0 & 0 & 0 & X_{-+} \\
-b & Y & -b & 0 & \ldots & 0 & 0 & 0 & 0 \\
0 & -b & Y & -b & \ldots & 0 & 0 & 0 & 0 \\
\rule{0ex}{3ex}
\vdots&\vdots& & \hspace{-3ex}\ddots\hspace{3ex} & \ddots &
\hspace{3ex}\ddots\hspace{-3ex} & &\vdots&\vdots \\
\rule{0ex}{3ex}
0 & 0 & 0 & 0 & \ldots & -b & Y & -b & 0 \\
0 & 0 & 0 & 0 & \ldots & 0 & -b & Y & -b \\
X_{+-} & 0 & 0 & 0 & \ldots & 0 & 0 & -b & X_{--}
\end{array}\right) ,
\end{equation}
where\footnote{%
  The construction of the domain wall propagator is generalized to the
  interacting theory in Sec.~\ref{chirN5}, following Ref.~\cite{Shamir:1998ww}.
}
\begin{subequations}
\label{YX}
\begin{eqnarray}
\label{YXa}
  Y &=& b^2(p) + \bp^2 + 1 \ ,
\\
\label{YXb}
  X_{++} &=& Y + P_L (m^2-1)
\\
\label{YXc}
  X_{--} &=& Y + P_R (m^2-1)
\\
\label{YXd}
  X_{+-} &=& X_{-+} = mb(p) \ .
\end{eqnarray}
\end{subequations}
The second-order operator is real symmetric, and writing
\begin{equation}
\label{OLR}
\O = P_R \O_+ + P_L \O_- \ ,
\end{equation}
it follows that $\O_+ = \car \O_- \car$ or, explicitly,
\begin{equation}
\label{OLRrel}
\O_+(s,s') = \O_-(N_5+1-s,N_5+1-s') \ .
\end{equation}
As before, the domain wall propagator can be expressed as
$\GDWF = \DDWF^\dagger G$, where $G$ is the inverse of $\O$.
By Eq.~(\ref{OLR}), $G$ admits a similar decomposition in terms of $G_\pm$,
and since $G_+ = \car G_- \car$, it is enough to construct $G_-$.

Noting that $\O$ is symmetric, $G_-$ must have the form
\begin{eqnarray}
\label{Gm}
G_-(s,s') &=& G_5(s,s') + A_- \, e^{-\a(s+s'-2)} + A_+ \, e^{-\a(2N_5-s-s')}
\\
&& + A_m(e^{-\a(N_5-1+s-s')}+e^{-\a(N_5-1+s'-s)}) \ .
\nonumber
\end{eqnarray}
Notice that now all four combinations of the homogeneous solutions
$e^{\a(\pm s \pm s')}$ are present.  As the inverse of $\O_-$,
the relation defining $G_-$ is
\begin{equation}
\label{OGm}
\sum_{s''}\O_-(s,s'')G_-(s'',s') - \d_{s,s'} = 0 \ .
\end{equation}
For $2\le s \le N_5-1$, this equation is satisfied for arbitrary
values of the amplitudes $A_-,A_+,A_m$.  Only for $s=1$ or $s=N_5$
is the left-hand side of Eq.~(\ref{OGm}) different from zero
for general $A_-,A_+,A_m$, which in turn allows us to determine
these amplitudes by imposing Eq.~(\ref{OGm}).
There are two matrix equations relating these amplitudes,
since the dependence of the left-hand side on $s'$ can be $e^{\pm\a s'}$.
Since the left-hand side does not vanish for both $s=1$ and $s=N_5$,
we obtain two 2-by-2 matrix equations
\begin{eqnarray}
\label{Amm}
\cc \left( \begin{array}{c} A_- \\ A_m \end{array} \right)
&=& \cn \left( \begin{array}{c} 1-b\,e^{-\a}-m^2 \\ -mb \end{array} \right) ,
\\
\label{Apm}
\rule{0ex}{4.5ex}
\cc \left( \begin{array}{c} A_m \\ A_+ \end{array} \right)
&=& \cn \left( \begin{array}{c} -mb \\ -b\,e^{-\a} \end{array} \right) ,
\end{eqnarray}
where
\begin{equation}
\label{Cmat}
\cc = \left( \begin{array}{ccc}
         b\,e^\a+m^2-1 & & mb \\ mb & & b\,e^\a
         \end{array} \right) .
\end{equation}
In the expression for $\cc$ we neglected
exponentially small corrections of order $e^{-N_5\a}$.
Notice that $A_m$ is over-constrained. Nonetheless, these equations have
a consistent solution\footnote{%
  Equation~(\ref{N5Amass}) corrects a mistake
  in Eq.~(37) of Ref.~\cite{Shamir:1993zy}.
}
\begin{eqnarray}
\label{N5Am}
A_- &=& \D^{-1}\cn(1-m^2)(e^\a-b) \ ,
\\
\label{N5Ap}
A_+ &=&  \D^{-1}\cn(1-m^2)(e^{-\a}-b)\ ,
\\
\label{N5Amass}
A_m &=& -2\D^{-1} \cn b m \sinh\a \ ,
\end{eqnarray}
where
\begin{equation}
\label{detC}
\D=b^{-1}\det\,\cc= e^\a (b\,e^\a-1)+m^2(e^\a-b) \ .
\end{equation}

In order to expose the light Dirac fermion we keep
only the leading behavior for $p^2,m^2\ll 1$, obtaining
\begin{equation}
\label{detCpsq}
\D^{-1} = \frac{M(2-M)(1-M)}{p^2+M^2(2-M)^2m^2} = M(2-M)(1-M) \cd^{-1} \ ,
\end{equation}
where $\cd=p^2+M^2(2-M)^2m^2$, and
\begin{eqnarray}
\label{psqAm}
A_- &=& M(2-M)\cd^{-1} \ ,
\\
\label{psqAp}
A_+ &=& -\frac{(1-M)^2}{M(2-M)}\,p^2\cd^{-1} \ ,
\\
\label{psqAmass}
A_m &=& -M(2-M)(1-M)m\cd^{-1} \ .
\end{eqnarray}
We may now obtain the propagator of the domain-wall quark field introduced
in Eq.~(\ref{dwq}).  Once again neglecting exponentially small corrections,
we find\footnote{
  Since $\bq_L(x) = \bj(x,1)P_L$,
  and the domain wall propagator is $\GDWF = \DDWF^\dagger G$,
  it follows that the contribution of $A_+$ to $\svev{q\,\bq}$
  is always exponentially small, and the right-hand side of Eq.~(\ref{qprop})
  receives contributions from $A_-$ and $A_m$ only.
}
\begin{equation}
\label{qprop}
\svev{q\,\bq} = -M(2-M)\frac{i{\sl{p}}+m_q}{p^2+m_q^2} \ ,
\end{equation}
showing the presence of a Dirac fermion with mass
\begin{equation}
\label{mq}
m_q = M(2-M)m \ .
\end{equation}
Remember that the normalized wave function of the RH zero mode near
the $s=1$ boundary is $\c(s)=(M(2-M))^{1/2}(1-M)^{1-s}$,
hence $|\c(1)|^2 = M(2-M)$.  Similar statements apply to
the LH zero mode localized near the $s=N_5$ boundary.
Thus, the factors of $M(2-M)$ are seen to originate from the support
of the wave functions of the zero modes on their respective boundary layers.

Our construction of the domain wall quarks
could be generalized in two related ways.  Instead of Eq.~(\ref{dwq}),
which defines the effective quark fields $q(x),\bq(x)$
in terms of the five-dimensional fields on the boundary layers only,
we could define the effective quark fields as some weighted average
of the five-dimensional fields $\j(x,s),\bj(x,s)$ on near-boundary layers.
Similarly, we could consider mass terms
that couple near-boundary layers on the two sides, and not just
the two boundary layers themselves, as we have done in Eq.~(\ref{finites}).
Such generalizations are less economic, and do not provide any advantage.
Hence we will always use the simplest domain wall quark field and mass term
introduced above.

The mass of the free domain wall quark is $m_q$
in the limit $N_5\to\infty$.  If we moreover set $m=0$, the RH and LH
components of the domain wall quark completely decouple,
and behave as if they each live on their own semi-infinite fifth direction.
This then reproduces the idealization discussed in Sec.~\ref{semi5}.

While Eqs.~(\ref{qprop}) and~(\ref{mq}) are important for the physical interpretation
of domain wall fermions, the concrete relation in Eq.~(\ref{mq})
between the parameter $m$ that couples the two boundary layers
and the mass $m_q$ of the domain wall quark is largely inconsequential,
because it will be modified in the interacting theory.
Similarly, the factor of $M(2-M)$ in Eq.~(\ref{qprop}) will
undergo renormalization in the interacting theory.

In any numerical lattice calculation using domain wall fermions,
$N_5$ is necessarily finite.  This raises the following question.
In this subsection we have dealt with two qualitatively different types
of parametrically small quantities.  One is $(1-M)^{N_5}$ or, more generally,
$e^{-\a(p)N_5}$, which results from the finiteness of $N_5$.
The other small parameters are $p^2$ and $m^2$, which will ultimately
be associated with some physical scale (expressed in lattice units) of the
interacting theory.  The question we should be asking ourselves is what if
these two sets of parametrically small quantities are comparable in size,
or more generally, what if finite-$N_5$ effects\footnote{%
  For the exponentially small terms relevant for the construction
  of the domain wall propagator, see appendix A of Ref.~\cite{Shamir:1998ww}.
}
are not negligible compared with the effect of an explicit domain wall
quark mass term?  As we will see, this situation is in fact common
in numerical lattice calculations.
The strategies developed to trace and control such effects
will play a central role in subsequent sections.
But first we have to introduce the interacting theory.

\section{\label{interacting} Interacting domain wall fermions}
In this section we discuss interacting domain wall fermions as introduced
in Ref.~\cite{Furman:1994ky}.  As usual, the gauge field degrees of freedom
live on the four-dimensional links.
The hopping terms in the Wilson kernel~(\ref{DW}) become
\begin{eqnarray}
\label{DKU}
D_K(x,y) &=&
\half \sum_\m \left[\d_{x+\hat\m,y} U_\m(x) - \d_{x-\hat\m,y} U^\dagger_\m(y) \right]
\g_\m\ ,
\\
\label{WU}
W(x,y) &=& \half \sum_\m
\left[2\d_{xy} - \d_{x+\hat\m,y} U_\m(x) - \d_{x-\hat\m,y} U^\dagger_\m(y) \right]
\ .
\end{eqnarray}
The link variables $U_\m(x)$ take values in a Lie group $G$, which is usually
assumed to be compact.  For QCD the group is $G={\rm SU}(3)$.
So far, this is standard.  What is special about domain wall fermions
is that the fermion field lives on a five-dimensional lattice.
Noting how the domain-wall fermion operator~(\ref{finites})
depends on the Wilson kernel, we immediately see that
the coupling of the fermion degrees of freedom to the four-dimensional
gauge field is independent of the fifth coordinate.
The continuum limit will therefore be a four-dimensional gauge theory.
We will keep choosing the domain wall height $M$ such that
each (five-dimensional) domain-wall fermion field gives rise to
a single light (four-dimensional) quark field.  As discussed in Sec.~\ref{semi5},
for the free theory we take $0<M\le 1$.  We will discuss how the
appropriate range of $M$ is modified in the interacting theory
in Sec.~\ref{wavef} and Sec.~\ref{quench}.

Considering for simplicity a lattice gauge theory with a single
domain wall fermion, the partition function is
\begin{equation}
\label{Z}
Z = \prod_{x\m} dU_\m(x)\, e^{-S_g/g_0^2}\, \det(\DDWF(m))\, \det^{-1}(\DDWF(1)) \ ,
\end{equation}
where $dU_\m(x)$ is an invariant group measure (for SU(N), the Haar measure).
$S_g$ is the gauge field action, a lattice discretization of the
continuum action, and $g_0$ is the bare coupling.  The fermion determinant
can be represented in the usual way as a Grassmann path integral,
\begin{equation}
\label{Z5}
\det(\DDWF(m)) = \prod_{x,s} d\j(x,s) d\bj(x,s)\, e^{-S_F} \ ,
\end{equation}
where as before, $\j(x,s),\bj(x,s)$ are the five-dimensional fermion fields.
The fermion action is $S_F = \bj \DDWF(m) \j$, where we have suppressed
the summations over all coordinates and indices of the fermion field.
$\DDWF(m)$ is the domain wall fermion operator of Eq.~(\ref{finites}),
where we have indicated explicitly its dependence on the parameter $m$
occurring in the upper-right and lower-left entries that couple
the two boundaries.

Apart from its anticipated dependence on the gauge and fermion fields,
the partition function~(\ref{Z}) includes an additional factor,
$\det^{-1}(\DDWF(1))$, usually called the ``Pauli-Villars determinant.''
Its role will be explained in Sec.~\ref{PV} below.
For now, we only note that it does not introduce any new light
degrees of freedom, because for $m=1$ the operator~(\ref{finites})
implements anti-periodic boundary conditions in the fifth direction,
and thus it describes a five-dimensional fermion field whose mass
is $O(1)$ in lattice units.

\subsection{\label{syms} Flavor symmetries and currents}
We next turn to the flavor symmetries of a lattice gauge theory coupled
to several domain wall fermions.
We will assume that there are $N_f$ domain-wall fermion
fields, all belonging to the same representation of the gauge group
(the fundamental representation of SU(3), for QCD).  We will also assume
for simplicity that the mass parameter $m$ in the
domain-wall fermion operator~(\ref{finites})
is the same for all fermion species, which implies that all the
domain wall quarks are degenerate.  The results can be generalized
straightforwardly to the case of non-degenerate masses.

The five-dimensional fermion action is invariant under global
$\U(N_f) = \SU(N_f)\otimes\U(1)$ symmetry.
The conserved five-dimensional currents have the following structure.
For $\m=1,\ldots,4,$ and all $1\le s \le N_5$,
\begin{eqnarray}
\label{jmu}
  j^a_\m(x,s) &=& \half\left(
        \bj(x,s) (1+\g_\m) U_\m(x) \l^a \j(x+\hat\m,s) \right.
\\
  && - \left.\bj(x+\hat\m,s) (1-\g_\m) U^\dagger_\m(x) \l^a \j(x,s) \right) .
\nonumber
\end{eqnarray}
The hermitian matrix $\l^a$ is one of the $\U(N_f)$ generators.
For the U(1) group, which is the fermion number symmetry, $\l^a$
is the identity matrix.  For $\SU(N_f)$, the generators $\l^a$ are hermitian
and traceless. Apart from the additional dependence on the fifth coordinate $s$,
Eq.~(\ref{jmu}) is recognized as the familiar vector current of Wilson fermions.
For the fifth component we define
\begin{equation}
\label{j5}
\hspace{-2.5ex}
  j^a_5(x,s) = \left\{ \begin{array}{ll}
    \bj(x,s) P_R \l^a \j(x,s+1)  - \bj(x,s+1) P_L \l^a \j(x,s)\ ,
    & 1\le s < N_5 \ ,
\\
    -m \bj(x,N_5) P_R \l^a \j(x,1) + m \bj(x,1) P_L \l^a \j(x,N_5)\ ,
    & s=N_5 \ . \rule{0ex}{3ex}
     \end{array}\right.
\end{equation}
Using the Noether procedure,
the five-dimensional continuity equation is
\begin{equation}
\label{5cons}
  \sum_\m\D_\m\, j^a _\m(x,s) = - \D_5\, j^a_5(x,s)\ .
\end{equation}
Here $\D_\m$ and $\D_5$ are the backward difference operators
\begin{equation}
\label{Delta4}
  \D_\m f(x,s) = f(x,s)-f(x-\hat\m,s) \ ,
\end{equation}
and
\begin{equation}
\label{Delta5}
  \D_5 f(x,s) =
     \left\{ \begin{array}{ll}
          f(x,s)-f(x,s-1) \ ,      &  1<s\le N_5 \ , \\
          f(x,1)-f(x,N_5) \ ,      &  s=1 \ . \rule{0ex}{3ex}
     \end{array}\right.
\end{equation}
This is the usual form of the backward difference operator
for periodic boundary conditions.  The special boundary conditions
in the fifth direction are taken care of by the dependence of $j^a_5(x,N_5)$
on $m$.  In particular, $j^a_5(x,N_5)$ vanishes for $m=0$.
The above expressions for $j^a_5(x,N_5)$ and for $\D_5$
are somewhat different from Ref.~\cite{Furman:1994ky}, but the expression
for $ \D_5\, j^a_5(x,s)$ remains the same for all $s$.
When carried out on the lattice,
the Noether procedure has a certain degree of freedom, and as a result,
relegating the $m$ dependence to $j^a_5$ (as we did here)
or to $\D_5$ (as was done in Ref.~\cite{Furman:1994ky})
are both valid options.\footnote{
  See Sec.~\ref{subsec:mcc} for a related discussion.
}

We will now use the five-dimensional conserved currents as the building blocks
for four-dimensional vector and axial currents.
There is a unique set of conserved vector currents, obtained by summing
the conserved five-dimensional current over all $s$ values,
\begin{equation}
\label{Vmu}
  \cv^a_\m(x)=\sum_{s=1}^{N_5} j^a _\m(x,s) \,.
\end{equation}
Conservation of these vector currents, $\sum_\m\D_\m \cv^a_\m=0$,
follows from from the five-dimensional continuity equation~(\ref{5cons}).

Unlike the vector currents, it is not possible to define exactly conserved
axial currents.  In order to define partially conserved axial currents,
we take advantage of the global separation of the RH and LH components
of the effective quark field~(\ref{dwq}),
which are supported on opposite boundaries of the
five-dimensional lattice. We will define the axial transformations to act {\it
vectorially} on each four-dimensional layer, but we assign opposite charges
to fermion degrees of freedom in the two half-spaces
\begin{eqnarray}
\label{axialpsi}
  \d^a_{A} \j(x,s)  &=& + i \e(s) \l^a \j(x,s) \ , \\
  \d^a_{A} \bj(x,s) &=& - i \e(s) \bj(x,s) \l^a \ ,
\nonumber
\end{eqnarray}
where\footnote{%
Recall we always take $N_5$ even.
}
\begin{equation}
\label{epss}
  \e(s) = \left\{ \begin{array}{rr}
              1\,,  &  1\le s \le N_5/2 \ , \\
              -1\,,  &  N_5/2 < s \le N_5  \rule{0ex}{3ex} \ .
           \end{array}\right.
\end{equation}
For the effective domain wall quark fields defined in Eq.~(\ref{dwq}),
the transformations~(\ref{axialpsi}) take the form
\begin{eqnarray}
\label{axialq}
  \d^a_{A} q(x)  &=& i \g_5 \l^a q(x) \ , \\
  \d^a_{A} \bq(x) &=& i \bq(x) \l^a \g_5 \ ,
\nonumber
\end{eqnarray}
which are recognized as axial transformations.

The virtue of this construction is that, just like the vector currents,
the partially-conserved axial currents take the form of suitable sums
over the five-dimensional conserved currents.  Explicitly,
\begin{equation}
\label{Amu}
  \ca^a_\m(x) = - \sum_{s=1}^{N_5} \e(s) j^a_\m (x,s) \ .
\end{equation}
For $m=0$ (see Eq.~(\ref{finites})), the non-invariance of the action under the
transformations~(\ref{axialpsi}) comes solely from the coupling between
the four-dimensional layers $s=N_5/2$ and $s=N_5/2+1$.
For $m\ne 0$, there is an additional contribution coming from the
direct coupling between the boundary layers $s=1$ and $s=N_5$.
As a result, the axial currents satisfy the
partial conservation equation\footnote{
  If we restore the lattice spacing, the coefficient
  of $J^a_{5q}(x)$ in Eq.~(\ref{divA}) becomes $2/a$.
}
\begin{equation}
\label{divA}
  \sum_\m \D_\m \ca^a_\m(x) = 2m J^a_5(x) + 2 J^a_{5q}(x) \ ,
\end{equation}
where
\begin{eqnarray}
\label{J5a}
   -m J^a_5(x) &=& j^a_5(x,N_5) \ ,
\\
\label{J5qa}
   J^a_{5q}(x) &=& j^a_5(x,N_5/2) \ .
\end{eqnarray}
The operator $j^a_5(x,N_5)$ can be expressed in terms of the
effective quark fields of Eq.~(\ref{dwq}).
Having factored out the $m$ dependence on the left-hand side of Eq.~(\ref{J5a}),
$J^a_5$ takes the form of the familiar pseudoscalar density
\begin{equation}
\label{J5adwq}
   J^a_5(x) = \bq(x) \g_5 \l^a q(x) \ .
\end{equation}
We see that, were it not for the extra term $J^a_{5q}(x)$ on its
right-hand side, Eq.~(\ref{divA}) would express the familiar
partial conservation of the axial current (PCAC).

For later use, the associated Ward-Takahashi identities (WTIs) are
\begin{eqnarray}
\label{WTI}
   \sum_\m \D_\m \vev{\ca^a_\m(x)\, O(y_1,y_2,\ldots) }
   &=& 2m \vev{J^a_5(x)\, O(y_1,y_2,\ldots) }
\\
  && + 2 \vev{J^a_{5q}(x)\, O(y_1,y_2,\ldots)}
\nonumber
\\
   &&  + i \vev{\d^a_A(x)\, O(y_1,y_2,\ldots)} \ .
\nonumber
\end{eqnarray}
The multi-local operator $O(y_1,y_2,\ldots)$ will be assumed to be a product of
the domain-wall quark fields $q(y)$ and $\bq(y)$, on which the local variation
acts as $\d^a_A(x) q(y) = i \d^4_{x,y} \g_5 \l^a q(y)$
and $\d^a_A(x) \bq(y) = i \d^4_{x,y} \bq(y) \l^a \g_5$.

\begin{boldmath}
\subsection{\label{chirN5} Chiral symmetry restoration for $N_5\to\infty$}
\end{boldmath}
In the rest of this section we distinguish between the singlet
and non-singlet currents by omitting the upper index $a$ for the
singlet currents. Postponing the discussion of the singlet axial current
$\ca_\m(x)$ to the next subsection, in this subsection we discuss
the non-singlet axial currents $\ca^a_\m(x)$.
We will show that, in an appropriate sense, the operator $J^a_{5q}$
on the right-hand side of Eq.~(\ref{divA}) vanishes for $N_5\to\infty$.
This is a result of the following theorem.

\medskip

\begin{quotation}
\noindent {\em Theorem: chiral symmetry restoration}.\
Consider the correlation function
of $J^a_{5q}$ with an arbitrary set of domain-wall quark fields,
\begin{equation}
\label{J5qaq}
\svev{J^a_{5q}(x)\,q(y_1)q(y_2)\cdots \bq(z_1)\bq(z_2)\cdots} \ .
\end{equation}
Then, for $N_5\to\infty$, this correlation function vanishes.
\end{quotation}

Less formally, thought of as an operator that acts on states created by
the boundary fields $q$ and $\bq$ only, the theorem states that
$J^a_{5q}$ vanishes for $N_5\to\infty$.
Assuming that the axial currents $\ca^a_\m(x)$ act on the same set of states,
it follows immediately that the PCAC relation is reproduced,
\begin{equation}
\label{divAlim}
   \sum_\m \D_\m \ca^a_\m(x) = 2m J^a_5(x) \ , \qquad N_5\to\infty \ .
\end{equation}

A proof of chiral symmetry restoration for $N_5\to\infty$ was first given
in Ref.~\cite{Furman:1994ky} using a second-quantized transfer matrix formalism.
Usually, the transfer matrix describes propagation in euclidean space
along the time direction \cite{Luscher:1976ms}, but here it was adapted
to describe propagation along the fifth direction
\cite{Narayanan:1993sk,Narayanan:1993ss,Furman:1994ky}.

The proof rests on three essential physical ingredients.
The first ingredient is that the gauge field is independent
of the fifth coordinate.  As a result, propagation along the fifth direction
is controlled by a second-quantized transfer matrix $\hcT=\hat{T}(\cu)$,
which is also independent of the fifth coordinate.
Here $\cu$ denotes the four-dimensional gauge field.
We may write $\hcT=\exp(-\hcH)$, regarding $\hcH$ as
a ``hamiltonian'' in $4+1$ dimensions.  Consider now momentarily
an infinite fifth direction where $-\infty < s < \infty$.
Let us excite an eigenstate of $\hcH$ with energy $E>0$ above the
second-quantized vacuum on a four-dimensional layer with fifth coordinate $s$,
and propagate it to another layer $s'$.
This yields a suppression factor $\exp(-E|s-s'|)$,
exhibiting translation invariance in the fifth direction.
The transfer matrix formalism can be generalized to the case of
domain wall fermions where $1\le s \le N_5$, and translation invariance
in the fifth direction is explicitly broken by the boundaries.
Using this formalism one derives expressions for $\det(\DDWF(m))$, as well as
for the correlation functions of domain wall fermions \cite{Furman:1994ky}.

The second ingredient is that, as we shall see, any violation
of the conservation of a non-singlet axial current in a correlation function
of the domain-wall quark fields, Eq.~(\ref{J5qaq}), requires fermion propagation
across the entire fifth dimension, hence a factor of $\sim \exp(-N_5 \hcH)$.

In order to understand the last crucial ingredient let us momentarily
assume that for a given background gauge field the ground state of $\hcH$
is separated from the first excited state by a gap $\D>0$ . In this case
the correlation function~(\ref{J5qaq}) is bounded by $\exp(-N_5 \D)$
and vanish for $N_5\to\infty$.  In reality, there are
special gauge-field configurations for which $\D=0$,
equivalently, the transfer matrix has an eigenvalue equal to one,
and the argument will have to be suitably refined.

In the rest of this subsection we will discuss the proof more fully
but still omit various technicalities.  The full details may be found
in Refs.~\cite{Furman:1994ky,Shamir:1998ww}.  The proof we will describe here
makes use of the physical ingredients explained above, but avoids
the technically elaborate second-quantized transfer matrix formalism
of Ref.~\cite{Furman:1994ky}.
The discussion is based on Ref.~\cite{Shamir:1998ww}, which uses a
``first quantized'' transfer matrix formalism instead.\footnote{
  For the relation between the first-quantized and second-quantized
  transfer matrices see Ref.~\cite{Furman:1994ky}.
}

In the standard treatment of a Grassmann path integral,
the contraction of two fermion fields gives rise to a propagator
which is the inverse of the Dirac operator occurring in
the (bilinear) fermion action.  In the free theory, we constructed
the domain wall propagator in Sec.~\ref{finite5}.  In brief, the
propagator can be constructed in a standard manner in terms of the inverse
of the second-order domain wall operator $\O$ (Eq.~(\ref{DDdagN5})),
which in turn can be decomposed into its chirality components (Eq.~(\ref{OLR})),
and the inverse $G_-$ of $\O_-$ was given explicitly in Eq.~(\ref{Gm}).

Because the gauge field is the same on all four-dimensional slices
(and it has no fifth component) the solution of Sec.~\ref{finite5}
can be generalized to the interacting case.
We begin by writing the Wilson kernel in $2\times 2$ block form as
\begin{equation}
\label{Dkernel}
   D = D(M;\cu)
   = \left(\begin{array}{cc}
   1-B    & C     \\
   -C^\dagger  & 1-B
       \end{array}\right) ,
\end{equation}
where
\begin{equation}
\label{B}
B=1+W-M \ .
\end{equation}
Both $B$ and $C$ carry a $2\times 2$ spinor index, and
$B$ is proportional to the identity matrix in spinor space.
The explicit form of $C$ can be inferred from Eqs.~(\ref{DW}),
~(\ref{dirac}) and~(\ref{DKU}).
In the presence of a gauge field the second-order operator
$\O=\DDWF\DDWF^\dagger$ is still given by Eq.~(\ref{DDdagN5}),\footnote{
  Notice that the definition of $\O$ in Eq.~(3.2)
  of Ref.~\cite{Shamir:1998ww} is slightly different.
}
while Eq.~(\ref{YX}) has a suitable generalization.
In particular, in the interacting case $Y$ is given explicitly by
\begin{equation}
\label{Yint}
Y = 2B + DD^\dagger \ .
\end{equation}
Next, introducing
\begin{equation}
\label{K}
   K = \left( \begin{array}{cc}
              B^{-1/2}      &  0       \\
              C^\dagger\, B^{-1/2} &  B^{1/2}
       \end{array}\right) ,
\qquad
   K^\dagger = \left( \begin{array}{cc}
              B^{-1/2}  &  B^{-1/2}\, C     \\
              0  &  B^{1/2}
       \end{array}\right) ,
\end{equation}
we construct two transfer matrices sharing the same eigenvalue spectrum,
\begin{eqnarray}
\label{T}
  T &=& KK^\dagger \ = \
  \left( \begin{array}{cc}
  B^{-1} &  B^{-1}\, C    \\
  C^\dagger\, B^{-1} &  B + C^\dagger\, B^{-1}\, C
       \end{array}\right) ,
\\
\label{Ttilde}
  \tT &=& K^\dagger K \ = \ \rule{0ex}{4.5ex}
  \left( \begin{array}{cc}
  B^{-1} + B^{-1/2} C C^\dagger B^{-1/2} & B^{-1/2} C B^{1/2} \\
  B^{1/2} C^\dagger B^{-1/2} & B
       \end{array}\right) .
\end{eqnarray}
We will encounter the transfer matrix $T$ in Sec.~\ref{DWF2GW} and Sec.~\ref{mobius}.
Here we will make use of the alternative version $\tT$.  A key result is
\begin{equation}
\label{YT}
\g_5 B^{-1/2}\, Y\, \g_5 B^{-1/2} = \tT + \tT^{-1} \ .
\end{equation}
In this subsection we restrict the domain wall height to $0<M<1$.
As a result, $B$ is a positive operator, and the same is true
for the transfer matrices.  We will consider the full range $0<M<2$
in Sec.~\ref{chiragain}.

We next introduce the spectral decomposition
\begin{equation}
\label{tT}
  \tT = \sum_i \sket{v_i} \l_i \sbra{v_i} \ ,
\end{equation}
where $\sket{v_i}$ are the eigenvectors. Note that the eigenvalues
$\l_i$ are all positive.
From the same ingredients we construct a third transfer matrix
\begin{equation}
\label{Q}
  Q = \sum_i \sket{v_i} \eta_i \sbra{v_i} \ ,
\end{equation}
where $\eta_i = {\rm min}\{\l_i,\l_i^{-1}\}$.
An immediate consequence is that the norm of $Q$ is bounded by one.
Equivalently, denoting the maximal eigenvalue of $Q$ by $\etamax$,
it follows that $\etamax\le 1$.  The virtue of $Q$ is that
it satisfies physical boundary conditions.
Namely, $Q^{|s-s'|}$ vanishes when the separation $|s-s'|$ tends to infinity.
This is true provided that no eigenvalue is exactly equal to 1,
and thus $\etamax<1$.
This feature, alongside with Eq.~(\ref{YT}) which exhibits the
algebraically equivalent roles of an eigenvalue $\l_i$ and its inverse,
allows us to construct the inverse of $\O$ as
$\g_5 B^{-1/2}\, \tilde{G}\, \g_5 B^{-1/2}$ where\footnote{
  For the full details, see Ref.~\cite{Shamir:1998ww}.
}
\begin{equation}
\label{Gint}
\tilde{G}(s,s') = \tilde{G}_5(s,s')
+ H_{+,+}(s,s') + H_{-,-}(s,s') + H_{+,-}(s,s') + H_{-,+}(s,s') \ .
\end{equation}
The translationally invariant part (compare Eq.~(\ref{G5})) is now given by
\begin{equation}
\label{G5int}
\tilde{G}_5(s,s') = Q^{|s-s'|} f(Q) \ ,
\end{equation}
where $f^{-1}(Q)= Q^{-1}-Q$.  Each of the four translationally non-invariant terms
has the general form
\begin{equation}
\label{HAQ}
H_{\pm,\pm}(s,s') = Q^{d(s)} A_{\pm,\pm} Q^{d(s')} \ ,
\end{equation}
where $d(s)$ is equal to $s$ or to $N_5+1-s$,
in other words, $d(s)$ is the distance to one of the boundaries.
The matrix amplitudes $A_{\pm,\pm}$ provide a suitable generalization
of the amplitudes $A_+$, $A_-$ and $A_m$ of Eq.~(\ref{Gm}) to the
interacting theory.  We note also that Eq.~(\ref{YT}) is the generalization
of Eq.~(\ref{alphap}) to the interacting theory.
Correspondingly, the eigenvalues $\eta_i$ of the matrix $Q$ generalize
the factors of $e^{-\a(p)}$ encountered in Sec.~\ref{free}.
In the absence of a gauge field $Q$ is diagonal in momentum space,
and its eigenvalues reduce to $e^{-\a(p)}$.

Having constructed the domain wall propagator in the interacting theory
we now make the following crucial observation.  Because of the traceless
flavor matrix $\l^a$ contained in the definition of $J^a_{5q}$, the contraction
of the $\j$ and $\bj$ fields in $J^a_{5q}$ with each other
always vanishes identically.  Hence, for the correlation function~(\ref{J5qaq})
to be nonzero, the field $\j$ in $J^a_{5q}$
must be contracted with a boundary field $\bq$.  A similar statement
applies to the $\bj$ field.  Since $J^a_{5q}$ lives in the middle of the
fifth direction, we are looking at two factors of $\sim Q^{N_5/2}$,
one from each fermion propagator.
Because the exponent is proportional to $N_5$, propagation
in the fifth direction will be dominated by those eigenvalues of $Q$
which are very close to its maximal eigenvalue $\etamax$.
Up to subleading power corrections, in a given background gauge field
all the correlation functions in Eq.~(\ref{J5qaq}) are thus bounded
by $c\,\etamax^{N_5}$, where the exponential factor $\etamax^{N_5}$
is universal, and only the prefactor $c$ depends on the
specifics of the correlation function.

We now come to the last step.
First, we momentarily assume that $\etamax=\etamax(\cu)$ admits
a non-trivial global bound\footnote{%
  This is attainable by imposing a constraint on the gauge field configuration
  space that does not alter the continuum limit
  \cite{Hernandez:1998et,Kikukawa:1999dk}.
}
for all gauge fields $\cu$.
Namely, $\etamax(\cu)\le \L_0$ for some $\L_0<1$.  In that case,
the ensemble average of any of the correlators~(\ref{J5qaq}) would be bounded
by $\L_0^{N_5}$, and would thus vanish for $N_5\to\infty$.

If no constraint is imposed on the gauge field configuration space,
there exist special configurations for which $\etamax(\cu)=1$,
in other words, the transfer matrix has an eigenvalue equal to one.
Let us show that this implies the existence of a zero mode
of the Wilson kernel $D=D(M;\cu)$ of Eq.~(\ref{Dkernel}) \cite{Furman:1994ky}.
Using Eqs.~(\ref{Yint}) and~(\ref{YT}), as well as $\g_5$-hermiticity
of the Wilson kernel, it follows that if $\tT\sket{v}=\sket{v}$ then
$\sbra{v} B^{-1/2} D^\dagger D B^{-1/2} \sket{v} = 0$.
Since $B$ is bounded, $B^{-1/2}$ cannot have a zero mode,
and therefore $B^{-1/2} \sket{v}$ is a zero mode of the Wilson kernel.
We will discuss the physical significance of the kernel's zero modes
in Sec.~\ref{kernel}.  Here we will only note that $D(M)$ does not have
any zero mode in the free theory for $0<M<2$, and that configurations for which
$D(M;\cu)$ has a zero mode are expected to be rare when the continuum limit
is approached.

Mathematically, having a zero mode means that $\det(D(M;\cu))=0$,
a constraint which is satisfied on a measure zero subset of the
gauge-field configuration space.  We will use this observation
to refine the proof of chiral symmetry restoration.  Choosing some $0<\e\ll 1$
we divide this space into a ``large'' and a ``small'' part.  The large part
consists of the configurations for which $\etamax(\cu)\le 1-\e$.
The small part, which is the rest, consists of configurations
for which $1-\e < \etamax(\cu)\le 1$.
Consider now the contributions of these two parts to some correlation function
in Eq.~(\ref{J5qaq}).  For $N_5\gg 1$ the contribution of the large part
is dominated by the eigenvalues close to $1-\e$,
and thus it scales as $O((1-\e)^{N_5})$.
The contribution of the small part is $O(\e)$ of course.
The contribution of the large part
vanishes for $N_5\to\infty$, leaving us in this limit with the contribution
of the small part only, which in turn implies an estimate of $O(\e)$
for any correlation function in Eq.~(\ref{J5qaq}).  Finally, since we can
repeat the same argument for arbitrarily small $\e$, it follows again that
all the correlation functions in Eq.~(\ref{J5qaq})
must vanish for $N_5\to\infty$.
This completes the proof of chiral symmetry restoration,
and thus that the non-singlet axial currents satisfy
the PCAC relation~(\ref{divAlim}).

We will expand the scope of the proof of chiral symmetry restoration
in Sec.~\ref{chiragain}.

\begin{figure}[t]
\begin{center}
\includegraphics*[width=5cm]{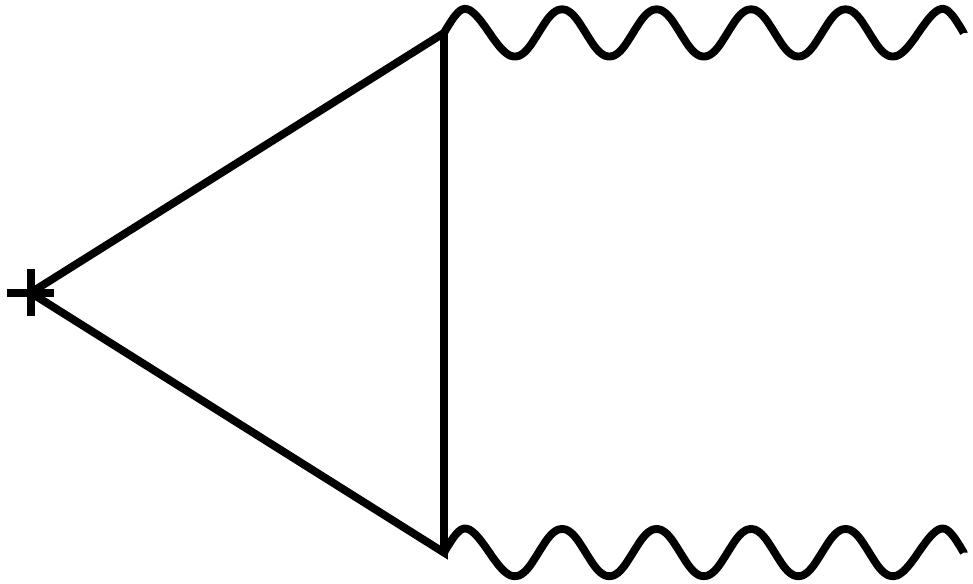}
\end{center}
\begin{quotation}
\floatcaption{anmlfig}%
{The triangle anomaly.  The vertex on the left represents the divergence
of the singlet axial current in the regularized theory.
For domain wall fermions, it is $J_{5q}$.
}
\end{quotation}
\vspace*{-4ex}
\end{figure}

\subsection{\label{anml} The axial anomaly}
Unlike the flavor non-singlet axial currents considered in the
previous subsection, whose conservation for $m\to 0$ is recovered
in the $N_5\to\infty$ limit, the singlet axial current is not conserved
when the same limits are taken.  This is as it must be, since
in the target continuum theory the singlet axial current is anomalous
(\Fig{anmlfig}).

At the technical level (see Eq.~(\ref{divA})),
the difference is that in the case of the singlet current,
the expectation value $\svev{J_{5q}}$ remains nonzero in the presence
of a background gauge field, even in the limit $N_5\to\infty$.
A one-loop calculation in lattice perturbation theory,
whose details we will not repeat here, finds that when the limit
$N_5\to\infty$ is taken, and up to corrections that vanish
in the continuum limit, the divergence of the singlet axial current is
\cite{Golterman:1992ub,Shamir:1993yf}
\begin{equation}
\label{dAanml}
\sum_\m \D_\m \ca_\m(x) = 2mJ_5(x) + \frac{ig^2 N_F}{4\p^2}\, \sum_{\m\n\l\r}
\e_{\m\n\l\r} \partial_\m A_\n(x) \partial_\l A_\r(x) \ .
\end{equation}
Here $g$ is the (bare) gauge coupling.
For simplicity, we have assumed that $A_\m$ is the vector potential
of an abelian background field.
(The relation between the link variable and the vector potential is
$U_\m(x) = e^{i g a A_\m(x)}$ where we have re-introduced the
lattice spacing $a$.  See for example Ref.~\cite{DeGrand:2006zz}.)
This result is valid for $N_F$ domain wall fermions in the fundamental
representation.  The last term on the right-hand side of Eq.~(\ref{dAanml})
is recognized as the axial anomaly, a result which is in agreement with
continuum calculations of the anomaly for $N_F$ Dirac fermions.

In Ref.~\cite{Golterman:1992ub}, which considered a geometrical limit
equivalent to the formal limit of a semi-infinite fifth direction,
it was moreover shown that
as the domain-wall height $M$ is varied, the coefficient of the anomaly
matches the number and chirality of the Weyl fermions in \Tab{zms}.

As originally observed by Kaplan \cite{Kaplan:1992bt,Kaplan:2009yg} the anomaly
emerges via the {\em Callan-Harvey mechanism}\ \cite{Callan:1984sa}.
Consider for simplicity a single domain wall fermion.  We sum
the four-dimensional components of the five-dimensional conserved current
over a limited range of the fifth coordinate, $\sum_{s=1}^{s_0} j_\m(x,s)$.
We assume that both $s_0$ and $N_5-s_0$ are large, and set $m=0$.
The only light degrees of freedom that contribute to $j_\m(x,s)$ in this sum
are those coming from the RH Weyl field localized near the $s=1$ boundary.
But when the global U(1) transformation associated with the
five-dimensional conserved current
is applied to this RH Weyl field only, it becomes anomalous.
The anomaly must therefore be reproduced by the four-divergence
$\D_\m\sum_{s=1}^{s_0} j_\m(x,s)$.  At the same time,
since the five-dimensional current is conserved (Eq.~(\ref{5cons})),
we have $\D_\m\sum_{s=1}^{s_0} j_\m(x,s)=-j_5(x,s_0)$,
which means that the anomaly is reproduced again by $-j_5(x,s_0)$,
also called the Chern-Simons current in this context
\cite{Kaplan:1992bt,Golterman:1992ub,Kaplan:2009yg}.

We may also consider the complement five-dimensional sum,
$\sum_{s=s_0+1}^{N_5} j_\m(x,s)$, whose four-divergence is equal to $+j_5(x,s_0)$.
This now correctly accounts for the anomaly arising from the LH chiral mode
near the $s=N_5$ boundary, including its opposite sign.
The remarkable effect is that $j_5(x,s_0)$ ``knows'' about the
light degrees of freedom near the boundaries,
even though it is located far from both of them, deep inside the
five-dimensional bulk which in itself supports only massive degrees of freedom.
Moreover, as we have seen, $\pm j_5(x,s_0)$ correctly accounts for the anomaly
due to the chiral field near each boundary.

When we sum $j_\m(x,s)$ over the entire range of the fifth coordinate
the two anomalous contributions cancel each other, and we recover
the conserved four-dimensional fermion-number current
(Eq.~(\ref{Vmu}) for the U(1) case).  If instead we consider the difference
of the two sums taking $s_0=N_5/2$ for definiteness,
we recover the U(1) axial current (Eq.~(\ref{Amu}) for the U(1) case),
where the anomalous contributions
of the RH and LH fields add up, yielding the axial anomaly
of a Dirac fermion in Eq.~(\ref{dAanml}).

Additional differences of a more technical nature exist between the singlet
and non-singlet axial currents.  These will be discussed in Sec.~\ref{mres}.

\subsection{\label{PV} The Pauli-Villars determinant}
In this subsection we complete the description of the domain-wall fermion
partition function~(\ref{Z}),
by explaining the role of the Pauli-Villars (PV) determinant.
Thought of as a set of $N_5$ four-dimensional fields, a domain wall
fermion field describes one light Dirac field and $N_5-1$ heavy Dirac fields
whose masses are $O(1)$ in lattice units.\footnote{%
  For the actual masses of the heavy fields, see \cite{Neuberger:1997bg}.
}
Integrating out all the $N_5-1$ heavy four-dimensional fields,
but not the light four-dimensional field, would generate
a {\em local} effective action for the lattice gauge field.
By this we mean that we expect the couplings between remote sites (or links)
in this effective action to decay exponentially, with a decay rate
that is $O(1)$ in lattice units.  The total action for the gauge field
(before including the contribution of the PV determinant) can
thus be expressed as
\begin{equation}
\label{Sgeff}
S_{\rm tot} \approx \frac{1}{g_0^2}\, S_g + (N_5-1) S_{\rm eff} + S'_{\rm eff}  \ .
\end{equation}
Here $S_g/g_0^2$ is the lattice discretization
of the continuum gauge field action introduced in Eq.~(\ref{Z}).
The contribution of the heavy fermion degrees of freedom that we have
integrated out splits into two terms.  The dominant term is $N_5 S_{\rm eff}$,
where $S_{\rm eff}$ is the effective action arising from integrating out
a single four-dimensional layer deep inside the five-dimensional bulk,
and the factor of $N_5$ arises from the approximate translation invariance
in the fifth direction away from the boundaries.  The rest of the
effective action, $S'_{\rm eff}-S_{\rm eff}$, is a correction that accounts for the influence
of the boundaries on the heavy degrees of freedom.

We immediately see the problem with Eq.~(\ref{Sgeff}).  A fixed, finite value
for $N_5$ would induce some finite renormalization of the bare coupling.
But since we are interested in the limit $N_5\to\infty$, this would entail
a diverging contribution to the effective action of the gauge-field.

The PV determinant remedies this problem.  The PV field is
a five-dimensional field with $m=1$ in Eq.~(\ref{finites}).  As noted already,
this implements anti-periodic boundary conditions in the fifth direction.
Hence, the PV field represents $N_5$ four-dimensional fields, all massive,
with opposite statistics to the domain-wall fermion field.
We thus expect that the PV contribution to the total gauge field
effective action will be
\begin{equation}
\label{SPV}
- N_5\, S_{\rm eff} \ .
\end{equation}
In the sum of Eqs.~(\ref{Sgeff}) and~(\ref{SPV}) the large contribution
proportional to $N_5$ drops out.
One is left with some $N_5$-independent contribution,
which is equivalent to a finite renormalization of the bare coupling.

The tight relation between the domain wall fermion and PV operators
has many useful consequences.  One such implication will be discussed
in the next subsection.
In modern numerical QCD calculations using domain wall fermions,
the intimate relation between the domain wall and PV operators is taken
one step further.  One writes
\begin{equation}
\label{detfPV}
\det(\DDWF(m))\, \det^{-1}(\DDWF(1)) = \det(\DDWF^{-1}(1)\DDWF(m)) \ ,
\end{equation}
and treats $\DDWF^{-1}(1)\DDWF(m)$ as a new domain-wall fermion operator.
This will be further discussed in the next subsection,
as well as in Sec.~\ref{mobius}.
Considered as the domain-wall fermion operator, $\DDWF^{-1}(1)\DDWF(m)$
couples lattice sites with arbitrarily large separation,
because the same is true for $\DDWF^{-1}(1)$. Nevertheless,
the Dirac operator $\DDWF^{-1}(1)\DDWF(m)$ is local in a similar sense
to what we have discussed above.
Because the operator $\DDWF^{-1}(1)$ has only lattice-scale excitations,
one expects that the coupling of remote sites via $\DDWF^{-1}(1)$
will decay exponentially with a lattice-scale rate.\footnote{
  For a discussion of the closely related issue of locality of
  exact or approximate GW operators, see Sec.~\ref{lcleff}.
}

\subsection{\label{DWF2GW} The effective Dirac operator for the light fermion}
When we integrate out the five-dimensional heavy fermion degrees of freedom,
this not only generates an effective action for the gauge field,
but also induces an effective Dirac operator for the light four-dimensional
fermion field. In this subsection we examine this effective Dirac operator
in more detail, both for finite $N_5$ and in the limit $N_5\to\infty$.
We will consider a single domain wall field, and for simplicity
we will first set $m=0$.
We will show how, with the help of the PV determinant, one obtains
an effective Dirac operator for the light four-dimensional field that satisfies
the Ginsparg-Wilson (GW) relation \cite{Ginsparg:1981bj}
in the limit $N_5\to\infty$.

We start by examining the global WTI for the domain-wall quark propagator
in a fixed gauge field background $\cu$ (the dependence on $\cu$ will
usually be omitted).  To this end we choose a bi-local operator
$O(y_1,y_2)=q(y_1)\bq(y_2)$ in Eq.~(\ref{WTI}),
considering the singlet axial current.  Upon summing over the four-dimensional
coordinates the left-hand side of Eq.~(\ref{WTI}) vanishes, and for $m=0$ we obtain
\begin{eqnarray}
\label{WTIqbq}
2 \sum_x \vev{J_{5q}(x)\, q(y_1)\bq(y_2)}
&=& \{\g_5,G_q(y_1,y_2)\} \ ,
\\
\label{effprop}
G_q(y_1,y_2) &=& \vev{q(y_1)\bq(y_2)} \ .
\end{eqnarray}
In the free theory, the expectation value $\svev{J_{5q}}$ vanishes for
$N_5\to\infty$ (for relevant technicalities see, \eg, Ref.~\cite{Shamir:1993yf}).
This implies that the left-hand side of Eq.~(\ref{WTIqbq}) does not have
a disconnected contribution.  As for the connected contribution,
the domain wall fields from which the ``midpoint'' pseudoscalar density $J_{5q}$
is built are contracted with the boundary fields $q$ and $\bq$.
Much like the non-singlet case (Sec.~\ref{chirN5}), this contribution
vanishes for $N_5\to\infty$ as well. Thus, the left-hand side
of Eq.~(\ref{WTIqbq}) as a whole vanishes in this limit.
We comment in passing that
the same conclusion applies in the background of perturbative gauge fields.
The physical reason is that the anomaly (Eq.~(\ref{dAanml})) can be expressed
as the divergence of a gauge non-invariant current whose integral
is the topological charge, which is zero in perturbation theory.
Hence $\sum_x \svev{J_{5q}(x)}$ again vanishes.  (This is true
up to the usual discretization effects associated with the definition of the
topological charge on the lattice.)

The conclusion is that for $N_5\to\infty$, the domain wall quark propagator
$\svev{q\,\bq}$ anti-commutes with $\g_5$ for $m=0$.  This result can be
confirmed in the free theory using the explicit form of the propagator
derived in Sec.~\ref{finite5}, as well as in the presence of a perturbative
background field
using the more general result derived in Ref.~\cite{Shamir:1998ww}.
In all cases, the conclusion simply follows from the fact that, for $m=0$,
the mixed-chirality components $\svev{q_R\,\bq_L}$ and $\svev{q_L\,\bq_R}$
require coupling between the two boundaries,
and thus vanish exponentially for $N_5\to\infty$.\footnote{%
For perturbative gauge fields the transfer matrix (Sec.~\ref{chirN5})
cannot have any eigenvalue equal to one.
}

The anti-commutativity of the domain-wall quark propagator $\svev{q\,\bq}$
with $\g_5$ is an expression of the restoration of chiral symmetry
at $m=0$ for $N_5\to\infty$.  Naively, it suggests that in this limit
we may define an effective domain-wall quark operator as
\begin{equation}
\label{Deff}
\Deff^{-1}=\svev{q\,\bq} \ .
\end{equation}
Being the inverse of the domain-wall quark propagator,
$\Deff$ anti-commutes with $\g_5$ as well.
However, this leads to a highly undesirable consequence!
According to the no-go theorems about
fermion doubling \cite{Karsten:1980wd,Nielsen:1980rz,Nielsen:1981xu,Pelissetto:1987ad,Karsten:1981gd}, a free lattice Dirac operator that anti-commutes
with $\g_5$ cannot describe a single massless Dirac fermion
if the dispersion relation is smooth.
In order to figure out what is going on, let us reexamine the free propagator
$\svev{q\,\bq}$ for $m_q=0$, in the limit $N_5\to\infty$.
In Sec.~\ref{finite5} we have calculated this propagator for $p\ll 1$
and general $m$, finding that for $m=0$ it has a pole at $p=0$
that represents one massless Dirac fermion (Eq.~(\ref{qprop})).
Returning to general $p$ (keeping $m=0$), the relevant term in the
free propagator takes the form of ${\sl\bp} A_-(p^2)$ up to
a proportionality constant.  The amplitude $A_-(p^2)$ has a $1/p^2$ pole
for $p\to 0$, whereas everywhere else in the Brillouin zone it is regular.
It follows that $\svev{q\,\bq}$ has a massless pole, ${\sl\bp}/p^2$, at $p=0$,
but at the other 15 corners of the Brillouin zone it has {\em zero}
eigenvalues.
Using Eq.~(\ref{Deff}), the effective Dirac operator $\Deff$ thus has {\em poles}
at all the corners of the Brillouin zone except for the origin.
While the presence of poles in $\Deff$ means that it is not subject to
the no-go theorems, this is clearly an undesirable situation that limits
the ability to interpret $\Deff$ as an effective low-energy operator.
For relevant discussion in the context of the no-go theorems, see
Refs.~\cite{Pelissetto:1987ad,Karsten:1981gd,Golterman:2023zqf,Golterman:2025boq}.

It should be clear that the issue we encounter here is not a fundamental flaw
of domain wall fermions. Indeed we already know that the domain wall operator
describes one light four-dimensional field alongside with $N_5-1$ heavy fields
with cutoff scale masses that decouple in the continuum limit.
Rather, the presence of poles in $\Deff$ is an artifact of our attempt to use
the inverse of $\svev{q\,\bq}$
as the effective Dirac operator for the light field.
As it turns out, in order to remedy this problem it is enough
to modify the definition of the effective Dirac operator by postulating that
its inverse is equal to the right-hand side of Eq.~(\ref{Deff}) up to a contact term.
For any $x\ne y$ the new free propagator will still be equal
to $\svev{q(x)\,\bq(y)}$, and will thus anti-commute with $\g_5$.
But this will no longer be true at coinciding points.
As we will see, the new Dirac operator satisfies the GW relation.

The construction works as follows \cite{Neuberger:1997bg,Kikukawa:2000ac}.
Starting from the representation of the product
\begin{equation}
\label{ZDW}
\ZDWF = \det(\DDWF(m=0))\, \det^{-1}(\DDWF(m=1)) \ ,
\end{equation}
as a path integral over five-dimensional
fermion and PV fields, we first integrate out most of the fermion and PV
degrees of freedom, obtaining (we suppress the dependence on $N_5$)
\begin{equation}
\label{ZqQ}
\ZDWF = \int dq d\bq\, dQ dQ^\dagger\, \exp\left(
-\bq \,\Deff\, q - Q^\dagger (\Deff+1) Q \right) \ ,
\end{equation}
where $\Deff$ is defined in Eq.~(\ref{Deff}).
The domain wall fermion degrees of freedom that we have integrated out include
the four-dimensional layers $s=2,3,\ldots,N_5-1$, as well as half of
the degrees of freedom on the two boundary layers, those that are {\em not}
used in the construction of the domain-wall quark field~(\ref{dwq}).
We treat the PV fields similarly, with the remaining $Q(x)$ and $Q^\dagger(x)$
four-dimensional PV fields constructed similarly to Eq.~(\ref{dwq}).
Notice that for the degrees of freedom that have been integrated out,
the domain-wall fermion and PV operators are identical, as none of these
degrees of freedom couple to the $m$-dependent terms in Eq.~(\ref{finites}).
Hence, the determinants resulting from this integration cancel each other.

The remaining degrees of freedom that occur on the right-hand side
of Eq.~(\ref{ZqQ}) are those that compose the domain-wall quark field,
as well as the corresponding PV degrees of freedom.
The difference between the two effective operators is simply because
the PV operator has the mass terms with $m=1$ coupling the two boundaries
in Eq.~(\ref{finites}), whereas for the domain-wall fermions we set $m=0$.
We now make a change of variables
\begin{equation}
\label{q2Psi}
\bJ = \bq \ , \qquad \J = (\Deff+1) q \ .
\end{equation}
Expressing the partition function in terms of the new variables gives
\begin{equation}
\label{ZGW}
\ZDWF = \int d\J d\bJ\, \exp\left(
-\bJ \,\DGW\, \J \right) \ ,
\end{equation}
where
\begin{equation}
\label{DGW}
\DGW = \frac{\Deff}{1+\Deff} \ ,
\end{equation}
and the change of variables~(\ref{q2Psi}) has absorbed the determinant
resulting from the integration over the remaining PV fields.
This result is valid for any $N_5$.

A first encouraging observation is that the fifteen poles that $\Deff$
admits for $N_5\to\infty$
cancel out between the numerator and the denominator in Eq.~(\ref{DGW}),
without affecting the existence of the relativistic massless state
at $p=0$.  Further understanding
of the physical nature of the new Dirac operator
comes from considering the new propagator.  From Eq.~(\ref{DGW}),
\begin{equation}
\label{GWprop}
\DGW^{-1} = \Deff^{-1} + 1 \ ,
\end{equation}
a result which is again valid for any $N_5$.  We have seen that
for $N_5\to\infty$, the propagator $\Deff^{-1}$ anti-commutes with $\g_5$.
Hence in this limit
\begin{equation}
\label{GW5}
\{\DGW^{-1},\g_5\} = \{\Deff^{-1},\g_5\} + 2\g_5 = 2\g_5 \ .
\end{equation}
As promised, in this limit
the new propagator $\DGW^{-1}$ still anti-commutes with $\g_5$,
{\em except} at coinciding points.  In fact, what we have found is
that for $N_5\to\infty$, the effective Dirac operator
$\DGW$ satisfies the GW relation,
\begin{equation}
\label{GWrel}
\{\DGW,\g_5\} = 2\DGW \g_5 \DGW \ .
\end{equation}

So far we have only used the general structure of the domain wall
operator~(\ref{finites}) and the restoration of chiral symmetry
for $N_5\to\infty$.  A lengthy derivation
\cite{Neuberger:1997bg,Kikukawa:2000ac} yields an explicit expression
for the approximate GW operator at finite $N_5$,
\begin{subequations}
\label{DGWN}
\begin{eqnarray}
\label{DGWNa}
\DGW(N_5) &=& \half\Big(1 + \g_5 \tanh\Big((N_5/2) H_T\Big)\Big)
\\
&=& \half \left(1 + \g_5 \frac{1-T^{N_5}}{1+T^{N_5}}\right) \ , \rule{0ex}{4ex}
\label{DGWNb}
\end{eqnarray}
\end{subequations}
where the transfer matrix $T$ is given in Eq.~(\ref{T}), and
\begin{equation}
\label{HlogT}
H_T = -\half \log T^2 \ .
\end{equation}
This definition ensures that $H_T$ is well defined in the presence of
both positive and negative eigenvalues of the transfer matrix $T$.\footnote{
  Negative eigenvalues of the transfer matrix are discussed in
  Sec.~\ref{chiragain}. Since we always take $N_5$ even, we are allowed
  to replace $T$ by the positive square root of $T^2$ in Eq.~(\ref{DGWNb}).
}
In the limit $N_5\to\infty$ we recover the familiar expression
for the GW operator,
\begin{equation}
\label{DGWlim}
\DGW = \half\Big(1 + \g_5\, \e(H)\Big) \ ,
\end{equation}
where $\e(x)$ is the sign function,
$\e(x)=+1$ for $x>0$ and $\e(x)=-1$ for $x<0$, and $H=H_T$.
The difference between this GW operator and the
standard overlap operator \cite{Neuberger:1997fp} is in the choice of $H$.
While here $H_T$ is defined in terms of the transfer matrix via Eq.~(\ref{HlogT}),
in the case of the standard overlap operator $H$ is chosen to be
the hermitian version of the Wilson kernel, $H_W = \g_5 D$ (see Eq.~(\ref{DW})).
We will further examine the relation between these two GW operators
in Sec.~\ref{kernel}, where we also discuss the special case
of zero eigenvalues of $H$.

The generalization to a nonzero mass for the domain wall fermion
is obtained by simply replacing $\bq \,\Deff\, q$ by $\bq(\Deff+m)q$
in Eq.~(\ref{ZqQ}) and repeating the subsequent steps.
Equation~(\ref{DGW}) can be inverted to give
\begin{equation}
\label{DGWinv}
\Deff = \frac{\DGW}{1-\DGW} \ ,
\end{equation}
and for $N_5\to\infty$ we find
\begin{equation}
\label{DGWm}
\DGW(m)
= \frac{1+m}{2} + \frac{1-m}{2} \g_5\e(H)
= \DGW(0) + m\Big(1-\DGW(0)\Big) \ ,
\end{equation}
where the massless GW operator $\DGW=\DGW(0)$ is given in Eq.~(\ref{DGWlim}).
For more details, see for example Refs.~\cite{Neuberger:1997bg,RBC:2014ntl}.

In summary, our starting point was the four-dimensional operator $\Deff$
defined by Eq.~(\ref{Deff}), which anti-commutes with $\g_5$, hence it is
invariant under ordinary chiral transformations.
But $\Deff$ also has some undesirable properties, which are necessary
in order to reconcile its chiral invariance with the no-go theorems.
Instead, we have reached a closely related operator that satisfies
the GW relation.  Any GW operator enjoys a modified form
of chiral symmetry \cite{Luscher:1998pqa}.  Here we will only note
that the chiral anomaly is correctly reproduced by the variation
of the lattice fermion measure under this modified chiral transformation.
For a detailed discussion of the GW relation, including its modified
chiral symmetry, we refer to the companion chapter of the {\em LQCD@50} book
by DeGrand \cite{DeGrand:2025ttq}.

We conclude with a technical comment.  The PV fields are bosonic variables,
and so in order to make the integration over them rigorously well defined
an analytic continuation is needed.  With $D_{PV}=\DDWF(m=1)$,
and denoting the PV fields as $\Phi$ and $\Phi^\dagger$, we have
\begin{equation}
\label{PVcont}
\det^{-1}(D_{PV}) = \lim_{\e\to 0} \int d\Phi d\Phi^\dagger
\exp\left(\Phi^\dagger \left(i\g_5\car D_{PV}-\e\right) \Phi \right) \ .
\end{equation}
Thanks to the $\g_5$-hermiticity properties discussed below Eq.~(\ref{finites}),
the eigenvalues of $i\g_5\car D_{PV}$ are pure imaginary,
and the added factor of $\e$ ensures convergence of the integral.\footnote{
  For a similar analytic continuation, see Ref.~\cite{Golterman:2005ie}.
}
All the results we have obtained above remain valid with this
more careful procedure.


\section{\label{resid} Residual breaking of chiral symmetry}
So far, the quantitative information we have on domain wall fermions
was mostly restricted to the free theory (Sec.~\ref{free}).  Beyond that,
the discussion of chiral symmetry restoration in Sec.~\ref{interacting}
gave us a glimpse into the dynamics of the interacting theory.
In particular, we observed that the transfer matrix can have eigenvalues
equal to unity.  Hence, the probability to encounter near-unity eigenvalues
of the transfer matrix in an ensemble of configurations is going
to affect the rate at which chiral symmetry is restored for $N_5\to\infty$.

In the rest of this introduction we will delve deeper into the physics of
domain wall fermions.  The main issue is the rate of restoration
of chiral symmetry with increasing $N_5$,
and the quality of chiral symmetry which is achievable in practice
in numerical simulations.  In this section we will introduce the notion
of the residual mass, which is a quantitative measure of the imperfect
chiral symmetry at finite $N_5$.  We will study the restoration
of chiral symmetry for $N_5\to\infty$, equivalently the vanishing
of the residual mass, within lattice perturbation theory.
Later, in Sec.~\ref{mresagain}, we will extend the discussion to non-perturbative effects (this will include the role of
the near-unity eigenvalues of the transfer matrix).

We begin in Sec.~\ref{wavef} with a calculation
of the dependence of the wave function of a domain-wall quark on the
fifth coordinate.  In the free theory, the wave function is just $(1-M)^s$.
In particular we noted that by setting the domain wall height to $M=1$,
this wave function is entirely supported on the relevant boundary layer
when $m=0$. We cannot expect this ideal situation to survive when the
gauge interactions are turned on, which leads us to study the wave function
in (tadpole improved) one-loop perturbation theory as the next step.

In Sec.~\ref{mres} we turn
to the most important physical effect that depends on the ``tail''
of the wave function. Much like (four-dimensional) Wilson fermions,
this is an additive correction to the mass of the domain-wall quark,
which is not proportional to the parameter $m$ introduced into the
domain-wall fermion operator~(\ref{finites}).  While the additive correction
is guaranteed to vanish for $N_5\to\infty$ (Sec.~\ref{chirN5}),
where chiral symmetry is restored,
in order to design numerical calculations using domain wall fermions
we need a pretty good idea what to expect for finite $N_5$.

Wilson fermions completely lack chiral symmetry at finite lattice spacing.
This implies that the bare Wilson mass is not renormalized multiplicatively.
In numerical lattice calculations the additive correction to the
bare Wilson mass is typically $O(1)$ in lattice units.
Large additive corrections similarly afflict other chirally sensitive
observables calculated using Wilson fermions.
By contrast, for domain wall fermions it is common to achieve
an additive mass correction smaller by several orders of magnitude.
Hence the name residual mass \cite{Blum:1997mz,Blum:2000kn}.
This is a dramatic improvement over Wilson fermions, that extends
to other chirally sensitive observables as well.  Unlike staggered fermions,
at the same time the (vector) flavor symmetry is fully preserved.

\subsection{\label{wavef} Wave function of domain wall fermions}
In Sec.~\ref{tadpole} we discuss tadpole improvement of the domain-wall quark
wave function, and in Sec.~\ref{oneloop} the genuine one-loop correction.

\begin{figure}[t]
\begin{center}
\includegraphics*[width=4cm]{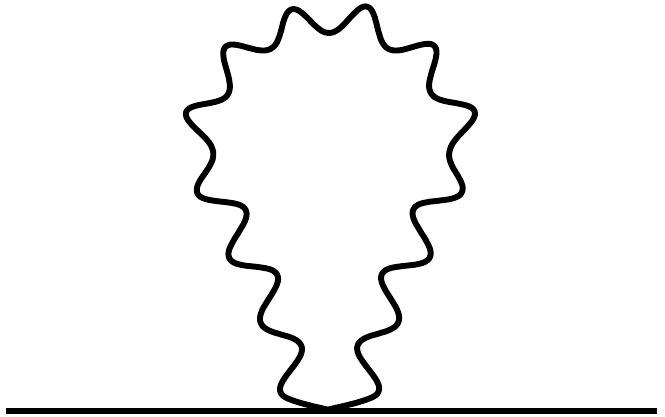}
\end{center}
\begin{quotation}
\floatcaption{tadpolefig}%
{Tadpole correction in one-loop lattice perturbation theory.
}
\end{quotation}
\vspace*{-4ex}
\end{figure}

\subsubsection{\label{tadpole} Tadpole improvement}
We start with the tadpole correction
\cite{Aoki:1997xg,Aoki:1998vv,Shamir:2000cf}.
In the limit $N_5\to\infty$,
hence in the absence of any mixing between the RH and LH zero modes,
the tadpole improved tree-level wave function of the RH zero mode is
(as in Eq.~(\ref{dwlatt}) we omit spinor indices and disregard
an overall normalization factor)
\begin{equation}
\label{dM}
\c_R(s) = (1+\d M-M)^s = (1-\Msub)^s \ ,
\end{equation}
where $\Msub = M-\d M$.  Diagrammatically, the leading tadpole correction
arises from the one-loop diagram shown in \Fig{tadpolefig}.
As such, the tadpole correction $\d M$ is formally an $O(g_0^2)$ effect.
We can expand the tadpole improved wave function around the tree level one,
\begin{equation}
\label{dMbinom}
(1+\d M-M)^s
= (1-M)^s \left( 1 + s\,R_M + \frac{s(s-1)}{2}\, R_M^2 + \cdots \right) ,
\end{equation}
where $R_M = \d M/(1-M)$.
Naively, it appears that $R_M$ is also an $O(g_0^2)$ quantity,
and in view of the usual rules of a perturbative expansion
we ought to truncate the expansion~(\ref{dMbinom}) at linear order in $R_M$.
In fact, this truncation must {\em not} be done, as we will now explain.

The first argument invokes the order of limits adequate for
domain wall fermions.  Since the fifth coordinate $s$ ranges from 1 to $N_5$,
the expansion parameter of Eq.~(\ref{dMbinom}) is effectively $g_0^2 N_5$.
If the chiral limit $N_5\to\infty$ is taken first, then $g_0^2 N_5$ diverges.
If the continuum limit $g_0^2\to 0$ is taken first, $g_0^2 N_5$ vanishes.
However, it is the chiral limit that we want to take first, in order
to restore chiral symmetry of the effective four-dimensional theory
already at finite lattice spacing (Sec.~\ref{chirN5} and Sec.~\ref{DWF2GW}).
Hence $g_0^2 N_5\gg 1$,
and we must always resum the expansion in~(\ref{dMbinom}).
This, of course, recovers the tadpole-improved tree level wave function
in Eq.~(\ref{dM}).

A comparison of Eqs.~(\ref{B}) and~(\ref{dM}) suggests that the physical origin
of the tadpole correction $\d M$ is the expectation value of
the (positive) Wilson term $W$ on the dominant eigenmodes of the
transfer matrix $Q$: the eigenmodes whose eigenvalues
are closest to unity (Sec.~\ref{chirN5}).  Equivalently, these are the
near-zero eigenmodes of the transfer matrix hamiltonian $H_T$, Eq.~(\ref{HlogT}).
Let us look for numerical evidence
that the additive mass renormalization $m_c$ of Wilson fermions
represents essentially the same physical effect as the additive correction
to the optimal domain wall height.  Why this is the case will be made clearer
in Sec.~\ref{kernel}.

When using Wilson fermions, in order to recover a massless quark
in the continuum limit we must tune the bare mass $m_0$ non-perturbatively
such that $m_0+m_c=0$.  Hence $m_0 = -m_c$, where the value of the
critical mass $m_c$ is usually in the range of 0.7 -- 1.0
for typical values of the bare coupling in simulations \cite{Lepage:1992xa}.
As for domain wall fermions, the values of the domain wall height $M$
that yield the best damping of the domain wall quark wave function
are usually found to be in the range of 1.7 -- 1.8.
As a heuristic way to determine the tadpole correction $\d M$
let us assume that this optimal domain wall height corresponds
to the most localized tree-level improved wave function,
for which $\Msub=1$ in Eq.~(\ref{dM}).  Equivalently, this implies $\d M = M-1$
after optimizing the value of $M$.  With the above range for the optimal $M$,
we obtain that $\d M$ is 0.7 -- 0.8,
nicely consistent with the range of the critical Wilson mass $m_c$.

Back to Eq.~(\ref{dMbinom}), the above procedure for estimating $\d M$
amounts to tuning the domain wall height $M$ such that $\d M + 1 - M = 0$,
hence $R_M=-1$.  While it is formally true that $\d M = O(g_0^2)$,
we see that $1-M$ is tuned to be of exactly the same size.
This leads to a stronger conclusion: as long as we apply this procedure,
it does not matter which limit will be taken first, the chiral limit
$N_5\to\infty$ or the continuum limit $g_0^2\to 0$.  Either way,
the expansion in Eq.~(\ref{dMbinom}) is not controlled by a small parameter,
and thus it must always be resummed to yield the tadpole improved
wave function~(\ref{dM}).

With $M$ in the range 1.7 -- 1.8, which means $1<M<2$,
the operator $B$ will have negative eigenvalues,
and the same is true for the transfer matrix.\footnote{%
  For example, in the free theory for $p=0$.
}
In practice, negative eigenvalues of $B$ are rarely encountered,
suggesting that on a realistic ensemble of gauge field configurations the
expectation value of the Wilson term that we have loosely described above
is fairly stable, and bounded below by $M-1$.
We will return to the connection between the critical Wilson mass
and the optimal domain wall height in Sec.~\ref{quench}.

\begin{figure}[t]
\begin{center}
\includegraphics*[width=8cm]{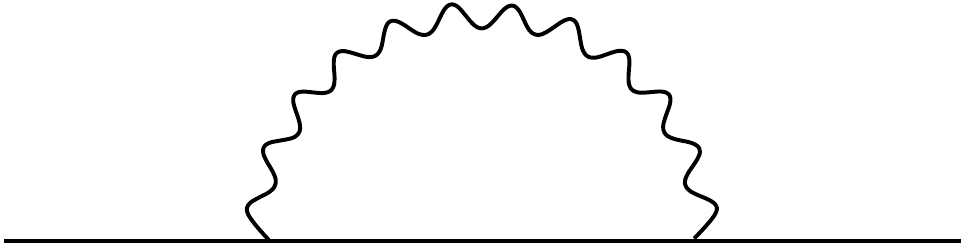}
\end{center}
\begin{quotation}
\floatcaption{sunset}%
{Setting sun diagram, the dominant contribution to the quantum wave function
  of the domain wall quark.
}
\end{quotation}
\vspace*{-4ex}
\end{figure}

\subsubsection{\label{oneloop} One loop wave function}
We now turn to the genuine one-loop correction to the domain wall quark
wave function which results from the ``setting sun'' diagram in \Fig{sunset}.
We will first present the final result,
and then explain the main steps of the calculation.\footnote{%
  For some discussion of two loop and higher corrections
  see Ref.~\cite{Shamir:2000cf}.
}
Following Ref.~\cite{Shamir:2000cf}, in this subsection we set $m=0$.

We define an effective wave function $\ceff(s)$ for the domain wall quark by looking at
the singular part of the dressed domain wall propagator $G_{s,s'}(p)$ arising
from the RH massless field near the $s=0$ boundary,
the part proportional to $P_R\, {\sl{p}}^{-1}$ (compare Eq.~(\ref{GF})).
Taking the limit $N_5\to\infty$, we may calculate the perturbative corrections
to the effective wave function using the semi-infinite setting of Sec.~\ref{semi5}.
The factor multiplying $P_R\, {\sl{p}}^{-1}$ is then expected to be
the product of the (index-less) effective wave functions at $s$ and $s'$.
We thus make the {\em ansatz}
\begin{equation}
\label{Gsnglr}
  G_{s,s'}(p) = \ceff(s)\, P_R\,\frac{1}{i{\sl{p}}(1+g_0^2\S_K)}\, \ceff(s')
  + \cdots\ ,
\end{equation}
neglecting terms whose $p$ dependence is regular.
Here $\S_K$ is a standard logarithmic correction that leads to
a wave-function renormalization of the domain wall quark field \cite{Aoki:1997xg}.
As a further simplification, we will calculate $\ceff(s)$ for $s,s'\gg 1$.
Within tadpole-improved one loop perturbation theory one finds\footnote{
  In this subsection we assume that the fifth coordinate takes values
  $s=0,1,2,\ldots$.
}
\begin{equation}
\label{wf1}
  \ceff(s) = \left(\Msub\left(2-\Msub\right)\right)^{1/2}
  \left(1-\Msub\right)^s
  + \c_1(s) = \d_{s,0} + \c_1(s) \ ,
\end{equation}
where the second equality is valid in the limit $\Msub\to 1$,
equivalently, after tuning $M$ to its optimal value $1+\d M$.
The genuine one-loop wave function is then
\begin{equation}
\label{dchi}
  \c_1(s) = - c g_0^2 \, \frac{1}{s^2\, 2^s} \ ,
\end{equation}
where $c \approx 0.8$.

This result is obtained as follows.  Tuning $M$ to its optimal value
in the tadpole improved wave function $(1+\d M - M)^s$ is effectively
the same as taking the limit $M\to 1$ in the tree level wave function $(1-M)^s$.
We similarly take the limit $M\to 1$ in the expressions for the
tree level propagator obtained in Sec.~\ref{semi5}.  On the external legs of the
self-energy diagram in \Fig{sunset} we thus have both $M\to 1$ and $p\to 0$.
As is easily inferred from the derivation in Sec.~\ref{semi5},
in this special limit the domain wall propagator can connect a given
four-dimensional layer only to itself or to one of its neighboring layers.

By contrast, the domain wall propagator inside the loop can carry any momentum,
and can therefore connect any two four-dimensional layers.
When one of these layers is the $s=0$ boundary or very close to it,
the amplitude is $\sim e^{-\a(p)s}$.  Moreover, the gauge field couples equally
to all the four-dimensional layers, and provides no extra suppression
as the separation in the fifth direction grows.  The outcome is that for
large separations in the fifth direction, the self-energy diagram
behaves like $\sim \etamax^s$, where $\etamax$ is
the maximum of $e^{-\a(p)}$ over the whole Brillouin zone.
For $M\to 1$ the maximum is $\etamax=1/2$, and it is obtained
at the four corners of the Brillouin zone where one momentum component
is equal to $\p$, while the rest are zero.  The exponential dependence
of the self-energy on $s$ thus goes like $1/2^s$.  The $1/s^2$ power correction
in Eq.~(\ref{dchi}) arises from the four-dimensional phase space
of a saddle-point integral around any one of the four degenerate maxima
of $e^{-\a(p)}$ \cite{Shamir:2000cf}.

The lesson from this calculation is the following.  Both the
purely tree-level and the tadpole improved wave functions can
be tuned to be entirely localized on the boundary.  But this is no longer
possible for the genuinely quantum part of the wave function.  The quantum
broadening of the wave function occurs because the virtual fermion
propagating inside the diagram can have arbitrary momentum
(with a matching momentum carried by the gauge field).
Hence, the least damped propagation in the fifth direction is always
controlled by the maximum of $e^{-\a(p)}$ over the Brillouin zone.
More generally, in an arbitrary gauge field background
the least damped propagation will originate from the largest
eigenvalues of the transfer matrix $Q$ (Eq.~(\ref{Q})),
as we will discuss in Sec.~\ref{mresagain}.

\begin{figure}[t]
\begin{center}
\includegraphics*[width=4cm]{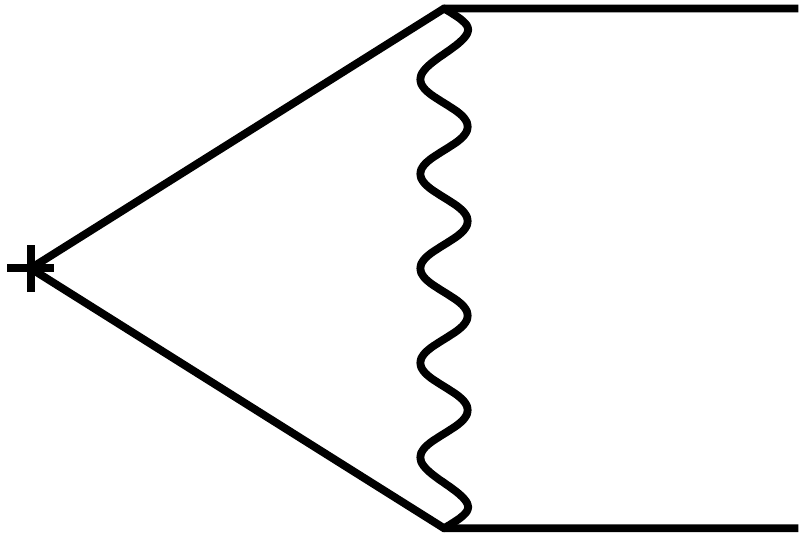}
\end{center}
\begin{quotation}
\floatcaption{mresfig}%
{Mixing of $J_{5q}^a$ with $J_5^a$, or $J_{5q}$ with $J_5$.
}
\end{quotation}
\vspace*{-4ex}
\end{figure}

\subsection{\label{mres} The residual mass}
There is no consistent regularization method that fully preserves
chiral symmetry.  The reason is that any consistent regularization
must reproduce the chiral anomaly.\footnote{%
  For the way the chiral anomaly arises for GW fermions,
  see, \eg, Refs.~\cite{DeGrand:2006zz,Niedermayer:1998bi}.
}
For example, in the continuum, and for a vanishing fermion mass,
the classical continuity equation of a flavor non-singlet axial current is
\begin{equation}
\label{Ja5mu}
\partial_\m J_{5\m}^a = 0 \ .
\end{equation}
When dimensional regularization is used, the continuity equation
develops a nonzero right-hand side
\begin{equation}
\label{Ja5mudr}
\partial_\m J_{5\m}^a = \D^a \ , \qquad d\ne 4\ .
\end{equation}
Here $\D^a$ is an {\em evanescent} operator:
an operator that formally vanishes for $d\to4$.
Thanks to the cancellation between a factor of $d-4$, arising
from the evanescent operator, and a factor of $1/(d-4)$,
arising from the loop integral, for $d\to 4$ one obtains
\begin{equation}
\label{Devn}
\D^a = (1-Z_5) \partial_\m J_{5\m}^a \ ,
\end{equation}
with finite $Z_5$.  This result
implies that the renormalized current (defined by generating
canonically normalized contact terms in WTIs) is $Z_5 J_{5\m}^a$.
In this case one can further show that in fact $Z_5=1$, namely,
the non-singlet axial currents are unrenormalized,
at least at one loop \cite{Collins:1984xc}.
However, this depends on further specific properties of
dimensional regularization that need not apply to other
consistent regularization schemes.

In the case of domain wall fermions, the counterpart of Eq.~(\ref{Ja5mudr})
is Eq.~(\ref{divA}), in which the quantum breaking of the conservation
of a non-singlet axial current $\ca^a_\m$ is represented by the mid-point
pseudoscalar density $J_{5q}^a$.
Re-introducing the lattice spacing $a$ and
generalizing Eq.~(\ref{Devn}) to $a>0$ we may write
\begin{equation}
\label{mresdef}
(1/a)J_{5q}^a(x)
= \mres J_5^a(x) + \frac{1-Z_\ca}{2} \sum_\m \D_\m \ca^a_\m(x) + O(a) \ .
\end{equation}
This equation expresses the action of the operator $J_{5q}^a$
on states created from the domain-wall quark fields $q$ and $\bq$
(see the discussion around Eq.~(\ref{J5qaq})).
The expansion on the right-hand side involves operators
with the same quantum numbers as $J_{5q}^a$ under the exact lattice symmetries.
In addition to the dimension-three operator $J_5^a$
and the dimension-four operator $\sum_\m \D_\m \ca^a_\m$,
there are corrections that involve operators of dimension five and higher
which are made out of the quark fields $q$ and $\bq$, and are
multiplied by positive powers of the lattice spacing.
This equation defines the residual mass $\mres$ as the coefficient of $J_5^a$
on the right-hand side. $\mres$ is thus a function of the bare coupling only.
We recall that $J_5^a$ is the standard pseudoscalar density
constructed from the quark fields on the four-dimensional boundaries.
Similarly, $(1-Z_\ca)/2$ is the coefficient of $\sum_\m \D_\m \ca^a_\m$ itself.
Equation~(\ref{mresdef}) applies within the Symanzik effective theory,
which is an expansion in powers of the lattice spacing.
The terms shown explicitly in Eq.~(\ref{mresdef})
are the ones that survive in the continuum limit.
Substituting this expansion back into Eq.~(\ref{divA}), it becomes\footnote{%
  In this subsection we are not concerned with multiplicative
  mass renormalization, and the associated renormalization
  of scalar and pseudoscalar densities,
  hence $m$, $\mres$ and $J_5^a$ are all bare quantities.
}
\begin{equation}
\label{divAren}
Z_\ca \sum_\m \D_\m \ca^a_\m(x) = 2(m+\mres) J^a_5(x) + O(a) \ .
\end{equation}
This result applies when considering any correlation function
involving only the boundary quark fields, as per Eq.~(\ref{J5qaq}).
The WTIs that follow from this partial conservation equation will have
canonically normalized contact terms, as in Eq.~(\ref{WTI}).

We would like to draw attention to a somewhat unusual feature
of Eq.~(\ref{mresdef}).  Suppose that our goal was to expand the
mid-point operator $J_{5q}^a$, which lives inside the five-dimensional bulk,
in terms of operators made out solely of the $q$ and $\bq$ fields,
from which the effective four-dimensional theory is built.  In this case
we should not make use of the partially-conserved current $\ca^a_\m$
in the expansion, because this current is not built solely
from the $q$ and $\bq$ fields. However, our ultimate goal
is to reach the expansion~(\ref{divAren}) for the partially-conserved
axial current itself, hence the presence of the $\sum_\m \D_\m \ca^a_\m$ term
on the right-hand side of Eq.~(\ref{mresdef}).  Our final result,
Eq.~(\ref{divAren}), indeed meets the expectation that all the operators
occurring on its right-hand side are built
from the $q$ and $\bq$ fields only.

In Sec.~\ref{wavef} we have seen that the quantum wave function
of the domain-wall quarks decays exponentially as we move in the $s$ direction
away from the boundaries.  When the domain wall height is tuned
to its optimal value, the actual falloff rate is $\etamax^s$.
Since $J_{5q}^a$ must connect
to quark and anti-quark operators on the boundaries,
this will always come with a factor of $\etamax^{2(N_5/2)}=\etamax^{N_5}$.
In the rest of this subsection we will elaborate on this observation.

Consider first the mixing of $J_{5q}^a$ with $J_5^a$ for finite $N_5$.
As we will shortly see, this mixing arises at the one loop level
from the diagram shown in \Fig{mresfig}.  Before we discuss this
one-loop diagram, however, let us consider the corresponding tree diagram
obtained by omitting the gauge field propagator in \Fig{mresfig}.
In this tree diagram, $J_{5q}^a$ will be connected to the quark fields
at the boundaries by two tree-level propagators in which we must take
the limits $M\to 1$ and $p\to 0$.  But, as we have seen in Sec.~\ref{wavef},
the tree-level propagator vanishes for this kinematics.

The first diagram that may contribute to the mixing of $J_{5q}^a$ with $J_5^a$
is thus the one-loop diagram of \Fig{mresfig}.  This works in essentially
the same way as in Sec.~\ref{wavef}.  While the external propagators
are constrained by both $M\to 1$ and $p\to 0$, the internal fermion propagators
can carry any momentum.  Once again, this implies that the loop integral
will be dominated by those virtual momenta for which $e^{-\a(p)}$ is close
to its maximal value $\etamax$.  The resulting exponential
suppression factor will be $\sim g_0^2 \etamax^{N_5}$ (this is true
up to subleading power-law corrections, compare Eq.~(\ref{dchi})).
We thus expect $\mres$ in Eq.~(\ref{divAren})
to be parametrically of this size, as far as perturbative corrections
are concerned.  The same estimate applies to the magnitude of $Z_\ca-1$,
when all quantities are evaluated in lattice units:
\begin{equation}
\label{mresZ}
a\mres,\, Z_\ca-1 \,\sim\, g_0^2 \etamax^{N_5} \ .
\end{equation}

\begin{figure}[t]
\begin{center}
\includegraphics*[width=7cm]{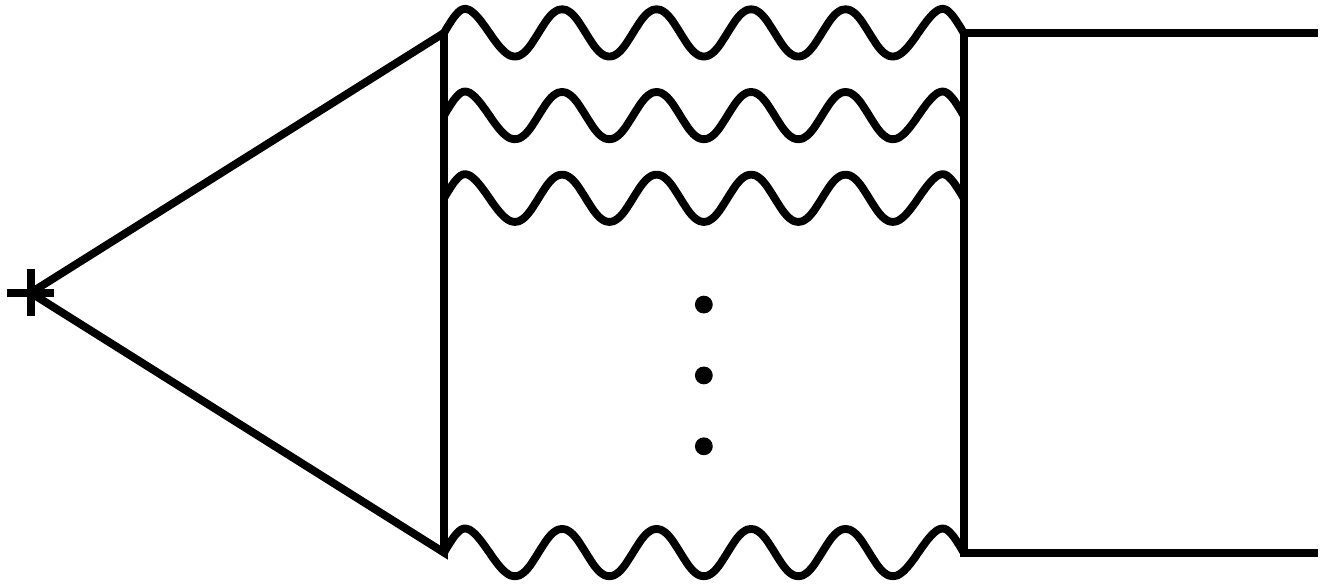}
\end{center}
\begin{quotation}
\floatcaption{mresingfig}%
{Mixing of $J_{5q}$ with $J_5$ only.
}
\end{quotation}
\vspace*{-4ex}
\end{figure}

Turning to the singlet axial current, we follow the same reasoning
while taking into account the axial anomaly (Eq.~(\ref{dAanml})).  This leads to
\begin{eqnarray}
\label{divAs}
Z_\ca^s \sum_\m \D_\m \ca_\m(x) &=& 2(m+\mres^s) J_5(x)
\\
&& + \frac{ig_0^2 N_F}{4\p^2}\, \sum_{\m\n\l\r}
\e_{\m\n\l\r} \partial_\m A_\n(x) \partial_\l A_\r(x) + O(a) \ ,
\nonumber
\end{eqnarray}
where the superscript $s$ indicates quantities pertaining to the singlet
axial current.  The question arises whether or not the parameter $\mres^s$
associated with the singlet axial current is equal to the
residual mass parameter $\mres$ of the non-singlet currents.
The answer is that they are different.
A class of diagrams that contribute to $\mres^s$
but not to $\mres$ is shown in \Fig{mresingfig}.
Similarly, $Z_\ca^s$ and $Z_\ca$ are not equal.

An alternative notion of the residual mass\footnote{%
  Historically, the residual mass was defined by Eq.~(\ref{mresp})
  when it was first introduced \cite{Blum:1997mz}.
}
may be introduced by directly comparing a physical
matrix element of $J_{5q}^a$ with the corresponding matrix element of $J_5^a$
\cite{Blum:2000kn,CP-PACS:2000fmi},
\begin{equation}
\label{mresp}
\mres' = \frac{1}{a}\frac{\sbra{0} J_{5q}^a \sket{\p^a}}
       {\sbra{0} J_5^a \sket{\p^a}} \ .
\end{equation}
The pion state is taken at $\vec{p}=0$.
Unlike $\mres$ of Eq.~(\ref{mresdef}), which is by definition
independent of the input quark mass $m$, Eq.~(\ref{mresp})
allows $\mres'$ to depend on $m$.
One can determine $\mres$ at one particular value of the bare coupling
by first obtaining $\mres'$ for several values of $m$,
and then extrapolating $\mres'$
to the limit of a vanishing input mass, $m\to 0$.
In practice, the differences between $\mres'$ and $\mres$
turn out to be at the few percents level (see, for example,
Fig.~7 of Ref.~\cite{RBC-UKQCD:2008mhs}).  Hence, both $\mres$ and $\mres'$ qualify
as a qualitative measure of the goodness of chiral symmetry at finite $N_5$.

Expressions such as Eq.~(\ref{divAren}), and their generalizations
to higher powers of lattice spacing $a$ within the Symanzik effective theory,
provide a crucial intermediate step in the application of
chiral perturbation theory for domain wall fermions.
For details, we refer to the original literature
\cite{Sharpe:2007yd,RBC-UKQCD:2008mhs,RBC:2010qam,RBC:2014ntl}.

\section{\label{kernel} The Wilson kernel}
The goal of this section is to explore non-perturbative effects associated
with the Wilson kernel.  These effects influence the rate of restoration
of chiral symmetry with increasing $N_5$, as well as the locality properties
of the effective four-dimensional operator for the domain wall quarks.
As it turns out, some ground work has to be done first.

We start in Sec.~\ref{aoki} with a seemingly unrelated topic:
the phase diagram of QCD with two flavors of dynamical Wilson fermions.
Three different types of phases are identified.  This includes
the Aoki phase in which isospin, the vector SU(2) symmetry,
is spontaneously broken to U(1) by a pion condensate; parity is
spontaneously broken as well.

The next task is to understand the relevance of this phase diagram for
dynamical domain wall fermions.
Crucially, from the perspective of the Wilson operator, or Wilson kernel,
this means that we are dealing with a quenched phase diagram.
What we mean by this is the following.  All expectation values will be defined
by the partition function of dynamical domain wall fermions.
For each quark flavor the Boltzmann weight contains the determinant of
the domain wall operator, $\det(\DDWF(m))$, with an appropriate value of $m$,
times the PV determinant $\det^{-1}(\DDWF(1))$ (see Eq.~(\ref{Z}));
the Boltzmann weight does not depend directly
on the determinant of the Wilson kernel.  We will refer to this setup
as {\em Wilson-quenched}, to distinguish it from the fully quenched case,
in which no fermion determinant at all is included in the Boltzmann weight.
The conceptually more involved Wilson-quenched phase diagram is discussed
in Sec.~\ref{quench}.  We identify the phase in which domain wall (or overlap)
fermions are to be defined.  We also further clarify the connection,
already discussed in Sec.~\ref{tadpole}, between the optimal domain wall
height $M$ and the critical Wilson-fermion mass $m_c$.

In Sec.~\ref{topo} we discuss the key role of the Wilson kernel's zero modes
in defining topological sectors.
In Sec.~\ref{rho} we extend the discussion to the kernel's low lying spectrum.
It is convenient to work with the hermitian version of the Wilson operator
\begin{equation}
\label{HW}
H_W = \g_5 D \ ,
\end{equation}
where $D$ is defined in Eq.~(\ref{DW}).  We present evidence that the
spectrum of $H_W$ has no gap everywhere in the super-critical region.
In terms of its spectral density $\r(\l)$, this means that
$\r(\l) \ne 0$ for small $\l$, except possibly for $\l=0$.
For the theory with dynamical Wilson fermions,
$\r(\l)\to 0$ for $\l\to 0$ in the super-critical region,
except inside the Aoki phase where $\r(0) > 0$.
By contrast, in the Wilson-quenched phase diagram
we argue that $\r(0) > 0$ throughout the entire super-critical region.
This conclusion appears to challenge the standard physical picture
of spontaneous symmetry breaking: the connection between
the spectral density $\r(0)$, the condensate, and the Goldstone theorem.

The resolution of this paradox is presented in Sec.~\ref{lcl}.
It is based on concepts familiar in condensed matter physics.
A key notion is the mobility edge $\l_c\ge 0$, which distinguishes
between {\em extended} eigenmodes of $H_W$ for $|\l|>\l_c$
and exponentially {\em localized} eigenmodes for $|\l|<\l_c$.
We will see how the mobility edge helps us identify the various phases,
and explain how the Goldstone theorem is avoided in a Wilson-quenched phase
when the eigenmodes with $\l\sim 0$ are localized.
The related important issue of locality of the effective four-dimensional
operator~(\ref{DGWN}), as well as of the overlap operator,
is addressed in Sec.~\ref{lcleff}.

In Sec.~\ref{mresagain} we turn to the implications for the residual mass.
The near-zero modes of the Wilson kernel give rise to a power-law component
of the wave function of the domain wall quarks.  The corresponding
contribution to the residual mass generically decays only like $1/N_5$.
While this decay rate is very slow, the magnitude of this component
can still be small provided that the spectral density of near-zero modes
is small.

In Sec.~\ref{remedies} we discuss methods for decreasing the slow
$1/N_5$ component of the residual mass by reducing the density
of the Wilson kernel's near-zero modes.
We conclude with the emerging general strategy for cost-optimal
numerical calculations with domain wall fermions.
Finally Sec.~\ref{conc} contains a brief summary comparing
domain wall fermions to GW fermions on the one hand and to ordinary
Wilson fermions on the other hand.  Methods for
further reducing the residual breaking of chiral symmetry that involve
changing the domain-wall operator itself will be discussed in Sec.~\ref{improve}.

\subsection{\label{aoki} Aoki phase}
In this subsection we review the phase diagram of QCD with two flavors
of dynamical Wilson fermions.  The following heuristic argument
suggests the existence of a phase with a pionic condensate
\cite{Aoki:1983qi,Aoki:1986xr}.  Consider the effective potential for pions,
and let $m_\p^2$ be the curvature of this potential at the origin.
According to continuum chiral perturbation theory, $m_\p^2 \propto m$,
where $m\ge 0$ is the quark mass (in the following we assume
that proportionality constants are positive).  By contrast, for Wilson fermions
there is an additive correction to the bare mass $m_0$,
and the continuum relation gets replaced by
\begin{equation}
\label{m0mc}
m_\p^2 \propto m_0 + m_c(g_0) \ ,
\end{equation}
with a critical mass $m_c(g_0)>0$ which is a function of the bare coupling.
As long as $m_0 + m_c(g_0) > 0$ the curvature, $m_\p^2$, is positive, and the
pions are massive.  When $m_0$ is tuned to $-m_c(g_0)$ the curvature vanishes,
and accordingly the pions become massless.  If we further decrease $m_0$,
we expect that the curvature $m_\p^2$ at the origin will become negative,
hence, the pion field will condense.

\begin{figure}[t]
\begin{center}
\includegraphics*[width=10cm]{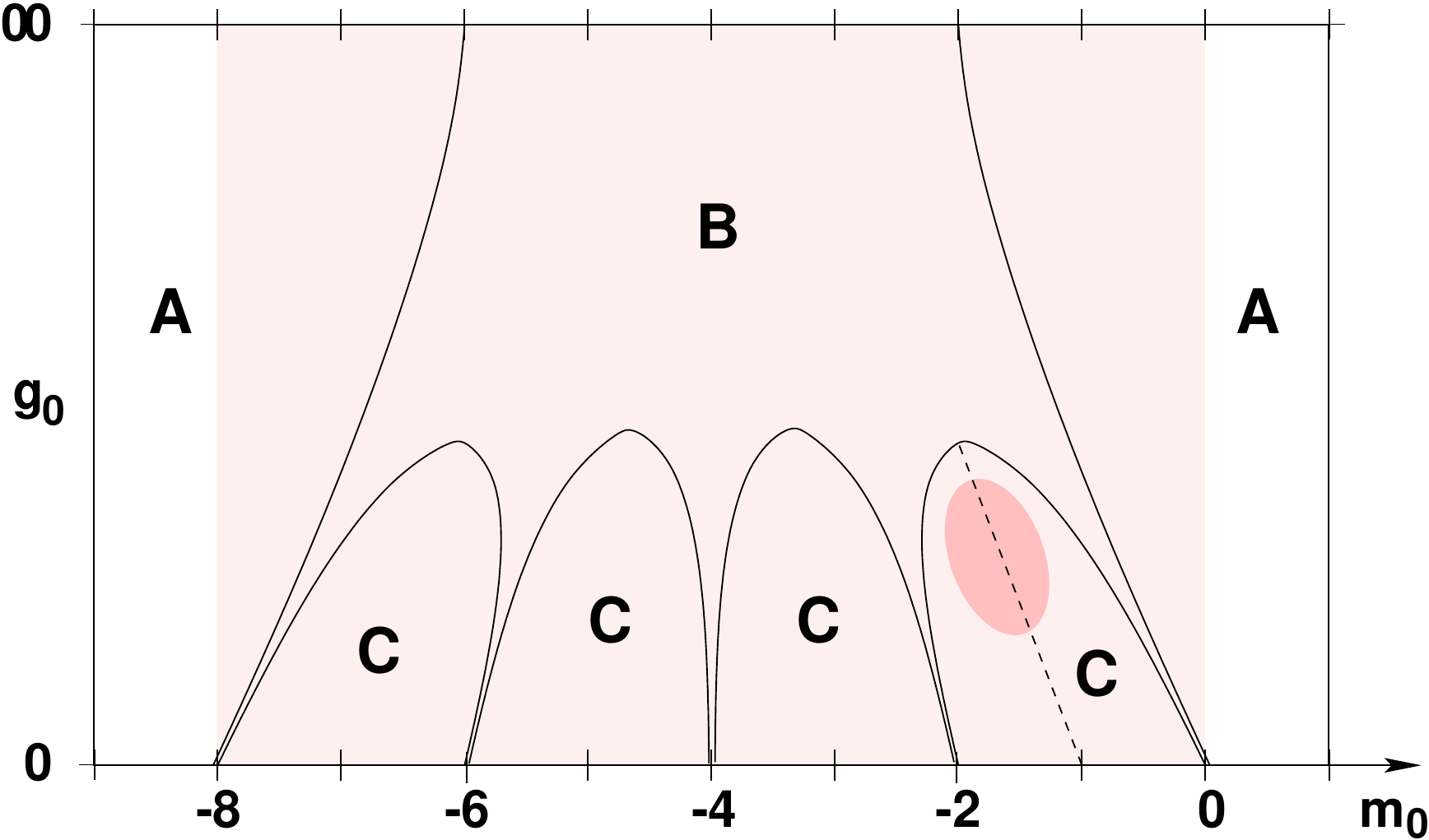}
\end{center}
\begin{quotation}
\floatcaption{aokifig}%
{Schematic phase diagram of lattice QCD with two dynamical Wilson fermions.
The (linear) horizontal axis is the bare mass $m_0$.
The vertical axis is the bare coupling $g_0$, with values extending
from zero to infinity.  The phase diagram is symmetric under reflection
with respect to the $m_0=-4$ line.  See text for further explanation.
}
\end{quotation}
\vspace*{-4ex}
\end{figure}

The phase diagram is depicted schematically in \Fig{aokifig}.
Before we describe the various phases we note the following general features.
First, the phase diagram is symmetric under reflection with respect
to the $m_0=-4$ line.  The reason is that the effect of the transformation
$\J(x) \to \e(x) \J(x)$ and $\bJ(x) \to -\e(x) \bJ(x)$, where $\J(x),\bJ(x)$
are the four-dimensional Wilson fermion fields
and $\e(x) = (-1)^{x_1+x_2+x_3+x_4}$,
can be undone by the replacement $m_0 \to -8-m_0$.

The second observation concerns the super-critical region $-8<m_0<0$,
the lightly shaded region in the figure.  Let us show that
only in this region can the Wilson operator have zero modes.
To this end, it is convenient to use the complex Wilson operator $D$
of Eq.~(\ref{DW}).  Consider the eigenvalue equation $D\J = \l \J$ where
$\J$ is a normalized eigenfunction and in general the eigenvalue $\l$
is complex.  Writing $D=A+iB$ with hermitian $A$ and $B$, it follows that
\begin{equation}
\label{supercrit}
2\Re\l = \sbra{\J} D^\dagger + D \sket{\J} = 2 \sbra{\J} A \sket{\J} \ .
\end{equation}
If $\J$ is a zero mode of $D$ (or of $H_W$),
then we must have $\sbra{\J} A \sket{\J} = 0$.
Recalling that $M=-m_0$, we have $A = -(W+m_0)$.
Since the spectrum of the Wilson term is bounded both above and below,
$0\le W \le 8$, it follows that for $m_0>0$ or $m_0<-8$ the expectation value
$\sbra{\J} A \sket{\J}$ cannot vanish, and thus $D$ (or $H_W$)
cannot have a zero eigenvalue.

We now turn to a description of the phase diagram.
The existence of a phase with a pion condensate, the Aoki phase,
was first proposed in Refs.~\cite{Aoki:1983qi,Aoki:1986xr},
based on strong coupling arguments.  This is phase B in the figure.
The A and C phases are massive.  Notice that the C phases are fully contained
within the super-critical region.  As we will see this plays
a crucial role for both domain wall and overlap fermions.
The borderlines of the super-critical region, $m_0=0$ and $m_0=-8$,
each belong to a different A phase, but nothing special happens
in the A phase when this line is crossed, except of course
in the continuum limit $g_0\to 0$.

Starting from some large positive $m_0$ in the A phase in the right part of
\Fig{aokifig}, as we move horizontally to the left at some generic value of
$g_0$, the mass of the three pions will be (roughly) described by Eq.~(\ref{m0mc}).
The A phase is isospin symmetric, and all three pions have the same mass.
When we reach $m_0=-m_c(g_0)$,
which is the phase transition line separating the A and B phases,
the pions become massless.  Upon entering the Aoki phase (B),
one of the pions condense, while the other two pions become massless
Nambu-Goldstone bosons (NGBs) associated with the spontaneous breaking of
the (vectorial) SU(2) isospin symmetry down to U(1).
It is conventional to take the condensate
pointing in the third isospin direction by adding a so-called twisted mass term
(see Sec.~\ref{lcl}).  The pion associated with the direction of the condensate
is not an NGB, and it becomes massive again inside the Aoki phase.

The boundaries of the Aoki phase in the strong coupling limit were
first determined prior to the understanding of the Aoki phase itself
in Ref.~\cite{Kawamoto:1981hw}. They are located at $m_0=-2,\ -6$,
as shown in \Fig{aokifig}.  As for the weak coupling limit,
the Aoki phase has the characteristic ``fingers,'' reaching
the critical points at $m_0=0,-2,-4,-6,-8$ on the $g_0=0$ axis,
where the free four-dimensional Wilson operator admits massless states.
The basic features of the fingers were described analytically using
Wilson chiral perturbation theory (WChPT) in Ref.~\cite{Sharpe:1998xm}.
Writing down the WChPT lagrangian appropriate for the rightmost finger,
which touches the $m_0=0$ point on the $g_0=0$ axis,
two possible scenarios were identified depending on the sign of a particular
low-energy constant.  (By symmetry, the same WChPT lagrangian accounts
for the finger that touches the point $m_0=-8$ as well.)
According to the first scenario, the width $\D m_0$
of the finger scales like $a \D m_0 \sim (a\L)^3$, where $\L$ is a physical
QCD scale.  According to the other scenario, the Aoki phase
doesn't reach the $g_0=0$ line.  Instead, the finger ends at some $g_0>0$,
beyond which the A phase on the right and the C phase on the left
are separated by a first-order transition line.  While both phases
are isospin symmetric, along their borderline the derivative
of the pion mass with respect to $m_0$ is discontinuous.

Before moving on, we comment in passing
that a different WChPT description is required for the fingers
associated with the $m_0=-2$ and $m_0=-6$ critical points on the $g_0=0$ line,
and a yet another WChPT description is required for the central
critical point on the symmetry axis $m_0=-4$.
The reason is the different numbers of continuum-limit quark species,
and hence pions, associated with these critical points.
It is thus possible in principle that, for some specific choice
of the complete lattice action, the $m_0=0$ and $m_0=-8$ fingers will follow
the first scenario, while the $m_0=-2$ and $m_0=-6$ fingers will follow
the second scenario, \etc.

\subsection{\label{quench} Wilson-quenched Phase diagram}
Returning to domain wall fermions, when we study the Wilson kernel
we are in principle interested in a three-dimensional phase diagram
spanned by the parameters $(g_0,M,m_0)$.  As before $g_0$ is the bare coupling
while $M$ is the domain wall height.  But we now consider the bare mass $m_0$
of the Wilson kernel as an independent parameter.  This is equivalent
to a {\em mixed-action} theory with dynamical (or sea) domain wall fermion
and valence Wilson fermions.  The actual Wilson kernel used by
the domain wall fermions is recovered by restricting to the plane
defined by $m_0=-M$.

As might be expected, the symmetry structure of this three-dimensional
phase diagram is more complicated than that of the pure Wilson fermions
of \Fig{aokifig}.  First, as a generalization of the symmetry discussed
in Sec.~\ref{aoki}, in $d$ dimensions the phase diagram of
pure Wilson fermions is symmetric around the $m_0=-d$ line.
Indeed, by itself the phase diagram of domain wall fermions
inherits the symmetry of five-dimensional Wilson fermions around
the $M=5$ axis, as reflected by \Tab{zms}.  But since the $(g_0,M,m_0)$
phase diagram treats independently the five-dimensional domain wall fermions
and the four-dimensional Wilson kernel, in contrast with \Fig{aokifig}
the ``physical plane'' defined by $(g_0,m_0=-M)$ might not have any
obvious symmetry structure.
Still, we will only be interested in the C phase between the
two rightmost fingers in \Fig{aokifig}.  Its boundary on the
$g_0=0$ axis is given by $-2 < m_0 < 0$, and so this phase defines the region
where a domain wall field gives rise to a single light quark
in the continuum limit.  For this C phase, it is reasonable to expect
that the geometry of the corresponding portion of the $(g_0,m_0=-M)$ plane
of the Wilson-quenched phase diagram will be qualitatively similar
to the way this phase has been depicted in \Fig{aokifig}.

We already discussed in Sec.~\ref{tadpole} the plausible connection
$\d M \sim m_c(g_0)$ between the tadpole correction $\d M$
for the domain wall height and the critical Wilson mass $m_c(g_0)$.
With the phase diagram in \Fig{aokifig} in mind we now propose
a broader view of this connection.  To this end, first, we will ignore
the small width of the fingers, identifying the whole rightmost finger
with the critical line $m_0=-m_c(g_0)$.  In order to reach a simple
physical picture we will further assume that the next finger is roughly
parallel, and thus, when approximated by a line, is given by $m_0=-2-m_c(g_0)$.
As we increase $g_0$, the appropriate range of domain wall height
is given by the corresponding section of the rightmost $C$ phase
in \Fig{aokifig}.  Moreover, under the simplifying assumptions we have made,
as long as $g_0$ is not too large this section is given by
(now making the usual identification $M=-m_0$)
\begin{equation}
\label{shiftM}
m_c(g_0) \,\leqx\, M \,\leqx\, 2+m_c(g_0) \ ,
\end{equation}
for which the tadpole improved wave function
\begin{equation}
\label{tadmc}
  (1+m_c(g_0)-M)^s \ ,
\end{equation}
is a bound state.
The dashed line in \Fig{aokifig} corresponds to the optimal choice
$M=1+m_c(g_0)$.  To complete the picture, the darker blob is meant to represent
the parameter range actually used in numerical calculations
with domain wall fermion.

The standard overlap operator \cite{Neuberger:1997fp,Niedermayer:1998bi}
is defined by Eq.~(\ref{DGWlim}) taking $H=H_W$ to be the usual Wilson kernel.
Similarly to domain wall fermions, in virtually all numerical work with
overlap fermions one requires that the overlap operator will describe
a single quark field in the continuum limit.  This implies that the bare mass
of the Wilson kernel must be chosen so as to place it inside
the rightmost C phase in \Fig{aokifig}, precisely the same region that
we have just identified as suitable for domain wall fermions.

We will discuss the phase diagram of domain wall fermions in more detail
in Sec.~\ref{phased}.

\subsection{\label{topo} Topological sectors}
A characteristic feature of massless continuum fermions is
the existence of exact zero modes in topologically non-trivial backgrounds.
According to the index theorem the net chiral charge of the zero modes
is proportional to the net topological charge of the background field.
The group theoretical proportionality constant is a positive integer,
conventionally equal to one for the fundamental representation
of $\SU(N_c)$.
By contrast, on the lattice the gauge field configuration space
does not have any intrinsic topological structure, as any configuration
is continuously connected to the trivial configuration $U_\m(x)=I$.
Various lattice discretizations of the continuum topological charge
have been used to estimate the topological charge of lattice configurations,
but obviously any such expression suffers from discretization errors
(compare Eq.~(\ref{dAanml})).

Nonetheless, overlap fermions support stable, exact zero modes
\cite{Neuberger:1997fp}.  More generally, this is true for
any GW fermion defined by Eq.~(\ref{DGWlim}) with some reasonable $H$.
In the free-field case, the Wilson kernel $H_W$ has an equal number of
positive and negative eigenvalues, and thus the sign function $\e(H_W)$
has equal numbers, both even, of $+1$ and $-1$ eigenvalues.  If a single
eigenvalue of $H_W$ crosses zero and changes sign, making the total number of
negative eigenvalues odd, this leads to the existence of a stable zero modes
of the overlap operator.  The stable zero modes all share the same chirality,
and their number times their $\pm$ chirality can be used to define
the topological charge of the background field
\cite{Neuberger:1997fp,Niedermayer:1998bi}.

There are plenty of reasonable choices for the hermitian kernel
used to the define the GW operator in Eq.~(\ref{DGWlim}).
The relation between the Wilson kernel $H_W$ of the standard overlap operator,
and the hamiltonian $H_T$ derived from the transfer matrix
of domain wall fermions, is particularly tight.
As already mentioned in Sec.~\ref{chirN5},
$H_W$ and $H_T$ share the same zero eigenmodes \cite{Furman:1994ky}.
Hence, the standard overlap operator, and the GW operator defined
by the $N_5\to\infty$ limit of domain wall fermions, always share
the same (chirality and) number of stable zero modes.

\begin{figure}[t]
\begin{center}
\includegraphics*[height=8cm]{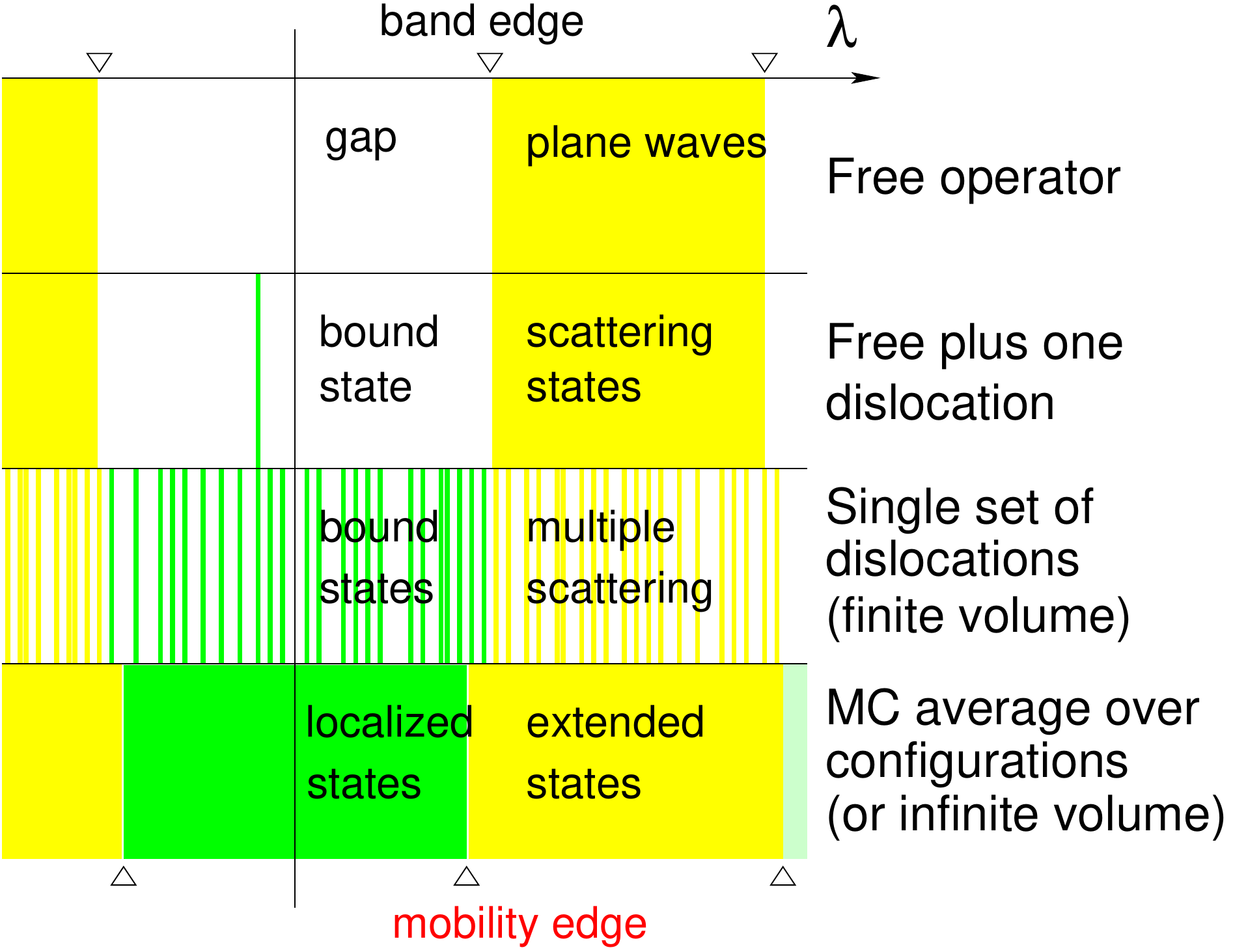}
\end{center}
\begin{quotation}
\floatcaption{lclfig}%
{How the spectrum of $H_W$ changes as one adds to the gauge field
(top to bottom) one dislocation, many dislocations,
and a random ensemble of dislocations.
}
\end{quotation}
\vspace*{-4ex}
\end{figure}

\subsection{\label{rho} Low-lying eigenstates and their spectral density}
By now we have encountered several important features of domain wall fermions
as well as overlap fermions which, one way or another,
are controlled by the low-lying spectrum of the Wilson kernel $H_W$
or of the closely related transfer matrix hamiltonian $H_T$.
In this subsection and the next one we will focus on the simpler
Wilson kernel $H_W$, assuming that the behavior of the
transfer matrix hamiltonian $H_T$ is qualitatively similar.
We start by organizing the available information about
the low-lying spectrum.  This will lead to a certain puzzle,
whose resolution will be the subject of the next subsection.

The free Wilson operator $H_W^0$, in which all links are set to $U_\m(x)=I$,
has a gap for all values of $m_0$ except at the five critical points
on the $g_0=0$ axis, $m_0=\{0,-2,-4,-6,-8\}$.  The gap $\ce$ is given by
the distance from $m_0$ to the closest critical point
(see, for example, Ref.~\cite{Golterman:2003qe}).  The five critical points
are visible in \Fig{aokifig}, while the gap for some value of $m_0$
(including super-critical values) is depicted in the top segment of
\Fig{lclfig}.  That said, we have noted in Sec.~\ref{aoki} that the bounds on
the Wilson term allow in principle for the existence of zero modes in the
entire super-critical region $-8<m_0<0$.  Furthermore, as we have just seen
in the previous subsection, the (simultaneous) zero modes
of $H_W$ and $H_T$ can be used to define the boundaries of topological
sectors of the lattice gauge field configuration space.  If we are to reproduce
topological sectors in the continuum limit, then evidently these zero modes
should exist on a subspace of the gauge field configuration space
of codimension one.  In addition,
in Sec.~\ref{interacting} and Sec.~\ref{resid} we found out
that the rate of restoration of chiral symmetry for $N_5\to\infty$
is controlled by the smallest eigenvalues of $H_T$.
And, as we will see later on in this section,
the low-lying eigenmodes of $H_W$ or $H_T$ are also important for the
locality properties of the overlap operator or the effective
four-dimensional operator for domain wall quarks~(\ref{DGWN}).

The accumulation of these observations, which were mostly known by the
late 1990s, prompted a more detailed research of the low-lying spectrum
of $H_W$.  Following up on earlier numerical work \cite{Edwards:1998sh},
interesting progress was made in Ref.~\cite{Berruto:2000fx}.
The question raised in this paper was whether super-critically,
namely having $-8<m_0<0$, is not just a necessary condition,
but also a sufficient condition for the existence of gauge field configurations
which support a zero mode of $H_W$.  The main result of Ref.~\cite{Berruto:2000fx}
was the following.  Consider a gauge field configuration equal to
the classical vacuum $U_\m(x)=I$ almost everywhere, except inside
a single hypercube of size $\ell^4$, whose link variables
can take any value in the gauge group.  We will refer to such configurations
as having a single {\em dislocation}.  Ref.~\cite{Berruto:2000fx} considered
in particular dislocations contained in a unit hypercube (\ie,
$\ell=a$) for the SU(2) gauge group.  Using a combination of
analytic considerations and numerical minimization methods, it was found that
the gauge links of the dislocation can be dialed so as to produce
a bound state whose eigenvalue $\l_1$ lies inside the gap of the
free Wilson operator, $-\ce < \l_1 < \ce$ (see the second segment of
\Fig{lclfig}).  Moreover, it is possible to tune the dislocation to support
a zero mode, $\l_1=0$, for roughly $-7 \leqx\, m_0 \,\leqx -\!1$.
In view of this result it is plausible that, by allowing the dislocation
to gradually occupy a larger and larger hypercube,
eventually single-dislocation configurations will be found that
support a zero mode of $H_W$ throughout the entire super-critical region.
Henceforth, we will assume that this is indeed the case.

The next logical step is to consider a large volume with $n$
dislocations placed far apart from each other.  Intuitively,
we expect each dislocation to support essentially the same bound state(s)
as before; the presence of other, remote, dislocations should have
a negligible effect (third segment of \Fig{lclfig}).
As for the continuous spectrum of the free operator,
in the presence of a single dislocation the plane waves turn into
scattering states, while in the presence of more than one dislocation
we have multiple scattering.

Taking the infinite volume limit we arrive at the following conclusion.
Based on the above evidence, we assume the existence of a bound state spectrum
of $H_W$ for any $\l$ inside the gap $\ce$ of the free operator for
appropriately chosen gauge field configurations, everywhere inside the
super-critical region.  It follows that the spectral density $\r(\l)$ of $H_W$
is nonzero for any $-\ce \le \l \le \ce$ in the super-critical region
of the Wilson-quenched phase diagram.
Of course, the spectral density is nonzero for $|\l|>\ce$ as well
(bottom segment of \Fig{lclfig}).\footnote{%
  The Wilson operator $H_W$ is a bounded operator, hence its entire spectrum
  is contained in some interval $-\lmax \le \l \le \lmax$,
  with $\lmax=\lmax(m_0)$.  Here we will not be concerned with the upper limit.
}
It is important to note that the same conclusion might or might not apply to
$\r(0)$ in the case of (two) dynamical Wilson fermions (Sec.~\ref{aoki}).
The Boltzmann weight then contains a factor of $\det^2(H_W)$,
that acts to suppress gauge field configurations with near-zero
eigenvalues.  But no such suppression factor is present in the case of
the Wilson-quenched phase diagram, hence it will have a nonzero $\r(0)$
throughout the entire super-critical region.
We will further discuss the differences between the dynamical Wilson
and quenched Wilson cases in the next subsection.

We have now arrived at the following puzzle.  Consider the Wilson-quenched
phase diagram for two valence Wilson fermions, which possess
an isospin symmetry.  The Banks-Casher relation asserts that the
pion condensate is equal to $2\p\r(0)$, hence the condensate is nonzero
whenever $\r(0)$ is.  A nonzero pion condensate implies
the spontaneous breaking of the (valence) isospin symmetry.  Moreover,
according to the usual picture of spontaneous symmetry breaking,
the pion condensate ought to be accompanied by two massless NGBs, which are
the other two pions.  Finally, all this is supposed to happen
{\em everywhere} inside the super-critical region, since we have argued
that $\r(0)$ is nonzero in the entire super-critical region of
the Wilson-quenched phase diagram.  As we will see in the next subsection,
some of these conclusions are in fact incorrect: for the Wilson quenched case
there is a way out that evades the Goldstone theorem.

\subsection{\label{lcl} Localization and mobility edge}
The key element missing from the discussion of the previous subsection is the
need to distinguish between {\em extended} and (exponentially) {\em localized}
eigenstates of $H_W$.  Understanding the different roles of the extended
and the localized spectra is the subject of this subsection.
To avoid inessential technicalities, we will assume an infinite volume
unless otherwise stated.

Let us revisit \Fig{lclfig}, whose second segment
describes the spectrum for a single dislocation, the case considered
in Ref.~\cite{Berruto:2000fx}.  We can write
\begin{equation}
\label{HWV}
H_W = H_W^0 + V \ ,
\end{equation}
where the potential $V$ is confined to the hypercube containing the dislocation.
Everywhere else, $H_W$ is equal to the free operator $H_W^0$.
Consider an eigenstates with an eigenvalue $\l$.  If $|\l|$ is smaller than
the gap $\ce$ of the free operator, this is a bound state that decays
exponentially away from the dislocation.  In the third segment of \Fig{lclfig}
we have bound states attached to several dislocations which are far from
each other.  Due to the exponential decay of all bound states,
remote dislocations will indeed cause an (exponentially) small disruption,
a feature we have already invoked in the previous subsection.
The bound states of both the single dislocation and the
several (but far apart) dislocations are examples of localized eigenstates.

For an ensemble of configurations that belong to some point of the
Wilson-quenched phase diagram the situation is the following.
Up to short-distance random fluctuations, the mode density of
an exponentially localized eigenstate $\J(x)$ is expected to behave like
\begin{equation}
\label{lcllength}
|\J(x)|^2 \sim \ell^{-4} \exp\left(-\frac{|x-x_0|}{\ell}\right) \ ,
\qquad |x-x_0|\gg 1 \ .
\end{equation}
Here $x_0$ is some ``center,'' while $\ell$ is the localization length.
As the eigenvalue $\l$ of a localized eigenstate grows starting from $\l=0$,
we expect the localization length to be a monotonically increasing function
of $\l$.  Eventually, we will reach a critical value $\l_c$,
called the mobility edge, where the localization length diverges,
and the eigenstate becomes an extended state.\footnote{%
  \label{negev}
  The situation is similar for negative $\l$.  We expect the mobility edges
  at positive and negative $\l$ to have the same absolute value.
  Also note that, because of boundedness of $H_W$ (see the previous footnote),
  a second pair of mobility edges may exist at a larger $|\l|$,
  where the spectrum turns back from extended to localized eigenstates.
  This situation is depicted in the bottom segment of \Fig{lclfig}.
  We will not be concerned with these additional mobility edges.
}
As long as $H_W$ can be expressed by Eq.~(\ref{HWV})
with a potential $V$ that is confined
to a compact region of the total (infinite) volume, the mobility edge $\l_c$
will be equal to the gap $\ce$ of the free operator.  This is the situation
depicted in the second and third segments of \Fig{lclfig}.
But for a realistic ensemble, $\l_c$ will in general be different from $\ce$.

Focusing our attention on the mobility edge, let us increase $g_0$
at some fixed super-critical $m_0$, moving vertically inside the rightmost
C phase in \Fig{aokifig}.  As the bare coupling $g_0$ is increased,
the randomness of typical gauge field configurations in an ensemble
will increase.  As a result, we expect
the mobility edge $\l_c$ to go down, eventually reaching zero.
We can define the Aoki phase as the region where $\l_c=0$,
and the spectral density $\r(0)$ comes from extended states.
This definition applies to both the dynamical Wilson and quenched Wilson cases.
In the A and C phases of the Wilson-quenched case $\r(0)$
remains nonzero,\footnote{
  For the A phase this is true in the super-critical region.
}
but it comes from localized states.  In the dynamical-Wilson
case, the factor of $\det^2(H_W)$ in the Boltzmann weight is sufficiently
effective to yield $\r(0)=0$ outside of the Aoki phase, in other words,
whenever $\l_c>0$.

Let us summarize the emerging understanding of the super-critical region.
We will consider infinite volume ensembles defined by
the partition function of (two) dynamical Wilson fermions, or by
the partition function of (one or more) dynamical domain wall fermions---the
Wilson-quenched case.  We start with the common features:
\begin{enumerate}
\item There is no spectral gap throughout the entire super-critical region
of the phase diagram, $-8<m_0<0$.  At a generic point in the super-critical
region the spectrum of $H_W$ consists of localized states below the
mobility edge, $|\l| < \l_c$, and extended states\footnote{
  See, however, footnote \ref{negev}.
}
above it, $|\l| \ge \l_c$.
\item The Aoki phase can always be defined by the vanishing of the
mobility edge, $\l_c=0$.  In the A and C phases, $\l_c>0$.
The Aoki phase is thus characterized by nonzero $\r(0)$
coming from extended states.
\item The main difference between the dynamical-Wilson and Wilson-quenched cases
is the following.  In the dynamical Wilson case, $\r(0)=0$ in the
A and C phases, whereas in the Wilson-quenched case $\r(0)\ne 0$
in the entire super-critical region. In the A and C phases $\r(0)$ comes
from localized states.
\end{enumerate}
The Wilson-quenched phase diagram was studied using effective field theory
(EFT) methods in Ref.~\cite{Golterman:2005ie}
(see also Ref.~\cite{Aoki:2001su}).  As explained there,
the low-energy EFT is only sensitive to the long-range degrees of freedom,
and thus it predicts the existence of a pionic condensate
only when it arises from extended states.  This is consistent
with the definition of the Aoki phase in point (2) above.
The main conclusion of Ref.~\cite{Golterman:2005ie} is that, for small enough
lattice spacing where the EFT applies, the phase structure of the
Wilson-quenched case is qualitatively similar to the unquenched case,
including the feasibility of the two scenarios discussed in Sec.~\ref{aoki}.

It remains to resolve the puzzle associated with the Goldstone theorem
\cite{Golterman:2003qe,Golterman:2003ui}.
We will show that in the Wilson-quenched case, if the pion condensate
$2\p\r(0)$ comes from localized states, it is possible to have
no massless NGBs
and indeed no long-range correlations at all in the valence sector,
as first shown in Ref.~\cite{McKane:1980fs}.

To this end we introduce a small twisted mass term that will control
any infrared divergences and provide the seed for the pion condensate
and hence for the orientation of isospin symmetry breaking,
\begin{equation}
\label{twmass}
D \Rightarrow D - i\m \g_5\t_3 = \g_5 (H_W - i\m\t_3) \ .
\end{equation}
Here $\t_i$ are the Pauli matrices acting on the isospin index of the
Wilson fermions.  Let us consider the momentum space WTI
\begin{equation}
\label{WTIlcl}
  \sum_\m p_\m\tG_\m(p) + 2\m \tG(p) = \svev{\p_3} \ ,
\end{equation}
which is true on the lattice up to $O((ap)^2)$ corrections.
The correlators $\tG(p)$ and $\tG_\m(p)$ are the Fourier transforms of
$\svev{\p_+(x)\,\p_-(y)}$ and $\svev{J_\m^+(x)\,\p_-(y)}$ respectively.
$J_\m^+(x)$ is a conserved isospin current, and the $\pm$ refer to operators
involving the Pauli matrices $\t_\pm$ that raise or lower isospin by one unit.

For $\m>0$ isospin symmetry is explicitly broken, and there are no
massless states.  This applies to hadrons made of both sea and valence
Wilson fermions, including any (would-be) NGBs.
We can therefore set $p=0$ Eq.~(\ref{WTIlcl}).
The $\tG_\m(p)$ term drops out, and we arrive at
\begin{equation}
\label{pipm0}
\tG(0) = \frac{\svev{\p_3}}{2\m} \ , \qquad \m>0 \ .
\end{equation}
This result is valid for both sea and valence Wilson fermions
in finite volume, as well as in the infinite volume limit.
However, the physical implications of Eq.~(\ref{pipm0}) are
qualitatively different in the two cases.

We first consider valence Wilson fermions.  We will assume, as discussed above,
that the spectral density $\r(0)$ is nonzero, and that it comes
from localized states, in other words,
the mobility edge is $\l_c>0$.  According to the Banks-Casher relation
the pion condensate is $\svev{\p_3}=2\p\r(0)\ne 0$,
hence the valence isospin symmetry is broken spontaneously already
in finite volume.  In fact it is natural that the same behavior will be found
in both finite and infinite volume, because localized states
are insensitive to the volume (as soon as the linear size of the system
is large compared to their localization length).
According to Eq.~(\ref{pipm0}), $\svev{\p_3} \ne 0$ implies that $\tG(0)$
diverges like $1/\m$.  Moreover, since by assumption
the mobility edge is $\l_c>0$,
for small enough $p$ one can further show
that \cite{Golterman:2003qe,Golterman:2003ui}
\begin{equation}
\label{pipmp}
\tG(p) = \frac{\svev{\p_3}}{2\m}\,\left(1+O\Big(p^2\ell_0^2\Big)\right) \ ,
\end{equation}
where $\ell_0$ is an average localization length that characterizes
the near-zero modes.  Hence, the $1/\m$ divergence of $\tG(p)$ extends to
a range of $p\ne 0$ as well.  This evades the Goldstone theorem:
Because $\l_c>0$ there are no long-range correlations for any $\m$,
including for $\m\to 0$.
$\tG_\m(p)$ is regular, and $p_\m\tG_\m(p)$ vanishes for $p\to 0$.
Finally, for $\svev{\p_3}\ne 0$ the
left-hand side of the WTI~(\ref{WTIlcl}) is dominated by
$2\m\tG(p)$, which stays finite when $p$ and/or $\m$ tends to zero
thanks to the $\m\to 0$ divergence of $\tG(p)$.

In order to complete the physical picture we revisit the proof
of the Goldstone theorem for the case of two dynamical Wilson fermions.
We begin by proving the absence of any $1/\m$ divergence in finite volume.
In the case at hand the rules of Grassmann integration imply, first, that
any {\em unnormalized} fermion correlation function
is a holomorphic function of the parameters of the fermion action.
We note that the same is {\em not} true in the Wilson-quenched case.
The formal reason is that while the valence Wilson fermions contribute
to the Boltzmann weight a factor of $\det^2(H_W)$, there is a matching
contribution of $\det^{-2}(H_W)$ coming from the ghost quarks,
and these two contributions cancel each other by construction.

In addition, the Boltzmann weight $\det^2(H_W)$ of the dynamical
Wilson fermions is always real and non-negative.  Moreover,
in finite volume $H_W$ can have any exact zero modes only on
a measure zero subset of the gauge field configuration space.\footnote{
  This property was already used in the proof of chiral symmetry restoration
  in Sec.~\ref{chirN5}, and it is true for any value of the Wilson mass parameter.
}
On the complementary subset we have $\det^2(H_W)>0$.  As a result,
the partition function is always strictly positive,
and bounded below by some $z_0>0$ for a range of values of $\m$
that includes $\m=0$ for any bare mass.  It follows that normalized
fermion correlation functions are holomorphic functions of $\m$ as well.
This rules out a $1/\m$ divergence in $\tG(p)$.
By Eq.~(\ref{pipm0}), the fermion condensate must therefore vanish
(at least) linearly with $\m$ in finite volume.
This recovers the familiar result that spontaneous symmetry breaking
cannot occur in finite volume.

In the thermodynamical limit (first taking the
infinite volume limit and then sending $\m\to 0$) we arrive at
\begin{equation}
\label{WTINGB}
  \sum_\m p_\m\tG_\m(p) = \svev{\p_3} \ .
\end{equation}
This is the Goldstone theorem: if $\svev{\p_3} \ne 0$ then $\tG_\m(p)$
must have a $p_\m/p^2$ pole.
In the dynamical-Wilson theory a fermion condensate can only
arise from extended states, and only in the infinite volume limit.
As we have just explained, any nonzero $\r(0)$ coming from localized states
in the infinite volume limit would have existed already in finite volume.
But this would entail a $1/\m$ divergence in $\tG(p)$,
which is impossible in the dynamical-Wilson case.
Thus, a nonzero $\r(0)$ necessarily comes from extended states, consistent with
the fact that NGBs describe long-range fluctuations of the order parameter.
This is the situation realized in the Aoki phase of the dynamical-Wilson theory.

In order to avoid any confusion, we note that all the ``sicknesses''
discussed above of the Wilson-quenched phase diagram
have to do with the correlation functions of {\em valence}
Wilson fermions in a lattice theory with dynamical domain wall fermions.
No similar issues arise in a unitary theory.  In particular,
the Goldstone theorem applies to every theory of dynamical domain wall fermions,
just as to every theory of dynamical Wilson fermions, and
spontaneous symmetry breaking can occur only in the thermodynamical limit.

The localization properties of the low lying spectrum of $H_W$
and the mobility edge were studied numerically in
Refs.~\cite{Golterman:2004cy,Golterman:2005fe,RBC:2008cmd}.
Strictly speaking, the distinction between localized and extended states
can be made precise only in infinite volume.  Nevertheless, there exist
good estimators for the localization length, and, as a result,
for the mobility edge, which can be applied in numerical simulations.
A notable numerical result is that, below the mobility edge,
the spectral density of the localized states drops rapidly with
decreasing $\l$, but remains non-zero.
The rapid drop is good news for residual chiral symmetry breaking
effects, as we will discuss in Sec.~\ref{mresagain} below.

\subsection{\label{lcleff} Locality of Ginsparg-Wilson operators}
Any GW operator cannot be ultra-local: its coordinate representation
$\DGW(x,y)$ does not vanish for arbitrarily large separation $|x-y|$ \cite{Horvath:1998cm}.
GW operators can be local, however, in the following sense.
Consider the standard overlap operator
$\DOV$ with the free Wilson kernel $H_W^0$, where $-2 < m_0 < 0$.
If both $|m_0|$ and $|2+m_0|$ are $O(1)$, we expect $\DOV(x,y)$
to decay exponentially with the separation,\footnote{%
  The distance $|x-y|$ can be for example the usual euclidean norm
  or alternatively the ``taxi driver's distance.''
}
\begin{equation}
\label{Dovxy}
\| \DOV(x,y) \| \sim \exp(-M_{ov} |x-y|)\ ,\qquad |x-y|\gg 1 \ ,
\end{equation}
with a decay rate $M_{ov}=O(1)$ in lattice units.
A similar behavior is expected for relatively smooth gauge fields.
This was proved rigorously in Ref.~\cite{Hernandez:1998et}, which showed
that a bound of the form of Eq.~(\ref{Dovxy}) holds provided that the (latticy)
field strength is everywhere constrained to be smaller than some constant.
A similar rigorous result was proved in Ref.~\cite{Kikukawa:1999dk}
for the effective four-dimensional operator for domain wall quarks,
the approximate GW operator~(\ref{DGWN}).
For the overlap operator, Ref.~\cite{Hernandez:1998et}
further showed that a similar bound holds also if the spectrum of $H_W$
contains a single exponentially localized eigenstate inside an otherwise
empty spectral gap.  These situations correspond
to the upper two segments of \Fig{lclfig}, except that the gap will in general
be slightly different from the gap of the free operator.

Constraints on the lattice field strength of the kind invoked in
Refs.~\cite{Hernandez:1998et,Kikukawa:1999dk} do not alter the continuum limit,
because any field strength of the order of some physical scale
becomes vanishingly small in lattice units in the continuum limit.
But it is unrealistic to impose similar constraints for ensemble generation
using Monte-Carlo methods.  For realistic ensembles,
information about the range of the overlap operator or of the effective
domain-wall quarks operator must rely on numerical results.
One option is to measure the range of these operators directly.
Alternatively, one can combine analytic considerations with
numerical data about the localized and extended spectrum of the kernel and the
mobility edge \cite{Golterman:2004cy,Golterman:2005fe,RBC:2008cmd}.

Understanding the effects that control the range of
(exact or approximate) GW operators requires slightly more refined
acquaintance with the localized spectrum.  As we increase $\l$,
starting from $\l=0$ and moving towards the mobility edge,
several effects take place. First, the spectral density $\r(\l)$ of the
localized eigenmodes, which is very small for $\l\sim 0$, grows rapidly.
Second, the localization length $\ell(\l)$ increases,
ultimately diverging at the mobility edge.
Third, the eigenmodes that we encounter for small $\l$
typically have a single center around which the mode density is localized.
But as we increase $\l$ past some $\bl$ where $0<\bl<\l_c$,
the localized modes start becoming multi-centered.
We should consider separately the contribution of different spectral ranges.
For definiteness we will discuss the overlap operator,
but essentially the same considerations apply also to the
effective four-dimensional operator for domain wall quarks
\cite{Golterman:2004cy,Golterman:2005fe,RBC:2008cmd,Sharpe:2007yd}.

The lowest range is $0 \le |\l| \le \bl$.  The modes in this range
are exponentially localized and single-centered.  Using Eq.~(\ref{lcllength})
and the triangle inequality their contribution is easily estimated as
\begin{eqnarray}
\label{lcllow}
  \svev{|\DOV(x,y)|}_{|\l|\le \bl}
  &\leqx& \int_{-\bl}^{\bl} d\l\, \r(\l)
  \exp\left(-{\frac{|x-y|}{2\ell(\l)}}\right)
\\
  &\leqx& \rule{0ex}{4ex}
  \bl \r(\bl) \exp\left(-{\frac{|x-y|}{2\ell(\bl)}}\right) .
\nonumber
\end{eqnarray}
The second estimate holds because both $\r(\l)$ and $\ell(\l)$ are
monotonically increasing, hence the integral is dominated by its
lower and upper bounds.

The second and third ranges have in common that the modes are either localized
but multi-centered ($\bl \le |\l| \le \l_c$) or extended ($|\l| \ge \l_c$).
For these ranges of eigenmodes, which are all separated from zero
by a minimal gap $\bl$, one could envisage the application of some
generalization of the techniques of Refs.~\cite{Hernandez:1998et,Kikukawa:1999dk}.
In practice, the growth of the spectral density towards the mobility edge and
beyond is so rapid that one expects the contribution of the extended modes
to dominate, with the mobility edge effectively playing the role of a gap
in the spectrum.  This contribution should be qualitatively similar
to the effect of integrating out a massive field whose mass is $\l_c$.
It is thus estimated as
\begin{equation}
\label{lclhigh}
  \svev{|\DOV(x,y)|}_{|\l|\ge \l_c}
  \approx \cc \exp\left(-\l_c |x-y|\right) ,
\end{equation}
where $\cc=O(1)$.  Numerical evidence \cite{Golterman:2005fe}
suggests that the contribution of the extended modes, Eq.~(\ref{lclhigh}),
indeed dominates over the entire contribution of the localized spectrum,
including both the single-centered and multi-centered modes.

These consideration also clarify what goes wrong when we enter
the Aoki phase.  With the vanishing of the mobility edge $\l_c$,
the (exact or approximate) GW operators we have considered become non-local.
As long as we are careful to stay inside the C phase, these operators are local,
nevertheless one should pay attention to their actual range,
as illustrated by the following examples.  Working with
(fully) quenched ensembles generated using different gauge actions,
Ref.~\cite{Golterman:2005fe} found for several ensembles with a cutoff
of $\sim 2$~GeV that the estimated range of the overlap operator
corresponded to a mass scale of $\sim 800$~MeV.
Depending on the QCD observable of interest,
this scale might or might not be high enough to be considered
as part of the discretization effects.  For a lattice cutoff of $\sim 1$~GeV,
depending on the gauge action, the estimated range of the overlap operator
corresponded to a mass scale of 250 -- 320~MeV, clearly way too low.
Of course, by today's standards, a lattice cutoff of 1~GeV
would usually be considered much too low anyway.

\subsection{\label{mresagain} The residual mass revisited}
We are finally in a position to identify the non-perturbative effects that
contribute to the residual mass, in addition to the perturbative effects
discussed in Sec.~\ref{resid}.  The residual mass is defined by the mixing of
the operator $J_{5q}^a$ with $J_5^a$, a local effect (see Eq.~(\ref{mresdef})).
Considerations similar to the ones of the previous subsection lead to the
phenomenological estimate \cite{Shamir:2000cf,Aoki:2001su,RBC:2008cmd}
\begin{equation}
\label{mresnp}
  \mres \approx \int d\l\, \r(\l)\, e^{-N_5|\l|} \ ,
\end{equation}
where we now refer to the spectrum of the transfer matrix hamiltonian $H_T$.
For a mode with eigenvalue $\l$, the factor $e^{-N_5|\l|}$ accounts for
the dependence on $N_5$, the length of the fifth direction.
The occurrence of $|\l|$ in the exponential echoes the replacement of the transfer matrix $T$
by its bounded version $Q$, discussed in Sec.~\ref{chirN5}.

The combined effect of the suppression factor $e^{-N_5|\l|}$
and the rapid damping of the spectral density below the mobility edge
is that the contribution of the spectral range $|\l| \geqx \l_c$
to Eq.~(\ref{mresnp}) is peaked near $|\l| = \l_c$.  A saddle-point integration
yields the contribution $\r(\l_c) e^{-N_5\l_c} / N_5$.
In addition, as $N_5$ gets larger and larger,
the contribution of the (localized) near-zero modes will eventually dominate
since it effectively lacks the exponential suppression with $N_5$.
This yields a second contribution
\begin{equation}
\label{mreszero}
\int_{-1/N_5}^{1/N_5} d\l\, \r(\l) e^{-N_5|\l|}
\,\approx\,
\int_{-1/N_5}^{1/N_5} d\l\, \r(\l)
\,\approx\, \frac{\r(1/N_5)}{N_5} \ .
\end{equation}
As in Eq.~(\ref{lcllow}), $\r(1/N_5)$ provides a better estimate than $\r(0)$,
because the spectral density of the localized modes increases rapidly
with $\l$.  Putting the two contributions together we arrive at
\begin{equation}
\label{mrespheno}
\mres \,\approx\, \frac{\r(\l_c)\, e^{-N_5\l_c}}{N_5} + \frac{\r(1/N_5)}{N_5} \ .
\end{equation}
For a slightly more refined estimate, see Ref.~\cite{RBC:2008cmd}.

The exponentially suppressed term in Eq.~(\ref{mrespheno}) in effect replaces
the perturbative estimate of Eq.~(\ref{dchi}).  Notice that (setting $s=N_5$)
the power law correction in Eq.~(\ref{dchi}) is $1/N_5^2$, while here
it is $1/N_5$.  The reason is that in Eq.~(\ref{dchi})
the power law correction arises from the phase space around
the maximum of $e^{-\a(p)}$ for the free theory.  By contrast,
here the power correction reflects the nonzero spectral density
for $\l \approx \l_c$.  For a related discussion
see Appendix~C of Ref.~\cite{Shamir:2000cf}.

As for the last term on the right-hand side of Eq.~(\ref{mrespheno}),
apart from the slow $1/N_5$ power law, some further damping with $N_5$
may be provided by $\r(1/N_5)$, because below the mobility edge
the spectral density $\r(\l)$ is always found to be
a monotonically increasing function of $\l$.  Since we are targeting
the rightmost C phase of \Fig{aokifig}, it being part of
the super-critical region of the Wilson-quenched phase diagram means
that in all cases $\r(0)$ is nonzero.  But its variation among
different ensembles can be very large \cite{Golterman:2005fe}.

In conclusion, the main difference between the perturbative
and non-perturbative estimates of $\mres$ originates from the kernel's
near-zero modes, whose contribution to $\mres$ is not suppressed
exponentially.  We first encountered this effect in Sec.~\ref{chirN5},
and now we have arrived at a semi-quantitative expression for
this contribution.  Luckily, it is possible to achieve a sufficiently small
spectral density for the kernel's near-zero modes that the slow
$\r(1/N_5)/N_5$ tail will not lead to an uncomfortably large $\mres$.

\subsection{\label{remedies} Remedies}
Now that we understand the physical mechanisms that control
the residual breaking of chiral symmetry, and $\mres$ in particular,
the natural next step is to look for ``knobs'' that will allow us
to reduce $\mres$ at fixed values of the fifth direction $N_5$
and of the lattice spacing $a$ (in physical units).
In this subsection we consider methods that involve changing of
the Boltzmann weight without changing the domain wall fermion operator itself.
In Sec.~\ref{improve} we will discuss methods based on changing
the domain wall fermion operator.

Perhaps the simplest and most obvious modification to try out is to use
different gauge actions.  There is extensive literature on this topic,
see in particular Refs.~\cite{CP-PACS:2000fmi,Edwards:2000qv,Aoki:2002vt,Svetitsky:2005qa}
Gauge actions that have received special attention include the
standard Wilson plaquette action; the Iwasaki action \cite{Iwasaki:1984cj},
which typically yields a smaller $\mres$;
and the DBW2 action \cite{QCD-TARO:1999mox}, which yields a yet smaller $\mres$,
all for the same $N_5$ and lattice spacing.  The intuitive explanation
for this behavior is that dislocations that can support near-zero modes
of the Wilson kernel are suppressed by the Iwasaki action, and even more so
by the DBW2 action.

Another idea is to mimic the way small eigenvalues are suppressed
in a theory of dynamical Wilson fermions, by adding to the Boltzmann weight
a factor of $\det(f(H_W))$ for a suitable function $f(H_W)$ of the Wilson kernel
\cite{Vranas:2006zk,Fukaya:2006vs,Renfrew:2008zfx}.
When the continuum limit $g_0\to 0$ is approached, we should let
the domain wall height $M=-m_0$ move along the dashed line in \Fig{aokifig}.
Hence $m_0$ is always far from both critical points $m_0=0$
and $m_0=-2$ that mark the boundary of the rightmost $C$ phase
on the $g_0=0$ axis, where the Wilson kernel will support massless fermions
in the continuum limit.  Any near-zero modes of the Wilson kernel $H_W$
are thus lattice artifacts; they are localized states lying well below
the mobility edge.  Hence, adding a factor of $\det(f(H_W))$ to the
Boltzmann weight does not modify the continuum limit.
This method goes under name of
{\em dislocation suppressing determinant ratio} (DSDR),
and the function of choice is usually taken to be
\begin{equation}
\label{DSDR}
f(H_W) = \frac{H_W^2+\e_f^2}{H_W^2+\e_b^2} \ ,
\end{equation}
where one takes $\e_f \ll 1$ and $\e_b \leqx 1$.
A naive choice that achieves the suppression of near-zero modes would be
simply $f(H_W)=H_W^2$.  The reason for the added $\e_f^2$ in the
numerator is not to overly suppress the near-zero modes and along with them
the transitions between topological sectors (we return to this point shortly).
The introduction of the denominator, with $\e_b \leqx 1$, is aimed
to limit the effect of the DSDR factor to the smaller kernel eigenvalues,
while avoiding a lot of ``noise'' from the majority of the eigenvalues
which are $O(1)$, and thus anyway irrelevant for the goal of the DSDR factor.

Large scale domain-wall fermion simulations nowadays usually employ
the following strategy.  The complete lattice action and (if present)
DSDR factor are designed such that, at most, roughly one half of
the bare mass of the light quarks comes from $\mres$.
With the current lattice cutoffs of order few GeVs, this implies
that $\mres$ has to be on the order of $10^{-3}$ or smaller.
The total bare mass is\footnote{%
  For simplicity we consider the isospin limit $m_d=m_u=m_\ell$.
}
\begin{equation}
\label{mtot}
\tm_\ell = m_\ell + \mres \ .
\end{equation}
The tunable mass parameter $m=m_\ell$ in the domain wall fermion
operator~(\ref{finites}) then provides the rest of the desired $\tm_\ell$,
which in turn is multiplicatively renormalizable.
Since $\mres<\tm_\ell$ by design, we need $m_\ell>0$.  This is important,
because it implies that the determinant of the domain wall operator
is always positive (Sec.~\ref{finite5}), which in turn
stabilizes the fermion inversions.

This strategy optimizes the cost of simulations, and works well
for most observables.  An exception is observables that suffer from
a power divergence, such as notably the fermion condensate.
For such observables, suitable subtractions or indirect methods
may be necessary (see, \eg, Ref.~\cite{Sharpe:2007yd}).
Cases include weak decays~\cite{RBC:2020kdj,RBC:2023ynh,Chao:2024vvl}
and the attendant non-perturbative renormalization of four-quark operators
or when the decay as computed on the lattice is slightly
off-shell~\cite{RBC:2023ynh}.

The kernel's (near-)zero modes have a welcome feature:
they are responsible for the transitions between topological sectors;
they also have an undesirable feature: their slowly decaying contribution
to $\mres$ and other residual chiral symmetry breaking effects.
The two facets of the near-zero modes need to be balanced.
We expect the spectral density of the kernel's near-zero modes
to decrease rapidly when the bare coupling $g_0$ is decreased towards
to the continuum limit, $g_0\to 0$.  This can be understood
by noting that a near-zero mode is expected to be attached
to a dislocation in the lattice gauge field, a situation idealized
by the second segment of \Fig{lclfig}.  A dislocation is a small region
of the lattice where many gauge links are far from the identity element.
Hence, the contribution of a dislocation to the lattice action of
the gauge field $S_g$ is given by some $C>0$ where $C=O(1)$.
This amounts to a multiplicative factor in the Boltzmann weight
(see Eq.~(\ref{Z})) with the form of $\exp(-C/g_0^2)$.
The dependence on $g_0$
explains why dislocations are strongly suppressed as we get closer
to the continuum limit.

For a low lattice cutoff, or equivalently larger $g_0$,
the spectral density of near-zero modes will be relatively large,
implying that there is a lot of topological
activity, but maybe also an uncomfortably large $\mres$.
Under these circumstances, adding the DSDR factor may help reducing
$\mres$ to smaller values, while still leaving enough topological activity.

By contrast, many of today's numerical simulations have such a high cutoff
(equivalently, small $g_0$) that topological transitions may be suppressed
too much.  In the case of domain wall fermions we expect that $\mres$
will then be so tiny that one could tolerate larger values.
Under these circumstances, it has been proposed to introduce
a dislocation {\em enhancing} determinant ratio,
which is basically just the inverse of the DSDR factor~\cite{McGlynn:2016ahb}.
This would increase the near-zero modes density,
and thus also the topological activity. While $\mres$ would grow, presumably
it could still be kept small enough. Investigations for $a^{-1}\approx 3$ GeV appear promising~\cite{McGlynn:2016ahb}, but a comprehensive study has yet to be done.

\subsection{\label{conc} Conclusion}
If perfect chiral symmetry was our deciding consideration, we would always take
the chiral limit $N_5\to\infty$ of domain wall fermions analytically,
which amounts to utilizing the associated four-dimensional GW operator.
In this limit the residual mass vanishes, and the existence of the massless
four-dimensional quark field is topologically stable in the sense
that we can choose the bare coupling $g_0$ together with the domain wall height
$M=-m_0$ anywhere inside the rightmost C phase of \Fig{aokifig}.

Let us consider a different question that we have not raised till now:
how do domain wall fermions behave when $N_5$ is very small?
The smallest $N_5$ for which the five-dimensional space has two
separate boundaries is $N_5=2$.  The one-loop wave function of the
effective four-dimensional quark field was calculated
in Sec.~\ref{oneloop}, and, for $N_5,s\gg 1$, it is given by Eq.~(\ref{dchi}).
For $N_5=s=2$ the approximations used in the derivation do not apply,
but what we can say without a detailed calculation is that
the wave function will be given by $c_2 g_0^2$ for some $c_2=O(1)$ number.
Reinstating the lattice spacing this result is $O(g_0^2/a)$, and
the same estimate applies to the residual mass as well.

For $N_5=2$ we thus have $\mres = O(g_0^2/a)$, with no further
suppression factors; this is the same parametric behavior as the additive
renormalization of the mass of ordinary Wilson fermions.
Hence, for very small $N_5$ domain wall fermions are qualitatively the same
as ordinary Wilson fermions.

The practical success of domain wall fermions thus depends on the remarkably
fast falloff of the wave function of the effective four-dimensional
quark field with the fifth coordinate,
so that already for modest values of $N_5$ (say in the range of 10 to 30
as a figure of merit) the wave function has dropped by
several orders of magnitude and, with it, $\mres$.  Perturbatively,
the exponential falloff is described by Eq.~(\ref{dchi}).  Non-perturbatively,
a phenomenological description for the behavior of $\mres$ is given
in Eq.~(\ref{mrespheno}), which includes the slow ``tail'' originating from
the near zero modes of the Wilson kernel.  While the spectrum of
localized modes goes down all the way to $\lambda=0$,
luckily the spectral density of the near-zero modes is very small.
The end result is that values of $a \mres$ on the order of $10^{-3}$
are common in numerical simulations.  As we have explained in the
previous subsection, this is what one needs in order that $\mres$ will be
smaller than the (bare) mass or the light quarks.

\section{\label{improve} Improved domain wall fermions}
The improvement program aims to reduce discretization effects
in lattice calculations, and achieve faster convergence
to the continuum limit.  The discretization effects are responsible
in particular for breaking the continuum symmetry group
to a much smaller lattice symmetry group.
As a result, reducing the discretization effects
typically also gives rise to faster restoration of the continuum symmetries.

For fermions, the continuum symmetries which are broken explicitly
on the lattice include the axial (non-singlet) flavor symmetries.\footnote{
  The only exception is GW fermions,
  and in particular, overlap fermions (see Sec.~\ref{DWF2GW}).
}
The basic version of domain wall fermions, introduced in sections~\ref{free}
and~\ref{interacting}, already has much improved chiral symmetry
in comparison with Wilson fermions.  The main goal of the improvement program
in the context of domain wall fermions is to achieve the same quality of
chiral symmetry (as measured, \eg, by the residual mass)
for smaller extent of the fifth coordinate, thereby
reducing the cost of numerical calculations with domain wall fermions.

In Sec.~\ref{mobius} we discuss M\"obius fermions, which have become
the method of choice for large-scale numerical calculations
using domain wall fermions.  In Sec.~\ref{suppl} we revisit two topics
already discussed in earlier parts using results from Sec.~\ref{mobius}.
We resume the discussion
of improvements in Sec.~\ref{optimal}, where we discuss a version
of domain wall fermions with couplings that depend on the fifth coordinate,
aimed to minimize the residual mass for any given $N_5$.
Sec.~\ref{newK} is devoted to attempts to obtain faster reduction of
chiral symmetry violations by using different four-dimensional kernels.
Finally Sec.~\ref{dflt} presents a method to reduce the residual
chiral symmetry violations via deflation.

\subsection{\label{mobius} M\"obius fermions}
In this subsection we discuss M\"obius fermions \cite{Brower:2012vk}.
We first explain in Sec.~\ref{WhyM} why M\"obius fermions offer
faster convergence to the chiral limit than the original version of
domain wall fermions.
The M\"obius fermion action is then introduced in Sec.~\ref{DM}
following the concrete implementation of Ref.~\cite{RBC:2014ntl}.
We also discuss the resulting effective four-dimensional operator.
Last, in Sec.~\ref{subsec:mcc} we derive the M\"obius conserved
vector currents and partially-conserved axial currents.

\subsubsection{\label{WhyM} Why M\"obius fermions}
M\"obius domain wall fermions~\cite{Brower:2012vk} are a generalization
of domain wall fermions that are more efficient for numerical computations.
In the limit $N_5\to\infty$, the original version of domain wall fermions
and M\"obius fermions both correspond to the same chirally symmetric
fermion action (this statement will be made precise below). In other words,
at finite $N_5$ they differ by small residual chiral symmetry breaking effects
on the order of the residual mass.

It is easiest to understand the above facts by first following the steps
in Sec.~\ref{DWF2GW} to show the relationship between domain wall fermions
in the $N_5\to\infty$ limit and overlap fermions, and the most general
lattice chiral fermions, those that satisfy the Ginsparg-Wilson relation,
Eq.~(\ref{GWrel}).

In Sec.~\ref{DWF2GW} we considered $\DGW(N_5)$, the approximate GW operator
obtained from domain wall fermions at finite $N_5$, and expressed it
in Eq.~(\ref{DGWN}) in terms of the transfer matrix $T$ of Eq.~(\ref{T}).
An alternative expression for this transfer matrix which will be convenient
for what follows is
\begin{equation}
\label{THS}
T = \frac{1-H}{1+H} \ ,
\end{equation}
where the hermitian kernel is $H=H_S=\g_5 D_S$, with
\begin{equation}
    \label{Dshamir}
    D_{S} = \frac{a_5 D_W(M)}{2+a_5 D_W(M)} \ .
\end{equation}
Following Ref.~\cite{RBC:2014ntl} in this subsection, the notation for
the Wilson kernel is $D_W(M) = -D(M)$ with $D(M)$ in Eq.~(\ref{DW}),
namely, $D_W(M) =  -D_K +W -M$.
We have re-introduced the lattice spacing in the fifth dimension, $a_5$
(which is typically set equal to 1 in numerical simulations).
For $a_5=1$ one can prove Eq.~(\ref{THS}) using Eq.~(\ref{T}) and\footnote{%
  As in Sec.~\ref{free}, we work in units of the four-dimensional
  lattice spacing.  For $a_5\ne 1$ see Ref.~\cite{Shamir:1998ww}.
}
\begin{subequations}
\label{Drels}
\begin{eqnarray}
\label{Drelse}
  2B^{\half} K^{-1} &=& \rule{0ex}{4.5ex}
       2 \left(\begin{array}{cc}
         B  &  0     \\
         -C^\dagger & 1
         \end{array}\right)
  \ = \ (2+D_W^\dagger(M)) + \g_5 D_W(M)  \ ,
\\
\label{Drelsf}
   2B^{\half} K^\dagger &=& \rule{0ex}{4.5ex}
       2 \left(\begin{array}{cc}
         1  & C     \\
         0  & B
         \end{array}\right)
   \ = \ (2+D_W^\dagger(M)) - \g_5 D_W(M) \ .
\end{eqnarray}
\end{subequations}
With Eq.~(\ref{THS}) at hand, $\DGW(N_5)$ can be reexpressed as
\begin{equation}
\label{DHS}
\DGW(N_5;H) = \half\left(1 + \g_5 \te(N_5;H)\right) \ ,
\end{equation}
where
\begin{equation}
\label{tildee}
\te(N_5;H) = \frac{(1+H)^{N_5}-(1-H)^{N_5}}{(1+H)^{N_5}+(1-H)^{N_5}} \ ,
\end{equation}
and $H=H_S$.  For $N_5\to\infty$, the approximate sign function $\te(N_5;H)$
tends to the sign function $\e(H)$.\footnote{
  For the precise statement, see Sec.~\ref{chiragain} below.
}

Notice that the Shamir kernel $D_S$ resembles a conformal transformation
on the Wilson operator $D_W(0)=-D_K+W$, with transformation parameters
that depend on $M$ and $a_5$. Similarly, a general conformal transformation
of $D_W(0)$ on the real axis can be written as
\begin{equation}
    \label{Dmobius}
    D_{M} = \frac{(b+c) D_W(M)}{2+(b-c) D_W(M)}=\alpha D_S \ ,
\end{equation}
where the three real parameters of the conformal transformation
are functions of the three parameters $b$, $c$, and $M$.
This transformation is also called a M\"obius transformation, hence the name.
The original formulation of domain wall fermions can be thought of
as a special case of M\"obius fermions, one with $b=a_5$ and $c=0$
and, usually, $a_5=1$ as well.\footnote{
  Another special case is $b=c$, where $D_M$ reduces to the familiar
  Wilson kernel of the overlap operator
  up to a multiplicative factor \cite{Borici:1999da}.
}
More interesting is the simple rescaling~\cite{Brower:2012vk}
$\alpha=(b+c)/a_5$ in the second equation
while simultaneously keeping $a_5=b-c$ fixed,
which we will discuss shortly, assuming $a_5>0$ and $\a>0$.
If moreover we let $b=b(s)$ and $c=c(s)$ depend on the coordinate in the
fifth dimension, more general possibilities follow,
like the Zolotarev domain wall fermions discussed in Sec.~\ref{optimal}.

The simple rescaling in Eq.~(\ref{Dmobius}) leaves the sign function in
Eqs.~(\ref{DGWlim}) and~(\ref{DGWm}) unchanged if $N_5\to\infty$.
However if $N_5$ is finite, the approximations to the sign function for $H_S$
and for $H_M=\g_5 D_M$ will differ. To see how much they differ
we should investigate the violations of the Ginsparg-Wilson relation
for $m=0$ and finite $N_5$.
An obvious way is to simply quantify the violation of the
Ginsparg-Wilson relation~(\ref{GWrel}) induced by finite $N_5$,
\begin{equation}
\label{GWviolation}
2 \gamma_5\Delta(N_5;H)\equiv\gamma_5 D_{GW}+D_{GW}\gamma_5-2 D_{GW}\gamma_5 D_{GW}
= \frac{1}{2}\gamma_5(1-\tilde\epsilon(N_5;H)^2) \ .
\end{equation}
Notice that $\sum_{x,y} \bq(x) 2 \gamma_5 [\Delta(N_5;H)]_{xy} q(y)$
is the variation of the four-dimensional effective action
$\sum_{x,y} \bq(x) [D_{GW}(N_5;H)]_{xy} q(y) \rule{0ex}{2.25ex}$
under the corresponding approximate L\"uscher chiral symmetry
$\d\bq(x) = \bq(x) \g_5$,
$\d q(x) = \sum_y \g_5[1-2D_{GW}(N_5;H)]_{xy}q(y)  \rule{0ex}{2.25ex}$
\cite{Luscher:1998pqa}. It can be shown \cite{Brower:2012vk} that
$\Delta(N_5;H) \rule{0ex}{2.25ex}$ is a quantity of order $O(\mres)$,
and hence $\tilde\epsilon(N_5;H) \rule{0ex}{2.25ex}$
differs from $\epsilon(H)$ by $O(\mres)$ for any $H$.

Given an eigenvalue $\l$ of $H_S$, at a qualitative level we identify
three spectral regions for the approximate sign function:
\begin{equation}
\label{eq:scaling}
  \tilde\epsilon(N_5;H_M) = \tilde\epsilon(N_5;\alpha H_S) \approx
  \left\{ \begin{array}{ll}
    N_5 \alpha \lambda\ , \qquad & \alpha |\lambda| \ll 1/N_5 \ , \\
    \rule{0ex}{3ex}
    \pm 1 \ , \qquad & 1/N_5 \ll \alpha|\lambda| \ll N_5 \ , \\
    N_5 / (\alpha \lambda)\ , \qquad & \alpha |\lambda| \gg N_5 \ .
    \rule{0ex}{3ex}
  \end{array}\right.
\end{equation}
Notice the symmetry under $\lambda\to 1/\lambda$ of these
spectral regions, which can be traced back
to the behavior of Eq.~(\ref{THS}) under $H\to 1/H$.
We see that $\tilde\epsilon(N_5;\alpha \lambda)$ is a good approximation
of the sign function $\epsilon(\lambda) = \pm 1$ provided that
$\alpha |\lambda|$ is neither too small nor too large.
By contrast, in the regions $\alpha |\lambda| \ll 1/N_5$
and $\alpha |\lambda| \gg N_5$ we obtain poor approximations of the
sign function, hence these spectral regions will dominate the residual
breaking of chiral symmetry, including in particular $\mres$.

We may take the number of modes satisfying $\alpha |\lambda| \ll 1/N_5$
as a rough measure of the residual chiral symmetry breaking coming
from the near-zero spectrum.  If the spectral density in the relevant range
is roughly constant, this number will be inversely proportional
to $\alpha N_5$.  At fixed $N_5$, the residual chiral symmetry breaking
can therefore be reduced by switching to the M\"obius kernel $H_M$ and
increasing $\alpha$.  Stated differently, we may keep the residual breaking
of chiral symmetry roughly the same by simultaneously lowering $N_5$
and increasing $\alpha$ keeping their product fixed.
This allows reaching a given target quality of chiral symmetry
for smaller $N_5$, hence at smaller cost, which explains
why M\"obius fermions are more efficient
for numerical computations~\cite{Brower:2012vk}.

The value of $\alpha$ cannot be increased indefinitely, however,
because this will populate the region $\alpha |\lambda| \gg N_5$
where $\tilde\epsilon(N_5;\alpha \lambda)$ again becomes
a poor approximation of $\epsilon(\lambda)$.  Finding the optimal value
of $\alpha$ requires numerical experimentation.  In practice,
$\alpha=2$ was found to be a reasonable choice \cite{RBC:2014ntl}.

\subsubsection{\label{DM} The M\"obius fermion operator}
Having established the relationship between the M\"obius version and the
original version of domain wall fermions in Sec.~\ref{WhyM}, we next turn to
the five-dimensional action and fermion propagators for M\"obius fermions.
We define physical four-dimensional quark fields just as before, through
the fields on the boundaries of the fifth dimension, Eq.~(\ref{dwq}).
With these relations and similar ones for propagators into the bulk
fifth dimension, we will also obtain conserved vector currents
and partially-conserved axial currents for M\"obius domain wall fermions.

The five-dimensional M\"obius action closely resembles
the domain-wall fermion action:
\begin{equation}
    \label{eq:5d mobius action}
     S_M^5=\bar\psi D^5_{M}\psi \ ,
\end{equation}
where $D_M^5=D_M^5(N_5,m)$ is
\begin{equation}
\label{eq5dDM}
\hspace{-3ex}
D_M^5 = \left(\begin{array}{ccccccccc}
\tilde D & -P_L & 0   & 0   &\hspace{1ex}\cdots\hspace{1ex}& 0 & 0 & 0 & m P_R \\
-P_R & \tilde D & -P_L & 0   & \cdots & 0 & 0 & 0 & 0 \\
0   & -P_R & \tilde D & -P_L & \cdots & 0 & 0 & 0 & 0 \\
\rule{0ex}{4ex}
\vdots&\vdots& & \hspace{-3ex}\ddots\hspace{3ex} & \ddots &
\hspace{3ex}\ddots\hspace{-3ex} & &\vdots&\vdots \\
\rule{0ex}{4ex}
0   & 0   & 0   & 0   & \cdots & -P_R & \tilde D & -P_L & 0   \\
0   & 0   & 0   & 0   & \cdots & 0   & -P_R & \tilde D & -P_L \\
m P_L & 0 & 0  & 0   & \cdots & 0   & 0   & -P_R & \tilde D
\end{array}\right) ,
\end{equation}
and
\begin{subequations}
\label{Dpm}
\begin{eqnarray}
    \tilde D &=& D_-^{-1} D_+,\\
    D_+ &=& b D_W + 1,\\
    D_- &=& 1-c D_W.
\end{eqnarray}
\end{subequations}
In this subsection we follow the RBC/UKQCD conventions \cite{RBC:2014ntl}
for the M\"obius operator,
which differ from Ref.~\cite{Brower:2012vk} by an explicit factor of $D_-$.
These conventions also differ from the conventions of Sec.~\ref{free}
for the domain wall operator, Eq.~(\ref{finites}),
in the flipping of the chiral projectors,
hence the boundaries supporting the LH and RH chiral modes are also flipped,
as well as by an overall minus sign,
including for the Wilson-Dirac operator as already mentioned above.

The four-dimensional projection of the five-dimensional M\"obius operator
can be related to the corresponding (approximate) GW operator.
For the derivation, see Appendix A of Ref.~\cite{RBC:2014ntl}.
The result is
\begin{equation}
\label{Dgw5}
  D_{GW}(N_5,m)
  = \big[ {\cal P}^{-1}D^5_{M}(N_5,1)^{-1}D^5_M(N_5,m){\cal P}\big]_{11},
\end{equation}
where the indices $[\cdot]_{ss'}=[\cdot]_{11}$ refer to the matrix structure
in the fifth direction,\footnote{
  As in Sec.~\ref{free} the fifth coordinate takes values $s=1,2,\ldots,N_5$.
}
and $D_{GW}(N_5,m)$
is the corresponding massive GW operator defined in Eq.~(\ref{DGWm}),
except with $\te(N_5;H_M)$ replacing $\e(H_M)$.
We have introduced a permutation matrix,
\begin{equation}
\label{eq5dP}
\hspace{-3ex}
{\cal P} = \left(\begin{array}{ccccccc}
P_L & P_R & 0 & \ldots & 0 & 0 & 0 \\
0 & P_L & P_R & \ldots & 0 & 0 & 0 \\
\rule{0ex}{3ex}
\vdots&\vdots& &
\hspace{-1ex}\ddots\hspace{1ex} & &\vdots&\vdots \\
\rule{0ex}{3ex}
0   & 0 & 0 & \ldots & P_L & P_R & 0 \\
0   & 0 & 0 & \ldots & 0 & P_L & P_R \\
P_R & 0 & 0 & \ldots & 0 & 0   & P_L
\end{array}\right).
\end{equation}
With the help of $\cp$, the four-dimensional quark fields introduced
in Eq.~(\ref{dwq}) can be reexpressed as
\begin{equation}
\label{qqbar}
q=(\cp^{-1}\j)_{s=1}\ , \qquad \bq=(\bj\car\cp)_{s=1}\ .
\end{equation}
We see that the role of the permutation matrix is to move
the RH quark field, which here lives originally near the $s=N_5$ boundary,
to the $s=1$ boundary.  The end result is that, in the operator
$D^5_M(N_5,m){\cal P}$, the two chiral components
of the light domain wall quark both live near the $s=1$ boundary.

With Eq.~(\ref{Dgw5}) at hand, it is a straightforward exercise to relate
$D_{GW}(N_5,m)$ back to the domain wall propagator for the effective
four-dimensional quark field, $\Deff^{-1}(N_5,m)$.  Generalizing Eq.~(\ref{GWprop})
to $m\ne 0$, we subtract a contact term from the propagator $D_{GW}^{-1}(N_5,m)$
and divide by a factor $(1-m)$, obtaining\footnote{%
  Let us denote the five-dimensional matrix occurring on the right-hand side
  of Eq.~(\ref{Dgw5}) by $\cm$, so that the equation becomes
  $D_{GW}(N_5,m)=[\cm]_{11}$.
  Using the UDL decomposition introduced in Ref.~\cite{RBC:2014ntl}
  it can then be shown that $D_{GW}^{-1}(N_5,m)=[\cm^{-1}]_{11}$.
}
\begin{eqnarray}
    D^{-1}_{\rm eff}(N_5,m) &\equiv& \frac{1}{1-m}(D_{GW}^{-1}(N_5,m)-1) \\
     &=& \frac{1}{1-m}\big[ {\cal P}^{-1}D^5_{M}(N_5,m)^{-1}
    \left(D^5_M(N_5,1)-D^5_{M}(N_5,m)\right){\cal P}\big]_{11} \ .
\nonumber
\end{eqnarray}
The subtraction term in parentheses is explicitly
\begin{equation}
[D^5_M(N_5,1)-D^5_{M}(N_5,m)]_{ss'} =
(1-m)(P_L\delta_{s,N_5}\delta_{s',1}+ P_R\delta_{s,1}\delta_{s',N_5}) \ ,
\end{equation}
hence $[(D^5_M(N_5,1)-D^5_{M}(N_5,m)){\cal P}]_{ss'}
= (1-m)[{\cal R}{\cal P}]_{ss'}$ for $s'=1$,
where $\car$ is the reflection on the fifth coordinate introduced in Sec.~\ref{finite5}.
This gives
\begin{equation}
\label{DeffM}
  D^{-1}_{\rm eff}(N_5,m)
  = \big[ {\cal P}^{-1}D^5_{M}(N_5,m)^{-1}{\cal R}{\cal P}\big]_{11} \ .
\end{equation}
Noting Eq.~(\ref{qqbar}), it follows that Eq.~(\ref{DeffM}) reproduces
the usual domain wall four-dimensional propagator, Eq.~(\ref{Deff}).
Thus, the four-dimensional domain wall or M\"obius propagator $\Deff^{-1}$
already encodes the usual contact term subtraction
of the GW propagator, which implies that for $m=0$ it anti-commutes with $\g_5$ in the limit $N_5\to\infty$.

\subsubsection{\label{subsec:mcc} M\"obius conserved currents}
To obtain formulae for the (partially) conserved vector and axial currents,
we have two options.  The first option is to invoke Noether's theorem,
as we have done in Sec.~\ref{syms}.  The essence of the method
is to promote the global symmetry transformation parameter $\o^a$
to a spacetime dependent one, $\o^a(x)$.  In the continuum,
because the action is invariant under the global flavor transformation
where $\o^a$ is constant, the variation must then take the form
$\d S = \int d^dx\, J_\m^a(x) \partial_\m\o^a(x)$,
which defines the current $J_\m^a(x)$ as the coefficient of $\partial_\m\o^a(x)$.
Conservation of the current then follows by partial integration,
while the ensuing Ward-Takahashi identities follow by performing the
corresponding change of variables that leaves the partition function invariant.

Unlike in the continuum, on the lattice there can be more than one way
of expressing $\d S$ as $\sum_x \o^a(x) D_\m J_\m^a(x)$ where $D_\m$
is a difference operator.  Indeed we have already made use of this freedom
in Sec.~\ref{syms} to express $j_5(x,s)$ and $\D_5$ in a slightly different way
from Ref.~\cite{Furman:1994ky} (keeping $\D_5 j_5(x,s)$ the same).
When the Dirac operator has an infinite range, as in the case
of the M\"obius operator of Eq.~(\ref{eq5dDM}),
it gets significantly more complicated to figure out how to best
express $\d S$ as a difference operator acting on a current.
The reason is that the current and/or the difference operator
must now also have an infinite range.

Instead, we will use here the alternative method.
We first promote the global symmetry to a local one by introducing
an external, fictitious flavor gauge field.  The conserved current
is then define as the infinitesimal response to this
gauge field~\cite{Brower:2012vk,RBC:2014ntl}, while current conservation
follows from gauge invariance of the partition function in the presence
of the external flavor gauge field.

We start with the conserved vector current (for the domain wall case,
see Eq.~(\ref{Vmu})).  We proceed as follows.
All (four-dimensional) links are replaced by
\begin{equation}
    U_\mu(x)\to U_\mu(x) \times e^{i\lambda^a A^a_\mu(x)},
\end{equation}
where $A^a_\mu(x)$ is the fictitious four-dimensional gauge potential
which will be set to zero at the end of our manipulations,
and $\lambda^a$ is a generator of the vector flavor symmetry.
The linearized variation of the flavor gauge field $A^a_\mu(x)$
under a gauge transformation is
$\delta A_\mu^a(x) = \o^a(x+\hat\m) - \o^a(x)$.
As a result, the difference operator occurring in the lattice
continuity equation will always be the nearest-neighbor backward
difference operator.

In order to obtain the conserved vector current we take the variation
of the five-dimensional M\"obius action with respect to $A^a_\mu(x)$.
Using Eq.~(\ref{Dpm}) we find
\begin{eqnarray}
  \label{eq:mcc}
  \cv_\mu^a(x) &\equiv& -i\delta_{A_\mu^a(x)} \bar \Psi D^5_{M}\Psi
\\
  &=& -i\sum_{s,z,y}
  \bar\Psi(z,s)\left[\delta_{A_\mu^a(x)}\tilde{D}\right]_{z,y}\Psi(y,s)
\nonumber\\
  &=& \sum_{s,z,y} \bar\Psi(z,s)\left[\frac{1}{D_-}c \lambda^a V_\mu(x)
  \frac{1}{D_-} D_+ + \frac{1}{D_-}b \lambda^a V_\mu(x) \right]_{z,y}\Psi(y,s)
\nonumber\\
  &=& \sum_{s,z,y} \bar\Psi(z,s)\left[\frac{1}{D_-}\lambda^a V_\mu(x)
  \left(c\frac{1}{D_-} D_+ +b\right)\right]_{z,y}\Psi(y,s)
\nonumber\\
  &=& (b+c)\sum_{s,z,y}  \bar\Psi(z,s)\left[\frac{1}{D_-}\lambda^a V_\mu(x)
  \frac{1}{D_-}\right]_{z,y}\Psi(y,s) \ ,
\nonumber
\end{eqnarray}
where the usual (finite-range) kernel for the Wilson fermion conserved current
acts within the four-dimensional layers,
and is given by (compare Eq.~(\ref{jmu}))
\begin{eqnarray}
  \l^a [V_\mu(x)]_{z,y} &=& -i[\delta_{A_\mu^a(x)}D_W]_{z,y}
\\
  &=& -\l^a\left(\frac{1+\gamma_\mu}{2}U_\mu(x)\delta_{z,x}\delta_{x+\hat\mu,y}
  -\frac{1-\gamma_\mu}{2}U^\dagger_\mu(x)\delta_{z+\hat\mu,x}\delta_{x,y}\right)\ .
\nonumber
\end{eqnarray}
In contrast to Refs.~\cite{Brower:2012vk,RBC:2014ntl}, Eq.~(\ref{eq:mcc})
is a very simple, compact result. In the case of domain wall fermions,
$b=1,\,c=0$, it reduced to the result found in Sec.~\ref{syms}
up to an overall minus sign.

At first sight the factors of $D_-^{-1}$ in (\ref{eq:mcc}) seem like
an added computational expense. However, we can multiply $D^5_{M}$ on the right
or left by $D_-$ before computing the propagators that sandwich the current
in a physical quark line diagram. Thus in practice the conserved current
can be implemented in essentially the same way as for Wilson (or domain wall)
fermions.  For an insertion of $\cv_\mu^a(x)$ into a physical quark line
we have, using Eqs.~(\ref{qqbar}) and~(\ref{eq:mcc}),
and denoting fermion contractions in a given
gauge field background with expectation values,
\begin{eqnarray}
\label{Cmu}
    && \hspace{-7ex} C_\mu^a(x,y,z) \ = \ \svev{q(z)\,\cv_\m^a(x)\,\bq(y)}
\\
    &=& (b+c)\sum_{s,z',y'} \svev{q(z)\,\bJ(z',s)}
    \left[\frac{1}{D_-}\lambda^a V_\mu(x)\frac{1}{D_-}\right]_{z',y'}
    \svev{\J(y',s)\bq(y)}
\nonumber\\
    &=& (b+c) \sum_{s,z',y'} [\cp^{-1}(D^{5}_{M})^{-1}]_{z,1;z',s}
    \left[\frac{1}{D_-}\lambda^a V_\mu(x) \frac{1}{D_-}\right]_{z',y'}
    [(D_M^5)^{-1}\car\cp]_{y',s;y,1}
\nonumber\\
    &=& (b+c) \sum_{s,z',y'} [\cp^{-1}(D_- D^{5}_{M})^{-1}]_{z,1;z',s}
    \left[\lambda^a V_\mu(x)\right]_{z',y'}
    [(D_M^5 D_-)^{-1}\car\cp]_{y',s;y,1}\ .
\nonumber
\end{eqnarray}
Since $D_-$ and $\tilde{D}$ commute,
$D_- \tilde{D} = \tilde{D} D_- = D_+$, it follows that $D_- D_M^5$
and $D_M^5 D_-$ are both finite-range operators.
These operators are also related via the $\g_5$-hermiticity properties
which are similar to those discussed in Sec.~\ref{finite5},
for example, $D_M^5 D_- = \g_5\car (D_- D_M^5)^\dagger  \g_5\car$.

In order to obtain the partially conserved axial current we first construct
the five-dimensional conserved current, which is defined in a similar manner
by taking the variation with respect to a five-dimensional gauge potential,
$(A^a_\mu(x,s),A^a_5(x,s))$.
The result for the four-dimensional components is identical to the above
except there is no sum over the fifth dimension,
\begin{equation}
  \label{eq:mcc 5d}
  j_\mu^a(x,s) = (b+c)\sum_{z,y}  \bar\Psi(z,s)\left[\frac{1}{D_-}\lambda^a V_\mu(x)
  \frac{1}{D_-}\right]_{z,y}\Psi(y,s) \ .
\end{equation}
Because of the different conventions we adopted for the M\"obius operator
(compare Eqs.~(\ref{finites}) and~(\ref{eq5dDM})) the current
in the fifth direction is slightly different from Eq.~(\ref{j5}).
Instead, we now have
\begin{equation}
\label{eq:j5Mobius}
  j^a_5(x,s) = \left\{ \begin{array}{ll}
    -\bj(x,s) P_L \l^a \j(x,s+1)  + \bj(x,s+1) P_R \l^a \j(x,s)\ ,
    & 1\le s < N_5 \ ,
\\
    m \bj(x,N_5) P_L \l^a \j(x,1) - m \bj(x,1) P_R \l^a \j(x,N_5)\ ,
    & s=N_5 \ . \rule{0ex}{3ex}
     \end{array}\right.
\end{equation}
As in Sec.~\ref{syms}, the only current which depends on the quark mass $m$
is $j^a_5(x,N_5)$.

The rest of the construction is the same as in Sec.~\ref{syms}.
The five-dimensional continuity equation is again Eq.~(\ref{5cons}).
The axial current is defined via Eq.~(\ref{Amu}) as before,
and its partial conservation equation is again given by Eq.~(\ref{divA})
with the right-hand side defined via Eqs.~(\ref{J5a}) and~(\ref{J5qa}).
Importantly, the calculation of the residual mass is same as for standard
domain wall fermions~\cite{Brower:2012vk,RBC:2014ntl} (see Sec.~\ref{mres}).

In Refs.~\cite{RBC:2014ntl,Boyle:2015vda} a somewhat different approach is followed
to construct the currents, but in the end they are equivalent to the ones
defined here.
Instead of taking the variation with respect to external gauge links
of the five-dimensional M\"obius operator, as we have done above
(see in particular Eq.~(\ref{eq:mcc})),
in Refs.~\cite{RBC:2014ntl,Boyle:2015vda} one takes the variation
of the (approximate) GW operator, Eq.~(\ref{Dgw5}).
Since this effective four-dimensional operator is expressed directly
in terms of the five-dimensional M\"obius operator, it is possible
to consider its variation with respect to both four-dimensional
and five-dimensional flavor gauge transformations; the latter
are needed in order to construct the partially conserved axial currents
using Eq.~(\ref{Amu}), which in turn leads to the domain-wall fermion
version of the PCAC relation in Eq.~(\ref{divA}).
The application of a (linearized) gauge transformation
to the external flavor gauge fields then yields the usual conservation law
(or Ward-Takahashi identity), and it is straightforward
to identify the currents themselves.  The procedure is somewhat involved,
and we refer the interested reader to Appendix~A of Ref.~\cite{RBC:2014ntl}
and the lattice proceedings Ref.~\cite{Boyle:2015vda}.

\subsection{\label{suppl} Further symmetry considerations}
In this subsection we revisit two topics already discussed earlier,
and expand their scope using ingredients introduced in our discussion
of M\"obius fermions in the previous subsection.
In Sec.~\ref{chiragain} we extend the scope
of the proof of chiral symmetry restoration given in Sec.~\ref{chirN5},
while in Sec.~\ref{phased} we establish the symmetry features of the phase diagram,
a subject we have already encountered in Sec.~\ref{aoki} and Sec.~\ref{quench}.
The discussion of improvements will be resume in Sec.~\ref{optimal}.

\subsubsection{\label{chiragain} Chiral symmetry restoration revisited}
In Sec.~\ref{chirN5} we proved the restoration of chiral symmetry for
$N_5\to\infty$ within the original formulation of domain wall fermions,
assuming a domain wall height $0<M<1$.
This implies that $B$ (Eq.~(\ref{B})) is a positive operator,
as are the various transfer matrices introduced there.
In the range $1<M<2$ the operator $B$ can have negative eigenvalues,
and the same is true for the transfer matrices.\footnote{
  Notice that when $B$ has negative eigenvalues, $K$ and $K^\dagger$
  have to be defined via suitable analytic continuations.
  Hence Eqs.~(\ref{T}) and~(\ref{Ttilde}) no longer imply that $T$ and $\tT$
  are positive.  But $T$ remains a hermitian matrix, and $T$ and $\tT$
  still share the same spectrum.
}
While we expect such negative eigenvalues to be rare (see Sec.~\ref{tadpole}),
in a proof of chiral symmetry restoration their role has to be addressed.

Here we will extend the proof of chiral symmetry restoration to the
entire range of domain wall height $0<M<2$, as well as to general values
of the M\"obius parameters $b$ and $c$.  As a measure of
chiral symmetry violation for finite $N_5$ we will use Eq.~(\ref{GWviolation})
with $H=H_M$.

As follows from the discussion in Sec.~\ref{WhyM},
in the limit $N_5\to\infty$
the approximate sign function $\te(N_5,\l)$ of Eq.~(\ref{tildee})
tends to $\e(\l)=\pm 1$ for arbitrary (real) $\l$, except for
$\l=0$ or $\l=\pm\infty$.  Let us examine these two exceptional cases.

In order for $H_M$ to have an eigenvalue $\l=0$, the numerator
in Eq.~(\ref{Dmobius}) must vanish, which implies that $D_W(M)$ has a zero mode.
This is the same condition we have already encountered in Sec.~\ref{chirN5}.
For an eigenvalue $\l=\pm\infty$ of $H_M$, the denominator in
Eq.~(\ref{Dmobius}) must vanish, which implies that $D_W(M)$ has an eigenvalue
equal to $2/(c-b)$, or equivalently, that $\det(2+(b-c)D_W(M))=0$.
Either one of these special cases can be satisfied only on
a measure zero subset of the gauge field configuration space.
Using this observation, we can now complete
the proof of chiral symmetry restoration in essentially the same way
as we did in Sec.~\ref{chirN5}, by showing that the ensemble average of
any correlation function that contains
an insertion of $1-\te(N_5;H)^2$ vanishes for $N_5\to\infty$.

According to Eq.~(\ref{THS}),
an eigenvalue $+1$ of the transfer matrix $T$ corresponds to a zero eigenvalue
of $H_S$, while an eigenvalue $-1$ of the transfer matrix corresponds
to eigenvalues $\l=\pm\infty$ of $H_S$.  These statements remain true
if we generalize the definition of $T$ by replacing $H_S$ with $H_M$
in Eq.~(\ref{THS}).  Eigenvalues $\pm 1$ of the transfer matrix
both give rise to zero eigenvalues of the transfer matrix hamiltonian $H_T$,
which is defined in terms of $\log T^2$ (see Eq.~(\ref{HlogT})).
We have already seen in Sec.~\ref{mresagain} that,
for the original formulation of domain wall fermions,
the dominant chiral symmetry violations for very large but finite $N_5$
come from the (near-)zero spectrum of $H_T$.
This behavior generalizes to M\"obius fermions, as it is consistent with
the roles of the spectral ranges in Eq.~(\ref{eq:scaling}):
transfer matrix eigenvalues near $+1$ correspond to
the range $\a\l\ll 1/N_5$ for the eigenvalues of $H_M$,
while transfer matrix eigenvalues near $-1$ correspond to $\a\l\gg N_5$.
Once again the underlying reason is the behavior of the transfer matrix
in Eq.~(\ref{THS}) and the approximate sign function in Eq.~(\ref{tildee})
under $H \to 1/H$.

\subsubsection{\label{phased} Symmetry of the phase diagram}
Symmetry features of the phase diagram were discussed in Sec.~\ref{aoki}
and Sec.~\ref{quench}.  Here we give a fuller discussion of the symmetry of
the phase diagram for the original formulation of domain wall fermions
with variable $a_5$, as well as for M\"obius fermions.
We verify that the corresponding chiral zero mode spectrum and effective
four-dimensional operators respect this symmetry.

We start with the Wilson operator in $d$ dimensions.
Using the same notation as in Sec.~\ref{mobius} for the Wilson operator,
and adding a superscript $d$ to denote $d$-dimensional quantities,
the $d$-dimensional Wilson operator is
\begin{equation}
\label{DWh}
D_W^d(M) = -D_K^d + W^d - M = D_h^d + d-M \ ,
\end{equation}
where as usual $M=-m_0$, and $D_h^d$ contains the hopping terms which can be
read off from Eqs.~(\ref{DKU}) and~(\ref{WU}).
It is easy to see that under the change of variables
\begin{equation}
\label{changev}
\j(x) \to \e_d(x) \j(x) \ , \qquad \bj(x) \to -\e_d(x) \bj(x) \ ,
\end{equation}
where $\e_d(x) = (-1)^{x_1+\cdots+x_d}$, the Wilson operator transforms as
\begin{equation}
\label{DMM'}
D_W^d(M;x,y) \to -\e_d(x) D_W^d(M;x,y) \e_d(y) = D_W^d(M';x,y) \ ,
\end{equation}
where
\begin{equation}
\label{M'}
M' = 2d-M \ .
\end{equation}
This result follows since the (nearest neighbor) hopping terms
are invariant under the transformation~(\ref{changev}), whereas
the same-site terms flip sign.  Hence the phase diagram is symmetric
with respect to the axis $M=-m_0=d$.

We next turn to the original formulation of domain wall fermions except
with variable lattice spacing $a_5$ in the fifth direction.
The transformation~(\ref{changev}) now takes $M$ to
\begin{equation}
\label{M'a5}
M' = 8+\frac{2}{a_5}-M \ ,
\end{equation}
hence the phase diagram is symmetric with respect to the axis $M=4+1/a_5$.

Let us check that the chiral spectrum of domain wall fermions is consistent
with this symmetry (see \eg\ Ref.~\cite{Shamir:1998ww}).
On a semi-infinite fifth dimension, the condition for having a chiral zero mode
at a given corner of the Brillouin zone is
\begin{equation}
\label{zmMn}
|1+a_5(2n-M)| < 1 \ .
\end{equation}
As in Sec.~\ref{semi5}, the number of components of the
4-momentum equal to $\p$ is $0\le n \le 4$,
while the remaining $4-n$ components are equal to zero.
In momentum space, the transformation~(\ref{changev}) interchanges
$p_\m=0$ with $p_\m=\p$.  Hence it induces the transformation $n\to n'$ where
\begin{equation}
\label{nn'}
n' = 4-n \ .
\end{equation}
Under the combined replacement $(M,n)\to(M',n')$ we have
\begin{equation}
\label{MnM'n'}
1+a_5(2n'-M') = -(1+a_5(2n-M)) \ .
\end{equation}
The overall minus sign on the right-hand side can be traced
to the factor of $(-1)^s$ contained in $\e_5(x,s)$.
Condition~(\ref{zmMn}) is seen to be invariant under this combined replacement,
which confirms that the zero modes spectrum respects the symmetry
of the phase diagram.  For $a_5=1$, \Tab{zms} is reproduced.

For M\"obius fermions, it is more convenient to use the original formulation
of Ref.~\cite{Brower:2012vk} with the M\"obius operator $\DBNO = D_- D_M^5$
(see Eq.~(\ref{eq5dDM})).
The M\"obius parameters can be expressed as $b = (a_5/2)(\a+1)$
and $c = (a_5/2)(\a-1)$.  The transformation~(\ref{M'a5})
of the mass parameter is now supplemented by a transformation of the
scaling parameter,
\begin{equation}
\label{aa'}
\a' = \a^{-1} \ .
\end{equation}
We then have
\begin{equation}
-\e_4 \left( 1 + a_5 \frac{1\pm\a'}{2}D_W(M') \right) \e_4
= \pm \a^{-1} \left( 1 + a_5 \frac{1\pm\a}{2}D_W(M) \right) .
\end{equation}
Using also that $\e_5 = \e_5(x,s) = \e_4(x)(-1)^s$ it follows that
\begin{equation}
\label{DBNO'}
\DBNO(M',\a') = -\a^{-1} \e_5 \DBNO(M,\a) \e_5 \ .
\end{equation}
Thus, in this formulation, the symmetry of the phase diagram under the
interchange $(M,\a) \to (M',\a')$ can be realized via the change of variables
\begin{equation}
\label{changeM}
\j(x,s) \to \e_5(x,s) \j(x,s) \ , \qquad
\bj(x,s) \to -\a^{-1} \e_5(x,s) \bj(x,s) \ .
\end{equation}
The original domain wall formulation is recovered for $\a=1$,
where the transformations reduce to what we already had before.

For the M\"obius operator, the condition for having
a zero mode at a given corner of the Brillouin zone is $|R|<1$ where
\begin{equation}
\label{RMobius}
R = \frac{1+b(2n-M)}{1-c(2n-M)}
= \frac{2+a_5(1+\a)(2n-M)}{2+a_5(1-\a)(2n-M)} \ .
\end{equation}
The transformations~(\ref{M'a5}) and~(\ref{nn'}) of $M$ and $n$
are again supplemented by the transformation~(\ref{aa'}) of the
scaling parameter $\a$.  We find
\begin{equation}
\label{RR'}
\frac{2+a_5(1+\a')(2n'-M')}{2+a_5(1-\a')(2n'-M')}
= - \frac{2+a_5(1+\a)(2n-M)}{2+a_5(1-\a)(2n-M)} \ .
\end{equation}
The minus sign on the right-hand side is similar to Eq.~(\ref{MnM'n'}).
The condition $|R|<1$ is thus invariant under the combined transformation
$(M,\a,n)\to(M',\a',n')$, thereby proving that the zero mode spectrum
respects the symmetry of the phase diagram.

Finally, let us check that the approximate GW operator $\DGW(N_5)$
of Eq.~(\ref{DGWNb}) also respects the symmetry of the phase diagram.
We will do this for the general M\"obius case since the original
domain wall formulation is obtained by setting $\a=1$.
The transfer matrix~(\ref{THS}) with $H=H_M=\g_5 D_M$ can be expressed as
(see Eq.~(\ref{Dmobius}) and notice also Eq.~(\ref{Drels}))
\begin{equation}
\label{Ta5alpha}
T = \left((2+a_5D_W^\dagger(M)) + \g_5 \a a_5D_W(M) \right)^{-1}
    \left((2+a_5D_W^\dagger(M)) - \g_5 \a a_5D_W(M) \right) \ .
\end{equation}
Transforming $M$ and $\a$ simultaneously gives
\begin{eqnarray}
\label{Ddagaa'}
&& -\e_4 \left((2+a_5D_W^\dagger(M)) \pm \g_5 a_5 \a D_W(M) \right) \e_4 \ =
\\
&& \hspace{8ex}
= \ \pm\g_5 (\a')^{-1}
\left((2+a_5D_W^\dagger(M')) \pm \g_5 \a' a_5D_W(M') \right) \g_5 \ ,
\nonumber
\end{eqnarray}
where we have use $\g_5$ hermiticity of $D_W$.  Using Eq.~(\ref{Ta5alpha})
it follows that
\begin{equation}
\label{Taa'}
\e_4 T(M,\a) \e_4 = -\g_5 T(M',\a') \g_5 \ .
\end{equation}
Finally using this result in Eq.~(\ref{DGWNb})
and noting that $N_5$ is always even we obtain
\begin{equation}
\label{DGWaa'}
\e_4 \DGW(N_5;M,\a) \e_4 = \g_5 \DGW(N_5;M',\a') \g_5 \ .
\end{equation}
Thus, the symmetry of the phase diagram under the combined interchange
$(M,\a) \to (M',\a')$ can be realized by transforming the
effective four-dimensional fields according to
$q(x) \to \e_4(x) \g_5 q(x)$, $\bq(x) \to \bq(x) \g_5 \e_4(x)$.

\subsection{\label{optimal} Zolotarev domain wall fermions}
The Zolotarev approximation of a GW operator is defined by
replacing the sign function $\e(H)$ in Eq.~(\ref{DGWlim}) with $\bar\e(N;H)$.
Here $\bar\e(N;x)=x R_N(x^2)$, where $R_N(\l)$ is the
Zolotarev rational approximation of $1/\sqrt{\l}$.
This approximation is optimal in that it satisfies
the requirement that for given $N$ and interval
$0 < \l_{\rm min} < \l < \l_{\rm max}$, $R_N(\l)$ is the
rational polynomial approximation of degree $N$ that minimizes
the maximal deviation $|R_N(\l)-1/\sqrt{\l}|$
over the interval $\l_{\rm min} \le \l \le \l_{\rm max}$.

The Zolotarev domain wall fermion operator \cite{Chiu:2002ir}
is defined by the requirement that the effective four-dimensional operator is
the approximate overlap operator obtained by replacing $\e(H)$ in Eq.~(\ref{DGWlim})
with $\bar\e(N_5;H)$, with $H=H_W$ the standard Wilson kernel.
We note that while $\l_{\rm max}$ can be taken to be some $O(1)$ number
that depends mildly on the domain wall height $M$ \cite{Chiu:2002ir},
the choice of $\l_{\rm min}$ is more tricky, and sensitive to the presence
of near-zero modes of the Wilson kernel \cite{Chen:2012jya}.

While Zolotarev domain wall fermions do in general lead to
a smaller residual mass at fixed $N_5$, this by itself does not necessarily mean that
they represent a better choice compared to other options,
and notably, compared to the M\"obius domain-wall fermions of Sec.~\ref{mobius}.

As we have explained in Sec.~\ref{remedies}
(see in particular Eq.~(\ref{mtot}))
assuming that the residual mass is on the order of $10^{-3}$ or less
in lattice units, in practice it is used as part of the light quark mass(es),
thereby ``turning the bug into a feature,'' as far as the residual mass itself
is concerned.  When comparing different domain-wall fermion formulations 
at equal residual mass, it is an open question which one will exhibit smaller 
chiral symmetry violations in other observables, particularly those sensitive 
to power divergences.  Also, the practical question is not quite how to achieve
a target value of the residual mass at the smallest possible $N_5$, but, rather,
how to achieve it at the smallest possible cost, as measured for example
by the average number of applications of the four-dimensional Wilson operator
required by the solver.  For more details, see Ref.~\cite{Brower:2012vk}
(see also Ref.~\cite{Chen:2012jya}).

\subsection{\label{newK} Changing the Wilson kernel}
Replacing the standard nearest-neighbor Wilson kernel in Eq.~(\ref{finites})
by an improved kernel was proposed by several authors.
We discuss here two proposals.

\subsubsection{\label{next} Improved exponential damping}
Motivated by the calculation of the one loop wave function of the
domain wall quark (Sec.~\ref{oneloop}), kernels that yield more efficient
exponential suppression of the wave function in perturbation theory
were discussed in Ref.~\cite{Shamir:2000cf}.

The basic idea is that
the wave function of the domain wall quark is controlled by the maximum
of $e^{-\a(p)}$ over the Brillouin zone, where in the case of the standard
Wilson kernel $\a(p)$ was determined by Eq.~(\ref{alphap}).
Let us consider a more general kernel,
\begin{equation}
\label{Wilsongen}
\tD(p) = i\sum_\m \g_\m f(p_\m) - r \tW(p) + M \ ,
\end{equation}
where $r>0$ is the Wilson parameter and $f(p_\m)$ and $\tW(p)$ will be specified
below.  Then Eq.~(\ref{alphap}) generalizes to
\begin{equation}
\label{alphapgen}
2\cosh\a(p) = \frac{1+\tb^2(p)+\sum_\m f^2(p_\m)}{\tb(p)} \ ,
\end{equation}
where now $\tb(p) = 1-M + r \tW(p)$.  By judiciously selecting
$f(p_\m)$, $r$, and $\tW(p)$ it may then be possible to reduce the maximum
of $e^{-\a(p)}$ over the Brillouin zone when the domain wall height $M$
is tuned to its optimal tadpole improved value (Sec.~\ref{tadpole}).

To keep the kernels relatively economic, only kernels with
beyond nearest neighbor couplings along straight lines were considered.
Specifically, the point $x$ was allowed to couple at most to $x\pm 2\hat\m$,
and then at most to $x \pm 3\hat\m$ in each direction.
The kernel with next-nearest neighbor couplings has $\tW(p)=W_2(p)$, where
\begin{equation}
\label{Wn}
W_n(p) = \sum_\m (1-\cos(p_\m))^n \ ,
\end{equation}
while $f(p_\m)$ was taken to be $\sin(p_\m)$ times a linear function
of $\cos(p_\m)$.  For the kernel with next-next-nearest neighbor couplings,
$\tW(p)=W_3(p)$ was taken, while $f(p_\m)$ was equal to $\sin(p_\m)$ times a
quadratic function of $\cos(p_\m)$.  The maximum of $e^{-\a(p)}$ over
the Brillouin zone was then determined numerically.

For the kernel with next-nearest neighbor couplings, it was found that
the maximum of $e^{-\a(p)}$ can be lowered from 0.5 (its value for the
standard nearest neighbor kernel,
see Sec.~\ref{oneloop}) to about 0.25, while maintaining a dispersion relation
with reasonably small $O(p^3)$ discretization effects.
For the kernel with next-next-nearest neighbor couplings,
the maximum of $e^{-\a(p)}$ can be lowered to roughly 0.1
if the dispersion relation is constrained to be $p+O(p^5)$
and the kernel's parameters are all $O(1)$.
In this example the exponential suppression factor of $1/2^s$
in the one-loop wave function~(\ref{dchi}) would thus be replaced
by $\sim 1/10^s$.

We leave it to the reader to estimate the relative cost of such improved kernels
compared to the standard Wilson kernel for fixed size of the residual
breaking of chiral symmetry.  Of course, the improvement
of the wave function predicted by perturbation theory would have to be
tested via nonperturbative numerical experimentation,
which, however, has not been done to date for these kernels.

\subsubsection{\label{WB} Hypercubic fermions}
An alternative strategy is to use an approximate solution
of the GW relation for the Wilson kernel.
This approach was proposed for both GW (or overlap) fermions
and for domain wall fermions \cite{Bietenholz:1998ut,Bietenholz:2006ni}.
It is easy to follow the logic of this approach in the case
of GW fermions.  Throughout, we will assume that all relevant operators
satisfy $\g_5$-hermiticity.

Given an exact or approximate GW operator $\tD$, one defines a new GW operator
by taking the kernel $H$ in Eq.~(\ref{DGWlim}) to be
\begin{equation}
\label{tH}
\tH = \g_5 (2\tD-1) \ .
\end{equation}
If $\tD$ is already an exact GW operator, it follows from the GW relation
together with $\g_5$-hermiticity that $\tH^2=1$.
Hence $\e(\tH)=\tH$, and using $\tH$ as the kernel
in Eq.~(\ref{DGWlim}) reproduces the original GW operator $\tD$.

If $\tD$ is only an approximate solution
of the GW relation then $\e(\tH)\ne\tH$.  However, for a good approximation
we still expect that $\tH^2\approx 1$.
Writing $\e(\tH) = \tH/|\tH| = \tH/(\tH^2)^{1/2}$ we then expect
that it should be easier to construct $(\tH^2)^{-1/2}$ numerically,
because the eigenvalues of $\tH$ should be closer to unity
in comparison with, say, the standard Wilson kernel.

For domain wall fermions, one expects that the convergence of $\te(N_5;H)$
to $\e(H)$ with increasing $N_5$ will be faster if in Eq.~(\ref{finites})
(and thus also in Eq.~(\ref{Dshamir})) one replaces the standard Wilson kernel
by an approximate GW operator $\tD$.  Equivalently, at fixed $N_5$,
the replacement of the standard Wilson kernel by an approximate GW operator
is expected to yield improved chiral behavior.

A particular family of approximate GW solutions goes under the name
of ``hypercubic fermions.'' One requires that the approximate GW operator
$\tD(x,y)$ be such that it couples two sites $x$ and $y$ only if there is
a unit hypercube to which both sites belong.  In comparison with
the overlap operator with the standard Wilson kernel,
the GW operator with a hypercubic fermion kernel given by Eq.~(\ref{tH})
is expected to have several advantages.  First, as already explained,
it should be easier
to achieve good numerical approximations of the sign function $\e(\tH)$.
In addition, one expects improved locality, rotational symmetry,
and scaling behavior.  The obvious price is that the hypercubic fermion
kernel $\tH$ connects each lattice site to $3^4$ neighboring sites
(itself included), which makes its application more costly.
For more details, including numerical tests,
see Refs.~\cite{Bietenholz:1998ut,Bietenholz:2006ni}.

\subsection{\label{dflt} Deflation}
The near-zero modes of the Wilson kernel
(or any other kernel for that matter) slow down the convergence
to the chiral limit of domain wall fermions.  For a GW operator,
near-zero modes of its kernel cause difficulties in the numerical evaluation of
the (operator) sign function $\e(H)$ in Eq.~(\ref{DGWlim}) or~(\ref{DGWm}).
Let $\hat\e(H)$ be some unspecified numerical approximation of $\e(H)$.
Typically, $\hat\e(\l)$ will provide a good approximation of
the sign function $\e(\l)$ when $\l$ lies within
some bounded interval $0 < \lmin \le |\l| \le \lmax$.
But when $|\l| < \lmin$ the approximation $\hat\e(\l)$ breaks down.\footnote{%
  See the chapter of the {\em LQCD@50} book on overlap fermions by DeGrand
  \cite{DeGrand:2025ttq}.
}

Of course, one could do better by lowering $\lmin$, so that the interval
where $\hat\e(\l)$ provides a good approximation reaches closer to zero.
But suppose that, for a typical gauge field configuration in an ensemble,
there are only a few kernel eigenvalues which are smaller
than some reasonable $\lmin$.  In that case, achieving a good approximation
of $\e(\l)$ for these near-zero eigenvalues by further lowering $\lmin$
can be very costly.  Deflation provides an alternative solution. In effect,
a small eigenvalue $\l<\lmin$ gets replaced by an $O(1)$ eigenvalue.
As long as the sign of $\l$ is preserved by the deflated eigenvalue,
$\e(\l)$ will be reproduced much more easily.

Deflation works as follows.  Starting from some kernel $H$
in Eq.~(\ref{DGWlim}), we replace $H$ by a deflated version
\begin{equation}
\label{dfltH}
\hH = H + \sum_{i=1}^n \left(\e(\l_i)-\l_i\right) \sket{i}\sbra{i}\ .
\end{equation}
Here $\sket{i}$ denotes an eigenvector of $H$ with eigenvalue $\l_i$.
The sum runs over the $n$ eigenvectors with smallest $|\l_i|$.
For simplicity we assumed that all the affected eigenvalues are deflated
to $\pm 1$.

In order to apply deflation to the overlap operator one first finds
the $n$ smallest eigenvalues of the Wilson kernel by some minimization method,
and constructs the projection of the sign function on this subspace,
$\sum_{i=1}^n \e(\l_i) \sket{i}\sbra{i}$.  Then effectively one needs to
approximate $\e(H_W)$ only on the orthogonal subspace, for which
a more comfortable value of $\lmin$ can be used.

The use of deflation for domain wall fermions was proposed
in Refs.~\cite{Edwards:2000qv,Jansen:2003jq}. For definiteness,
we follow Ref.~\cite{Jansen:2003jq}.  The domain wall operator is modified
in such a way that in the approximate GW operator at finite $N_5$, Eq.~(\ref{DHS}),
$H_S$ gets deflated according to Eq.~(\ref{dfltH}).  Now, an eigenvalue $\l_i$
yields in the approximate sign function of Eq.~(\ref{tildee}) a factor of
\begin{equation}
\label{approsl}
\te(N_5;\l_i)
= \frac{(1+\l_i)^{N_5}-(1-\l_i)^{N_5}}{(1+\l_i)^{N_5}+(1-\l_i)^{N_5}} \ .
\end{equation}
The effect of the deflated kernel~(\ref{dfltH}) is to trade $\te(N_5;\l_i)$
with $\te(N_5;\pm 1)$ for the relevant eigenvalues.
Clearly, if $|\l_i|\ll 1$ then the convergence of $\te(N_5;\pm 1)$
to $\e(\l_i)$ will be much faster than the convergence
of $\te(N_5;\l_i)$, thereby suppressing any chiral symmetry violating effects
from the corresponding eigenvector.
For small-scale numerical tests of these deflation methods
see Refs.~\cite{Edwards:2000qv,Jansen:2003jq}.

We recall (Sec.~\ref{kernel}) that both overlap and domain-wall fermions
live in the super-critical phase between the two rightmost
``fingers'' in \Fig{aokifig}, in which the mobility edge $\l_c$
is strictly positive.  At fixed value of the lattice spacing,
a larger $\l_c$ will yield a more local (exact or
approximate) GW operator.

When deflation is applied to the Wilson kernel of the overlap operator,
in principle this doesn't change the overlap operator itself;
it merely affects the numerical approximation used.
By contrast, modifying the kernel of domain wall fermions by deflation,
as described above, gives rise to a different domain-wall fermion operator
for any finite $N_5$.  However, as long as the mobility edge is $O(1)$,
all the deflated near-zero modes will be exponentially localized on the
lattice scale.  Hence, locality of the full domain-wall fermion operator,
and of the effective four-dimensional operator of Eq.~(\ref{DGWN}),
are both maintained.

\section{\label{epilogue} Epilogue}
When Kaplan introduced domain wall fermions \cite{Kaplan:1992bt}, his
original goal was to find a method to define anomaly-free chiral gauge theories
on the lattice -- a long-standing problem in lattice gauge theory.
However, it gradually became clear that the presence of
an opposite-chirality fermion on the ``far'' wall
represents a major obstacle for this program.

In the following years, important progress towards the construction of
chiral gauge theories was made by Narayanan and Neuberger
\cite{Narayanan:1993ss,Narayanan:1994gw},
but the seriousness of the open issues remained debated \cite{Golterman:1995gh}.

The next major step came after the invention of the overlap operator
by Neuberger \cite{Neuberger:1997fp}, and the realization that
the overlap operator satisfies the Ginsparg-Wilson relation,
which in turn leads to a modified form of chiral symmetry
on the lattice \cite{Luscher:1998pqa}.  Using these algebraic ingredients
as a starting point, L\"uscher
successfully constructed anomaly-free abelian chiral gauge theories, while
requiring one new algebraic constraint on the fermion spectrum beyond
the familiar anomaly-cancellation condition \cite{Luscher:1998du}.
As for nonabelian chiral gauge theories, he was able to define them
to all orders in lattice perturbation theory
\cite{Luscher:1999un,Luscher:2000zd}.
Going beyond lattice perturbation theory it was shown that
Witten's global anomaly is reproduced \cite{Bar:2000qa},
but it is unknown if additional non-perturbative obstructions
occur within this formulation.

Kaplan has continued to work on his original proposal with various
collaborators, aiming to find a way to extend the dynamical four-dimensional
gauge field into the fifth dimension in such a way that the
opposite-chirality fermion on the far wall would hopefully be decoupled
from the resulting four-dimensional theory.  This line of research
remains active to date \cite{Grabowska:2015qpk,Kaplan:2023pxd,Kaplan:2023pvd}.
In brief, the emerging picture is that this goal can be achieved
in the topologically trivial sector, if the topological charge is defined
in terms of the zero modes of the overlap operator (see Sec.~\ref{topo}).
For example, one can use gradient flow to extend the four-dimensional gauge
field from the boundary into the five-dimensional bulk, and for $N_5\to\infty$
the gauge field will die out before reaching the far wall.
However, the same is not true in topologically non-trivial sectors:
the chiral fermion on the far wall and/or the bulk
degrees of freedom cannot be fully decoupled
\cite{Grabowska:2015qpk,Golterman:2024ccm,Kaplan:2024ezz}.
Whether or not such a construction can account for the observed universe
remains to be seen.\footnote{%
  For other approaches to the construction of
  lattice chiral gauge theories see for example Ref.~\cite{Golterman:2025boq}
  and references therein.
}

As for QCD and similar vector-like theories, the presence of the two chiralities
is a welcome feature: quarks are Dirac fermions that contain
both a RH and a LH component.  Alongside with Wilson and staggered fermions,
which were introduced in the early days of lattice gauge theory,
domain wall fermions have become
one of the standard methods for large-scale lattice QCD calculations.

This introduction has focused on the formulation of domain wall fermions
and not on the physical results obtained with it.\footnote{
  We did not cover chiral perturbation theory for domain wall fermions.
  For references, see
  Ref.~\cite{Sharpe:2007yd,RBC-UKQCD:2008mhs,RBC:2010qam,RBC:2014ntl}.
  We also did not cover algorithms for ensemble generation.
  For an exact one-flavor algorithm for domain wall fermions,
  see Refs.~\cite{Chen:2014hyy,Jung:2017xef}.
}
The only observable
we have considered in detail is the residual mass, which serves as a measure
of the good (yet imperfect) chiral properties of domain wall fermions.
That a residual mass exists does not represent in itself a problem
because quarks are not massless, so long as the residual mass is smaller than
the (bare) mass required for the light quarks.

Since the first works~\cite{Blum:1996jf,Blum:1997mz} it was clear that
domain wall fermions are indispensable for light quark physics
where chiral symmetry (alongside flavor structure) is important.
This is particularly true
when operators protected by chiral symmetry in the continuum
mix with chiral symmetry-breaking operators in lattice calculations
using discretizations that explicitly break chiral symmetry.
For example, in studies of neutral kaon mixing and decay,
which are important for understanding CP violation in the Standard Model,
such artifacts are difficult or practically impossible to remove,
while calculations with domain wall fermions go through
like their continuum counterparts as far as mixing and renormalization
are concerned once the residual mass has been taken
into account~\cite{Aoki:2010pe,RBC:2020kdj}.
Power divergences, if present, are suppressed relative to Wilson fermions
by $O(a \mres)$.

As described in the last section, state of the art lattice QCD simulations
are now done with the M\"obius action, as well as
with the original version of domain wall fermions
and with Zolotarev domain-wall fermions, for a wide variety of
topics~\cite{RBC:2014ntl,RBC:2020kdj,RBC:2023pvn,RBC:2024fic},
including chiral symmetry restoration at
non-zero temperature~\cite{Cheng:2009be,Tomiya:2016jwr}
and heavy meson (quark) decays~\cite{Aoki:2023qpa,Flynn:2023nhi,Boyle:2017kli}.
So far, most computations using domain wall fermions for charm
and bottom quarks have been performed on 2+1 flavor ensembles,
so they are partially-quenched. These calculations have shown that
domain wall fermions can be used effectively for charm quark physics
if the lattice spacing is sufficiently small,
$a^{-1}\geqx 2$ GeV~\cite{Boyle:2017kli,Flynn:2023nhi,Bai:2023lkr}.
Dynamical simulations with 2+1+1 flavors of quarks, including charm
(and even bottom), with heavier-than-physical light quarks,
have begun~\cite{Chiu:2020tml, RDMLattice2024}
and have been used for important measurements like contributions to
the muon anomaly~\cite{RBC:2023pvn,RBC:2024fic}. More ensembles with dynamical charm
and physical masses will appear soon~\cite{RDMLattice2024}.
For calculations at non-zero temperature with four flavors
at the physical point using Zolotarev domain-wall fermions,
see Refs.~\cite{Chen:2022fid,Chiu:2023hnm}.

Overlap fermions have not been widely used
in dynamical simulations ($i.e.$, gauge field ensemble generation).
The reasons include their cost, as well as
the need for a complicated special algorithm in order to change
the topological charge in a dynamical overlap simulation.
We note an early dynamical overlap simulation by JLQCD which was restricted to
the topological charge $Q=0$ sector~\cite{JLQCD:2008jiv}.\footnote{
  See the companion chapter of the {\em LQCD@50} book
  by DeGrand \cite{DeGrand:2025ttq}.
}

Compared to lattice QCD simulations with Wilson or staggered fermions,
the obvious shortcoming of domain wall fermions is the added cost of the
extra fifth dimension.  However, experience suggests that domain wall fermions
often display smaller discretization effects for the same lattice spacing.
This improved scaling can offset much of their extra cost
since in general the cost of lattice calculations grows with a large power
of the inverse lattice spacing as the continuum limit is approached.
While it is difficult to carry out a systematic study of this question,
it is suggestive that the very good chiral symmetry of domain wall fermions
together with their simple flavor structure must be responsible for
this behavior.  For example, recent calculations of the hadronic contributions
to the muon anomalous magnetic moment appear to bear out
these expectations~\cite{RBC:2023pvn,RBC:2024fic,Aubin:2022hgm,Ce:2022kxy,FermilabLatticeHPQCD:2024ppc}.

\mynnnext
{\bf Acknowledgments.}
We thank Maarten Golterman for extensive discussions and comments.
We also thank Ting-Wai Chiu, Tom DeGrand and David Kaplan for comments.
TB is partially supported by the US Department of Energy under grant DE-SC0010339.
YS is supported by the Israel Science Foundation under grant no.~1429/21.

\bibliographystyle{jhep}  
\bibliography{dwf.bib}

\end{document}